\RequirePackage{fix-cm}
\documentclass[smallextended]{svjour3}       % onecolumn (second format)
\smartqed  % flush right qed marks, e.g. at end of proof

\usepackage[utf8]{inputenc}

\usepackage{graphicx}
\usepackage{amsmath}
\usepackage{amsfonts}
\usepackage{pifont}

\usepackage[version=4]{mhchem}
\usepackage{siunitx}
\usepackage{longtable,tabularx}
\usepackage{subfig}
\usepackage{float}
\usepackage{gensymb}
\usepackage{subfig}

\usepackage{hyperref}
\usepackage[numbers,sort,compress]{natbib}

\usepackage{color}

\usepackage{algorithm}
\usepackage[noend]{algpseudocode}

\newcommand{\hh}{\mathcal{H}_2}
\newcommand{\ghh}{\mathcal{H}_{2'}}

\newcommand{\defi}{:=} 
\newcommand{\idd}{\mathrm{d}}
\newcommand{\expm}[1]{\mathrm{e}^{#1}}
\newcommand{\inv}{^{-1}}
\newcommand{\trans}{^\mathsf{T}}
\newcommand{\trace}{\mathrm{trace}}
\renewcommand{\Re}{\mathbb{R}}

\usepackage{setspace}
\journalname{Theor. Comput. Fluid Dyn.}

\begin{document}
\title{Data-Driven Selection of Actuators for Optimal Control of Airfoil Separation}
%\subtitle{Do you have a subtitle?} % Insert a subtitle or remove this line
%
\titlerunning{Data-Driven Selection of Actuators for Optimal Control of Airfoil Separation}

\author{Debraj Bhattacharjee, Bjoern Klose, Gustaaf B. Jacobs, and Maziar S. Hemati}

\authorrunning{D. Bhattacharjee, B. Klose, G.B. Jacobs, and M.S. Hemati} % if too long for running head

\institute{D. Bhattacharjee \at
              Electrical and Computer Engineering, University of Minnesota, Minneapolis, MN 55455
              \and
              B. Klose and G.B. Jacobs \at
              Aerospace Engineering, San Diego State University, San Diego, CA 92182
                            \and
              M.~S. Hemati \at
              Aerospace Engineering and Mechanics, University of Minnesota, Minneapolis, MN 55455\\
              Tel.: +612-625-6857\\
              Fax: +612-626-1558\\
              \email{mhemati@umn.edu}
}

\date{Received date and accepted date}

\maketitle

%\title{Optimal Actuator Selection for\\ Airfoil Separation Control }

% \author{Debraj Bhattacharjee%
%     \thanks{Graduate Student, Electrical and Computer Engineering, University of Minnesota, AIAA Member.}
% 	\ and    Maziar S. Hemati%
%    \thanks{Assistant Professor, Aerospace Engineering and Mechanics, University of Minnesota, AIAA Senior Member.}
%     \\
%   {\normalsize\itshape
%   University of Minnesota, Minneapolis, MN 55455, USA} \\
%  Bjoern Klose%
% 	\thanks{Graduate Student, Department of Aerospace Engineering, San Diego State University.}
% 	\ and Gustaaf B. Jacobs%
% 	\thanks{Professor, Department of Aerospace Engineering, San Diego State University, AIAA Associate Fellow.}\\
%   {\normalsize\itshape
% San Diego State University, San Diego, CA 92182, USA}
%}
%\affil{San Diego State University, San Diego, CA 92182, USA}

% \begin{document}

% \maketitle

\abstract{
%   Flow separation can degrade aerodynamic performance in many engineering systems, including aircraft and wind turbines.
%   %
%   This has motivated the development of various active flow control strategies for
% separation mitigation.
% % Active control strategies such as open-loop forcing using synthetic jets have been relatively prevalent. A large number of these techniques use trial and error based tuning to force the flow at a favorable frequency. 
% %
% % Other studies have focused on understanding the physics of the flow to identify specific frequencies which result in flow reattachment.  
% While notable performance improvement can be achieved with active flow control, this improvement may not correspond to the very best performance that is achievable.
% % to the optimal performance of the fluid system, since the actuator positions have largely remained fixed. 
% %
% Indeed, most studies focus on optimizing the control action for a given actuator configuration; whereas, actuator placement is intimately tied to achievable performance.
%
%  In this paper,
We present a systematic approach for determining the 
optimal actuator location for separation control
from input-output response data, 
gathered from numerical simulations or physical experiments.
%
%\msh{discuss ingredients of the algorithm: ERA, state-space, minimizing energy, etc.}
%
The Eigensystem Realization Algorithm is used to extract state-space descriptions from
the response data associated with a candidate set of actuator locations.
These system realizations are then used to determine the actuator location among the set
that can drive the system output to an arbitrary value with minimal control effort.
The solution of the corresponding minimum energy optimal control problem
is evaluated by computing the generalized output controllability Gramian.
We use the method to analyze high-fidelity numerical simulation data of the lift and separation-angle responses to a pulse of localized body-force actuation from six distinct locations on the upper surface of a NACA 65(1)-412 airfoil.
% We use the method to analyze high-fidelity numerical simulation data of a NACA 65(1)-412 airfoil. to determine the optimal location for applying localized body-force actuation a NACA 65(1)-412 airfoil
% for controlling lift and separation angle.
%
We find that the optimal location for controlling lift is different from the optimal location for controlling separation angle.
In order to explain the physical mechanisms underlying these differences,
we conduct controllability analyses of the flowfield by leveraging the dynamic mode decomposition with control algorithm.
These modal analyses of flowfield response data reveal that excitation of coherent structures in the wake benefit lift control; whereas, excitation of coherent structures in the shear layer benefit separation-angle control.
%
%These results can provide guidance for designing actuation and control strategies in the future.
% \msh{rewrite these last two sentences. need to be more specific with the conclusions.}
% Modal analyses of velocity-field data from the same numerical simulations are conducted to acqurie  guidance  on  how  actuation  effects  the  physics  of  the  flow  at  the optimal locations and provide intuition in designing control strategies.
% %
% The results obtained in this study indicate that optimality depends on the definition of the final objective. 
%
% \msh{the remaining sentences require some work\dots }
%  A modal analysis is conducted on the velocity field-data obtained from these numerical simulations to identify dominant modes in response to the actuating pulse. It is observed that the dominant modes, contributing to the reduction in flow separation, have connections with the separated shear layer as well a line that separates the fluid, also known as the separation line. The results presented here combined with a knowledge of the dominant modes of the system provide a foundation to devise a method of identifying the actuator location in the most optimal sense. 
}

%\section*{Nomenclature}
%
%\noindent(Nomenclature entries should have the units identified)
%
%{\renewcommand\arraystretch{1.0}
%\noindent\begin{longtable*}{@{}l @{\quad=\quad} l@{}}
%$A$  & amplitude of oscillation \\
%$a$ &    cylinder diameter \\
%$C_p$& pressure coefficient \\
%$Cx$ & force coefficient in the \textit{x} direction \\
%$Cy$ & force coefficient in the \textit{y} direction \\
%c   & chord \\
%d$t$ & time step \\
%$Fx$ & $X$ component of the resultant pressure force acting on the vehicle \\
%$Fy$ & $Y$ component of the resultant pressure force acting on the vehicle \\
%$f, g$   & generic functions \\
%$h$  & height \\
%$i$  & time index during navigation \\
%$j$  & waypoint index \\
%$K$  & trailing-edge (TE) nondimensional angular deflection rate\\
%$\Theta$ & boundary-layer momentum thickness\\
%$\rho$ & density\\
%\multicolumn{2}{@{}l}{Subscripts}\\
%cg & center of gravity\\
%$G$ & generator body\\
%iso	& waypoint index
%\end{longtable*}}

\section{Introduction}
Flow separation can degrade performance in many engineering systems,  through reduced lift, increased drag, and decreased efficiency.
%
% An improvement in the performance of an airfoil can be observed through flow reattachment or reduction of the 	separation bubble.
% %
% Therefore, 
% In an effort
To alleviate the effects of 
flow separation on aerodynamic performance,
active flow control has been considered since the inception of the field of aerodynamics~\cite{gadelhakBook,williams2009}.
% various active flow control strategies 
% have garnered considerable attention in the 
% recent past.
%~\cite{hemati2016improving}.
%
%In particular,

% Recently, 
Open-loop flow control strategies
based on various actuator technologies~\cite{cattafesta2011}---such as plasma
actuators~\cite{plasma_1_2007,little2010high,mabe2009}, fluidic oscillators~\cite{cerretelli2009,gregory2012,woszildo2011,seo2018,ostermann2018}, and synthetic jets~\cite{glezer1998,glezer2002,hemati2016improving,deem2018,seo2018effect}---have been shown to effectively alter separated flows, and in some cases to even
% result in
yield complete reattachment.
%
%An excellent review of actuators used for active flow control can be found in~\cite{cattafesta2011}.
%
% Among such schemes, open-loop control strategies have been most prevalent, owing to the relative simplicity in the implementation of such methods compared to closed-loop control strategies.
%
% In particular, Zero Net Mass Flux (ZNMF) actuators, examples of which include plasma actuators \cite{plasma_1_2007,little2010high} and  synthetic jets \cite{jeong_th,glezer1998,glezer2002}, have indicated that the fluid flow can be altered in a desirable manner by forcing the flow at a favorable frequency, usually tuned by trial and error techniques. An instance of this can be found in \cite{seo2018effect}, where the effect of  synthetic jet frequency modulation schemes  and their effect in reducing flow separation was studied. It was shown that the proposed amplitude modulation scheme imparted more energy to the frequency targeted for separation control than other modulation schemes, thus, resulting in better control of flow separation.
%
%However, the configuration of controllers formulated in the aforementioned methods rely overly on physical insights of the system and may not be robust to disturbances and variations in model paramters \cite{deem2017}.  
%Separation control technologies have benefited from an extensive set of parametric studies~\cite{seifert1996,seifert1999,glezer2005}.
%
In~\cite{seifert1993},
oscillatory forcing 
was found to improve control authority
for separation control.
Several studies have observed that actuating at the dominant shear layer frequency is effective for mitigating flow separation~\cite{yarusevych2006,postl2011,marxen2015,yarusevych2017}.
%
%, other studies have found that the most effective actuation frequency for separation mitigation varies with actuator location.
%
%For example, the study in~\cite{raju2008}
Other studies have reported that separation mitigation is most effective when actuation is applied at the separation bubble frequency, not the shear layer frequency~\cite{raju2008}.
Further, it was shown that nonlinear flow interactions can result in lock-on effects that influence the optimal forcing frequency~\cite{mittal2006resonant,mittal2005numerical}.
%Several studies have endeavored to understand the nonlinear dynamics associated with a fluid flow in an attempt to be able to formulate more effective control strategies~\cite{mittal2006resonant,mittal2005numerical}.  
%
% Although such methods provide useful insight about nonlinear interactions in flows, such as lock-on between frequencies, a fundamental limitation is that the results obtained using these methods are a function of the probe locations \cite{tu2011koopman}. % MSH: this line seems irrelevant to our actuator selection work
%

Recent investigations have sought to identify candidate actuation frequencies more objectively
% from global descriptions of the separation dynamics
using operator-based and data-driven modal analysis techniques---such as linear stability analysis, resolvent analysis, and dynamic mode decomposition~(DMD)~\cite{marxen2015,yehJFM2019,taira2019,hemati2016improving,deem2017,hemati2017biasing}.
Actuation designed based on these analyses yielded improved open-loop controller designs;
however,
%through actuator forcing at specific
%frequencies associated with dynamically important modes of the flow.
%
% While these studies provide foundations for
% methods that reduce flow separation,
the actuator positions considered were fixed and may not necessarily translate to
the optimal performance achievable in terms of separation control.

The positioning of actuators and sensors is known to play a central role
in determining achievable control performance.
In most scenarios, using all available actuators and sensors
will yield the highest performance for a given system.
However, through judicious selection and placement,
it is possible to achieve optimal control
performance using fewer actuators and sensors.
To this end, systems theoretic optimization approaches for sensor and
actuator placement have been proposed in a number of studies.
The effect of white noise disturbances on actuator and sensor
placement for the Ginzburg-Landau system was investigated in~\cite{chen2011h},
where numerical optimization was used to minimize
the actuator effort and perturbation magnitude in an $\mathcal{H}_2$ sense.
In~\cite{chanekar2017optimal},  a branch-and-bound procedure was proposed to
determine the optimal actuator placement with constraints on the number of actuators.
Further, sensor selection for flow reconstruction has been considered in~\cite{manohar2018,clark2018},
and for feedback flow control in~\cite{yao2019}.

Despite all of these advances,
the optimal selection problem has remained relatively
unexplored within the context of separation control.
Placing the actuators in locations that 
are intuitively optimal~\cite{jeong_th,torres_th}
may not be optimal for control performance.
%
% A systematic and principled approach for determining the optimal actuator
% location for separation control would be beneficial.
% %
%
%
Therefore, the selection from a set of
candidate locations using systematic criteria may assist in identifying the actuator
location with the highest performance index.
Further, such an approach would ensure that the resulting selection would be
feasible in practice---as the candidate set would be constructed to adhere to
physical and economic constraints on the type and placement of actuators. 

 %
  %
% To accomplish the objective of optimal actuator placement, one would, also, require a thorough understanding of phenomena in non-linear fluid flows such as flow separation. 
%%
% While some strategies have leveraged knowledge of unsteady flow physics to limit flow separation, 
% the influence of actuator and sensor placement on separation control has remained relatively unexplored.
%
In separation control, actuator placement
is usually strongly correlated with the location of the separation point.
In steady flows, this location of flow separation from a no-slip wall is 
well-known to be identified exactly by Prandtl’s condition for separation in the Eularian 
frame, through a point of zero skin friction and a negative friction gradient in the wall-tangential direction. 
However, flow separation from a no-slip wall can also be considered in the Lagrangian frame by understanding fluid tracers breaking away from a wall.
While much work has been done on unsteady separation (see~\cite{simpsonAnnRev2001}),
only recently was it shown that
the dynamics of unsteady flow separation are better 
analyzed in a Lagrangian frame, wherein the Lagrangian separation point is fixed for 
a periodic flow~\cite{haller2004exact}.
In \cite{haller2004exact}, it was shown that the time-dependent \emph{separation angle} $\theta(t)$ of the Lagrangian unstable manifold can be computed using pressure and skin-friction data.
In \cite{haller2004exact} and \cite{klose2019kinematics}, it was further shown that particles near a separation point are drawn towards 
an unstable manifold---i.e,~an attracting line in the flowfield.
%an attracting material line that constitutes the separation line.

% Such attractors can also be identified in the flow field (away from the boundary) by extracting ridges in the Finite-Time Lyapunov Exponent (FTLE) field. The FTLE field identifies the local contraction or expansion of the field over a finite time. A material line or Lagrangian Coherent Structure~(LCS) can then be found to exist wherever 
% the contraction is maximum (i.e.~a ridge)~\cite{haller2001distinguished,shadden2005definition}. 
% %
% %

% In separation control applications, the placement of control devices is often guided by several factors, 
% including separation point location and free-stream conditions.
%
%  Several studies suggest placing the controller in locations that 
% are intuitively optimal~\cite{jeong_th,torres_th}.
% %
% However, such arrangements may not be feasible in practice due to physical, computational, and/or economic constraints~\cite{dhingra2014admm}.
% %
% Therefore, the selection of a subset of these actuator locations through a set of systematic criteria may assist in identifying the actuator location with the highest performance index.
%
In \cite{kamphuis2018pulse}, the separation angle  and lift response was recorded to a flow pulse
% This was done
at six candidate actuator locations.
%
% A linear approximation of the separation line can then be computed
% by utilizing the separation angle and the separation point. 
 %
 It was
%  further
demonstrated that an increase in separation angle leads the
% material
separation line to become concave, for any location upstream of the separation point.
In turn, the concavity of the
% material
separation line results in flow reattachment.
In contrast, a decrease in separation angle results in an increase in the separation region. The increase in separation angle coincided with an increase in lift and a reduction in drag. Thus, a pulse location yielding 
a greater increase in lift corresponded to a greater degree of reattachment, 
providing guidance on actuator selection for separation control.
These qualitative analyses on actuator selection would
benefit from a systematic and quantitative approach grounded
in optimal control theory.
Further, a purely data-driven approach would ensure
that the actuator selection method can be applied within the context of both numerical simulations and physical experiments.

% as was the case in the heuristic data analysis of~\cite{kamphuis2018pulse}.

% The rich literature in this field and the unique nature of the problem, wherein the only available information is data obtained from simulations, provides motivation to undertake a optimal actuator selection study for separation control and obtain maximum achievable performance by leveraging knowledge of traditional techniques in separation control. Of course, the definition of optimality would depend on the objective of the formulation and a relevant measure would have to be decided accordingly.

In this paper, we present a data-driven technique for determining the optimal actuator location for driving a quantity of interest (e.g.,~lift or separation angle) with minimal control effort.
The only requirement for the approach is a collection of input-output response data associated with a set of candidate actuator locations, making the approach attractive for both numerical and experimental studies.
The Eigensystem Realization Algorithm~(ERA)~\cite{juang1985eigensystem} is used to extract a system model that describes the dynamic response data.
This description is then used to solve a minimum energy optimal control problem, which yields an
objective measure for comparing the relative performance of each
actuator location
% for
in controlling the quantity of interest.
The specific measure we propose is based on the generalized output controllability Gramian, which is valid for both stable and unstable systems.
This 
% fact 
makes for a versatile approach that can be applied to general systems.

\begin{figure}[ht!]
\begin{center}
% 
     % \subfloat [NACA 65(1)-412 airfoil. The positions $x/c = \{.1, .2, .3, .4, .5, .6\}$, on the upper surface of the airfoil correspond to the various actuator locations. These are indicated by the red arrows.]{
     % \begin{minipage}{.5\textwidth} 
     % \centering
     % \includegraphics[width=\linewidth]{act_fig/airfoil}
     % \end{minipage}}  
     %     \subfloat [Lift and Separation Angle Responses]{
     % \begin{minipage}{.5\textwidth} 
     % \centering
     % \includegraphics[width=\linewidth]
     % {plots/responsesAllLocations/liftResponse.pdf}
     % \vfill
     % \includegraphics[width=\linewidth]
     % {plots/responsesAllLocations/sepAngResponse.pdf}
     % \end{minipage}}
  \includegraphics[width=.9\linewidth]{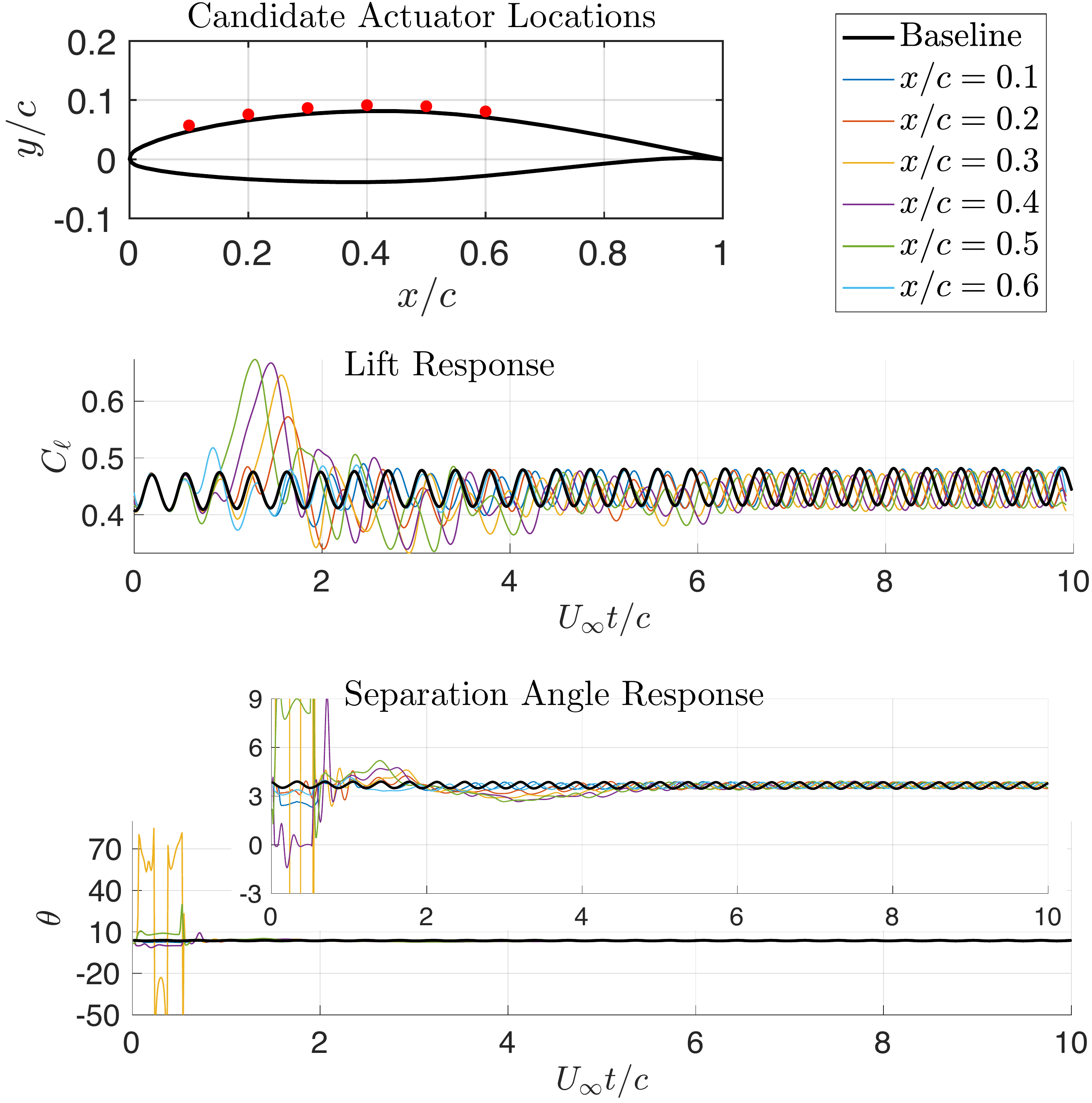}
\caption{ Lift and separation angle response data due to a pulse of localized body-force actuation at each of six candidate locations on a NACA 65(1)-412 airfoil.  High-fidelity numerical simulation data courtesy of~\cite{kamphuis2018pulse}.}
\label{fig:naca_airfoil}
\end{center}
\end{figure}

The optimal actuator selection method is applied to high-fidelity numerical data from~\cite{kamphuis2018pulse}, corresponding to the lift $C_\ell(t)$ and separation angle $\theta(t)$ responses due to a pulse of localized body-force actuation at six distinct locations on the upper-surface of a NACA 65(1)-412 airfoil with angle of attack $\alpha=\ang{4}$ and chord-based Reynolds number $Re_c=20,000$ (see Figure~\ref{fig:naca_airfoil}).
%
%The associated flowfield for the uncontrolled baseline and the pulse-response for actuator location $x/c=.4$ are shown in Figures \ref{fig:numsol_base} and \ref{fig:numsol_act4}, respectively.
%
The optimal actuator locations for controlling lift and separation angle are found to be different.
As such, we introduce a DMD-based controllability analysis to identify flow structures
that are most sensitive to the actuation.
This analysis sheds light on physical mechanisms that explain these differences in the actuator selection results.
The results suggest that the separation angle can be controlled more easily than lift,
provided that actuation is applied at the optimal location.

The paper is organized as follows:
In Section~\ref{sec:methods}, we present the optimality measure and necessary mathematical
machinery for conducting a data-driven analysis.
%the methodology for determining optimality among candidate locations and how such strategies generalize for alternative system classes.
In Section~\ref{sec:results}, the method is applied to analyze the data in Figure~\ref{fig:naca_airfoil} to determine the optimal actuator location for controlling lift and separation angle on a NACA 65(1)-412 airfoil.
We also introduce and use a DMD-based controllability analysis of the flowfield response to identify physical mechanisms that can explain the actuator selection results.
Conclusions are presented in Section~\ref{sec:conclusion}.

%

% \begin{figure}[ht!]
% \centering
% \includegraphics[width=0.6\columnwidth]{act_fig/airfoil}
% \caption{NACA 65(1)-412 airfoil. The positions $x/c = \{.1, .2, .3, .4, .5, .6\}$, on the upper surface of the airfoil correspond to the various actuator locations. These are indicated by the red arrows. \mshdb{update this figure to include two sub-plots (on top of one another and to the right of this airfoil plot), one of lift response and one of separation angle response. Each subplot should include the unforced response and the pulse response from each of the 6 actuator locations (labeled as ``Baseline'' and ``x/c=0.1'' etc. in the legend). Uncontrolled response should be in solid black. Pulse responses from each location should be in different shades of red (dark to light, starting with x/c=0.1).}
% %\msh{Put a marker at each of these locations?  Also be clear to indicate that the actuators are on the upper surface of the airfoil.}
% }
% \label{fig:naca_airfoil}
% \end{figure}

%% Figure as described above

% \begin{figure}[h!]
% \begin{center}
% \includegraphics[width=.8\textwidth]{act_fig/baseline}
% \caption{ NACA 65(1)-412,  $\alpha=4^\circ $,  $Re_c=20,000$, Vorticity field at $U_\infty t/c = 0.60$  for the baseline case. \mshdb{is there a way to incorporate the all of these flowfield plots into the output response plots above?  it also seems more relevant to show the $x/c=0.2$ and 0.3 cases instead of 0.4, since those turn out to be optimal.}}
% \label{fig:numsol_base}
% \end{center}
% \end{figure}

% \begin{figure}[h!]
% \begin{center}
% %\vspace{-1cm}
% \includegraphics[width=.8\textwidth]{act_fig/act_4}
% \caption{ NACA 65(1)-412, $\alpha=4^\circ $,  $Re_c=20,000$, Vorticity field at (a) $U_\infty t/c = 1.4$, (b) $U_\infty t/c = 2.0$ and (c) $U_\infty t/c = 2.4$ for the case with a pulse actuation at $x/c=0.4$. }
% \label{fig:numsol_act4}
% \end{center}
% \end{figure}

\section{Methodology}
\label{sec:methods}
Consider a finite-dimensional state-space realization $G=(A,B,C)$ representing the dynamic response from a single actuator input $u(t)$ to a single output quantity of interest $y(t)$:
\begin{equation}
\begin{gathered}
	\dot{x}(t) = Ax(t) + Bu(t)\\
    y(t) = Cx(t).
    \end{gathered}
    \label{eqn:ct_lti}
  \end{equation}  
  Here, $x\in\Re^n$ is the $n$-dimensional state vector.
  We assume that $G=(A,B,C)$ is minimal (i.e.,~it is both controllable and observable).
  It will be shown how such a realization can be determined from data in Section~\ref{sec:minreal}.
  %, $u(t)\in\Re^p$ is the input vector, and $y(t)\in\Re^q$ is the output vector.
%
% A control law $u^{\text{opt}}(t)$ that 
% alters lift $C_\ell(t)$ or separation angle $\theta(t)$ with minimum control effort is closely related
% to the problem of driving the system state to an arbitrary point in state-space using minimum control energy~\cite{Toscano2013}; 
% thus,
  
We seek the control input $u^{\text{opt}}(t)$ that drives the system state from the origin to an arbitrary point in state-space with minimal control energy over an infinite time-horizon\footnote{Although finite time-horizons can be considered, we choose to focus on the infinite time-horizon case in order to maintain objectivity in the optimality measure; the solution to the finite time-horizon problem is dependent on the final time, which can be undesirable because the final time can always be chosen to influence the outcome of the optimality measure.}.
%
%(Note: this is equivalent to driving the system from the origin to an arbitrary state with minimal control energy.)
%
%
%
This optimal control problem can be solved by standard methods and is commonly referred to as the \emph{minimum control energy problem}~\cite{Toscano2013}:
\begin{equation}
\begin{gathered}
\text{minimize}\qquad  J = \int_{0}^{\infty} u\trans(\tau)u(\tau)\idd\tau\\
\text{subject to}\qquad  
\dot{x}(t) = Ax(t) + Bu(t)\\
\hspace{0.25cm} x(0) = 0\\
\hspace{0.65cm} x(\infty) = x_f,
\end{gathered}
\end{equation}
which admits a solution if the system is controllable.
The minimal input energy associated with the optimal control is given by,
%The minimum control energy associated with this optimal control law follows,
\begin{equation}
J^\text{opt} = x_f\trans W_{c}\inv x_f,
\label{eq:optctrb}
\end{equation}
where the controllability Gramian
\begin{equation}
W_{c}\defi\int_0^\infty \expm{A\tau}BB\trans\expm{A\trans \tau}\idd \tau
\end{equation} 
is the stabilizing solution to the Lyapunov equation,
\begin{equation}
AW_c + W_cA\trans + BB\trans = 0
\end{equation}

To determine the actuator location that yields the minimum control energy, 
we can simply compare the relative sizes of $W_{c}$ corresponding to
the dynamics of each actuator location---a larger $W_c$ being more
controllable and requiring less input energy to control.
Note that the controllability Gramian $W_c$ is not invariant under similarity transformation.
This is an important point to consider when system realizations $G=(A,B,C)$ are obtained from data,
as will be discussed in Section~\ref{sec:minreal}.
In such instances, care must be taken when formulating measures of optimality directly based on $W_c$.
Some suitable choices that are invariant under similarity transformation are, e.g.,~$\det(W_{c}), \trace(W_c)$.

To gain an intuition for the optimal solution, we can view the quadratic form in~\eqref{eq:optctrb} 
as defining an ellipse  that contains all points in state-space that can be \emph{reached 
from the origin} using no greater than unit input energy, $X = \{x_f\in\Re^n\,|\,x_f\trans W_{c}\inv x_f\le1\}$.
%
%(By the equivalence of controllability and reachability for continuous-time systems, this ellipse equivalently contains all of the states that can be \emph{reached from the origin} with unit energy or less.)
%
The most controllable directions in state-space require the least control energy to traverse and are related to the eigendirections associated with 
the largest eigenvalues of $W_{c}$; the least controllable directions in state-space require 
the most control energy to traverse and are related to the eigendirections associated 
with the smallest eigenvalues of $W_{c}$.
%
%

% The goal here is to have maximum controllability in terms of a given quantity of interest (e.g.,~lift or separation angle).
Although, $W_c$ provides intuition about the most controllable directions in state-space, in practice, the quantity of interest may not directly correspond to these states; instead, the quantity of interest corresponds to a specific linear combination of these states: $y(t) = Cx(t)$.
Hence, rather than considering the state controllability Gramian $W_c$ directly, 
we can instead work with a suitably weighted version of $W_c$,
\begin{eqnarray}
W_{oc}&\defi&\int_0^\infty C\expm{A\tau}BB\trans\expm{A\trans \tau}C\trans\idd \tau\\
&=& CW_{c}C\trans,\label{eq:outputctrb}
\end{eqnarray} 
which is simply the \emph{output controllability Gramian}~\cite{kreindler1964}.
Output controllability is a more natural measure of optimality because it is invariant under similarity transformations, and thus constitutes a system property that is coordinate independent.
%
% Another advantage of using the output controllability Gramian over the state controllability Gramian is that it is invariant to the state-space realization, making it suitable for use with realizations determined from, e.g., impulse response data by the eigensystem realization algorithm---a consideration that will be essential to our approach, as will be described later.
%
%
This choice is particularly appealing because measures based on $W_{oc}$ admit numerous other interpretations, 
beyond those afforded by the minimum control energy perspective.
For instance, the output controllability Gramian is directly related to the $\hh$-norm of a stable LTI system as,
\begin{equation}
 \|g(t)\|_2 = \sqrt{\int_0^\infty g(t)\trans g(t)\idd t}=\sqrt{W_{oc}}\label{eq:h2trace}
\end{equation}
where $g(t)\defi C\expm{At}B$ is the impulse response.
Further, we can arrive at a frequency-domain interpretation of this measure by invoking Parseval's theorem~\cite{skog2005},
\begin{eqnarray}
  \|g(t)\|_2 = \|G(s)\|_2 &\defi& \sqrt{\frac{1}{2\pi}\int_{-\infty}^\infty(G(-j\omega)\trans G(j\omega))\idd\omega}\\
  &=& \sqrt{\frac{1}{2\pi}\int_{-\infty}^\infty|G(j\omega)|^2\idd\omega}
\end{eqnarray}
where $G(s)$ denotes the transfer function from the input to the output.
Hence, the {$\hh$-norm} can be interpreted as the average system gain 
over all forcing frequencies.
Consistent with the minimum control energy interpretation, 
this indicates that a system with a larger $\hh$-norm will 
tend to yield a larger output for the same input signal.
The {$\hh$-norm} also admits a stochastic interpretation from the lens of linear quadratic Gaussian~(LQG) control~\cite{skog2005}:~all else equal, a system with a larger {$\hh$-norm} will yield a larger output power
in response to a unit intensity white noise input.
%
% Based on all these interpretations, it should be evident that 
% output controllability and the {$\hh$-norm} can provide an indication of the effectiveness of an actuator location 
% in influencing the system output with an arbitrary input.

\subsection{Generalizibilty to unstable systems }
\label{sec:unstable}
% At this point, it is worth discussing generalizations of the above 
The $\hh$ optimality measure can be generalized to
unstable systems.
This generalization is useful if we are interested in comparing actuator locations for general systems, 
which may or may not be stable.
Of course, in the context of unstable systems, neither the state controllability Gramian 
nor the output controllability Gramian will necessarily be bounded;
thus, these optimality measures are ill-suited for 
comparing general flow control configurations that may exhibit
unstable dynamics. 
However, by taking a frequency-domain perspective of the state controllability Gramian, 
we can arrive at a \emph{generalized controllability Gramian} $P$ that is bounded for unstable systems~\cite{zhou1999}:
\begin{equation}
P = \frac{1}{2\pi}\int_{-\infty}^\infty\left(j\omega I-A\right)\inv BB\trans \left(-j\omega I-A\trans\right)\inv\idd \omega
\end{equation}
The generalized controllability Gramian is also related to the minimum control energy problem, as shown in Theorem~5 of~\cite{zhou1999}.
Specifically, when the system under consideration is controllable,
{$x_o\trans P\inv x_o = \inf\{\|u\|_2^2\,|\,x(0)=x_o,\,x(-\infty)=0,\,x(\infty)=0\}$}.
As with $W_c$, a larger $P$ indicates that less control energy is required to drive the state to the origin from an arbitrary initial state
(i.e.,~the system is ``more controllable'').
In other words, the generalized controllability Gramian $P$ has an equivalent interpretation as the conventional
controllability Gramian $W_c$, but extends the interpretation to the context of unstable systems.
Indeed, when the system under consideration is stable, the generalized controllability Gramian is equivalent to the standard controllability Gramian (i.e.,~$P=W_c$).

Conveniently, for a stabilizable and detectable system, the generalized controllability Gramian~$P$ can be 
computed directly from a state-space realization of the system.
The procedure follows directly from Theorem~2 in Zhou et al.~\cite{zhou1999}, which amounts to solving for the stabilizing solution~$X$ to the algebraic Riccati equation,
\begin{eqnarray}
	XA + A\trans X - XBB\trans X &=& 0
        % AY + YA\trans - YC\trans CY &=& 0.
\label{eq:gengram1}
\end{eqnarray}
followed by a computation of the generalized controllability Gramian $P$ as the solution to the Lyapunov equation, % and $Q$ are solutions to
\begin{eqnarray}
(A+BF)P + P(A+BF)\trans + BB\trans &=& 0,
% Q(A+LC) + (A+LC)\trans Q + C\trans C &=& 0.
\label{eq:gengram2}
\end{eqnarray}
where $F=-B\trans X$.
For stable systems,  $X=0$ and, therefore, $P=W_c$.

For the purpose of determining a measure of 
optimality for actuator placement, 
here we will define the the \emph{generalized $\hh$-norm} (denoted $\ghh$) in analogy 
with Eq.~\eqref{eq:h2trace}, but now using the notion of generalized output controllability 
~$CPC\trans$ instead of the conventional output controllability $CW_cC\trans$.
\begin{equation}
  \|G\|_{2'} = \sqrt{CPC\trans}
  \label{eq:genh2trace}
\end{equation}
This measure is related to the output controllability of the system and is often times more desirable, as the end goal is to effectively control the output. Another attractive feature of this measure is that it is invariant to system realizations and is therefore not dependent on the method in which system realizations are obtained.

In the remainder of this paper, $ \|G\|_{2'} $ will be used as a measure for determining the optimal actuator location among a set of candidate actuator locations. In our case, this measure is computed for all the candidate locations using the minimal realization obtained from pulse response data, as will be described in the next subsection.

\subsection{Minimal realizations from pulse response data}
\label{sec:minreal}
An imperative step in determining optimality among the candidate set of actuator locations is obtaining mathematical models for the dynamic response from actuator input~$u(t)$ to the quantity of interest~$y(t)$ for each candidate configuration. Once such system models are obtained, analyses corresponding to optimality can be conducted.  
 The field of system identification deals with obtaining mathematical models for a system based on data observations obtained from the system. 
 In general, such data is usually sampled at discrete instants of time in a large variety of applications. Hence, discrete-time system models show higher suitability for system identification methods. Identified models can be transformed subsequently to continuous-time as needed for further analysis.
 Here, we describe one such method for determining a minimal discrete-time system realization from empirical pulse response data.
These discrete-time state-space realizations are
 then converted to continuous-time realizations---in 
 the form of~(\ref{eqn:ct_lti})---by means of Tustin's approximation~\cite{astrom1984}.

Consider the discrete-time state-space realization $\hat{G}=(\hat{A},\hat{B},\hat{C})$:
\begin{equation}
\begin{gathered}
x(k+1)=\hat{A}x(k)+\hat{B}u(k)\\
y(k)=\hat{C}x(k) \label{eqn:dt_lti}
\end{gathered}
\end{equation}
where $x\in\Re^n$ is the state vector, $u\in\Re$ is the actuator input,
$y\in\Re$ is the output quantity of interest, and $k\in\mathbb{Z}$ is the sampling time index. %and $\hat{A}$, $\hat{B}$, $\hat{C}$, and $\hat{D}$ are the discrete-time state-space matrices.
The response of the quantity of interest to a pulse input yields a sequence of scalar Markov parameters,
\begin{align}
h_k&=\left\{\begin{array}{lr}0 &\text{ for } k=0\\ \hat{C}\hat{A}^{k-1}\hat{B} & \text{ for } k\ge1.\end{array}\right. 
\end{align}
% and we have taken the direct feedthrough term to be zero.
%
%Here, we assume the direct feedthrough term $\hat{D}=h_0=0$.
%
% Pulse response data of a discrete-time system is the response of the system subject to a unit pulse signal as input, and this set of values are commonly known as Markov parameters of the system.  
For each candidate actuator location, we appeal to the Eigensystem Realization Algorithm~(ERA)~\cite{juang1985eigensystem} to compute a minimal realization of the system $\hat{G}=(\hat{A},\hat{B},\hat{C})$ directly from this pulse response data $h_k$.
To do so, we define two Hankel matrices composed of the Markov parameters,
\begin{equation}
H_{0}=\left[
\begin{array}{c c c c}
h_1& h_{2} &   \cdots & h_{n_o} \\ 
h_{2} & h_{3}  & \cdots &  h_{n_o+1} \\
\vdots &\vdots & \ddots & \vdots\\
h_{n_c}  & h_{n_c+1} & \cdots&  h_{n_c+n_o}
\end{array}
\right],
\qquad 
H_{1}=\left[
\begin{array}{c c c c}
h_2& h_{3} &   \cdots & h_{n_o+1} \\ 
h_{3} & h_{4}  & \cdots &  h_{n_o+2} \\
\vdots &\vdots & \ddots & \vdots\\
h_{n_c+1}  & h_{n_c+2} & \cdots&  h_{n_c+n_o+1}
\end{array}
\right].
\label{eqn:HankelGen}
\end{equation}
Next, compute the Singular Value Decomposition~(SVD) of $H_0= U \Sigma V^*$, 
then store the $r$ largest singular values in a matrix $\Sigma_r$ and 
the corresponding left- and right-singular vectors in 
the matrices $U_r$ and $V_r$, respectively.
Finally, a minimal realization $(\hat{A},\hat{B},\hat{C})$ can be computed as,
\begin{align}
\hat{A}& \defi \Sigma_r^{-\frac{1}{2}}U_r^*H_1V_r\Sigma_r^{-\frac{1}{2}} \\
\hat{B} & \defi \, \text{First column of } \Sigma_r^{\frac{1}{2}}V_r^* \\
\hat{C}&\defi \, \text{First row of } U_r\Sigma_r^{\frac{1}{2}}
\end{align}
A complete description of ERA can be found in \cite{juang1985eigensystem}.
As was shown in the previous subsection, the $\ghh$-norm optimality measure associated 
with each actuator location can then be computed directly from this 
ERA-based minimal realization.

Our choice of utilizing pulse response data for system identification is quite natural since Markov parameters have the property of being unique for a given system and are often referred to as the ``signature'' of the system model \cite{2014okid}. 
In the event that other forms of input-output data are available through simulations/experiments, methods such as Observer/Kalman Filter Identification~(OKID) may be used to extend the applicability of ERA to general input-output response data~\cite{juang1993okid}. %\cite{chang2013observer,2014okid}.
We note that ERA introduces some elements of subjectivity to the optimal selection process,
since various ERA algorithm parameters, such as $n_c$, $n_o$ can be chosen to alter the specific realization;
however, additional precautions can be taken to ensure that the realization is sufficiently insensitive to these algorithmic parameters and that multiple ERA realizations based on the same pulse response data yield consistent optimal actuator rankings.
Indeed, this will be the case for all of the results that are reported in Section~\ref{sec:results}. %., as will be described in the subsequent sections.

We further note that ERA is applicable for both stable and unstable systems~\cite{flinoisJFM2016}.
%, we apply it to the available pulse response data gathered from each of the actuator locations under consideration. 
%
%
% A minimal realization can be obtained using this method based on sensor measurement from an pulse response simulation. 
% %
%For the data-sets that we will consider here, the system response appears to be mildly unstable.
%
%This finding is consistent with modal analysis of the flowfield conducted using dynamic mode decomposition, to be discussed later in Section~\ref{sec:modal}
%
In principle, it is possible to compute the output controllability Gramian by direct integration of 
pulse response data;
however, performing a direct integration of pulse response data for unstable systems (or of unconverged responses in general) over an infinite time-horizon is not possible.
Appealing to generalized Gramians computed via ERA system realizations
overcomes this challenge.

\section{Results}
\label{sec:results}

% \msh{Show response data from [36] in this paper. It needs to be included in this paper for completeness. We cannot ask the reviewer/reader to pick up another paper to be able to read the current one.

% Also, would be nice to summarize the algorithm for a specific case here. Something like 

% \begin{enumerate}
% \item Starting from the discrete data set \vec{C_l} vs. \vec{t} as shown in Figure....
% \item use ERA to generate a approximation to the data set over an infite time horizon 
% using formulae (18-20)
% \item  determine the Grammain with (x)
% \item determine the generalized Grammian according to x
% \item \dots
% \end{enumerate}
% }

We apply the approach described in Section~\ref{sec:methods} to the numerical
pulse response data shown in Figure~\ref{fig:naca_airfoil} to determine the optimal actuator
location for controlling lift and separation angle. 
For clarity, we first outline
the specific steps involved in determining
the optimal actuator location when the quantity of interest is the lift $C_\ell$:
\begin{enumerate}
\item \emph{Collect Data.}  Collect sampled pulse response data $C_\ell^i(k)$ for each of the $i=1,\dots,N$ candidate actuator locations.  Also, collect the uncontrolled baseline lift response $C_\ell^0(k)$ and compute its mean $\overline{C_\ell^0}$.
  \item \emph{Form Markov Parameters.} Subtract the uncontrolled baseline mean from each pulse response signal to obtain the associated sequence of Markov parameters $h_k^i=C_\ell^i(k)-\overline{C_\ell^0}$. 
  \item \emph{Identify System Realizations.} Perform ERA on each sequence $h_k^i$ to obtain a discrete-time system realization $\hat{G}^i=(\hat{A}^i,\hat{B}^i,\hat{C}^i)$.  Convert this realization to a continuous-time realization $G^i=(A^i,B^i,C^i)$ via Tustin's approximation.
  \item \emph{Compute  $\ghh$ Optimality Measures.} For each system realization $G^i$, compute the generalized controllability Gramian $P^i$ from Equations~\eqref{eq:gengram1} and \eqref{eq:gengram2}.  From this, compute the optimality measure $\|G^i\|_{2'}=\sqrt{C^iP{C^i}\trans}$ for each actuator location.
    \item \emph{Select Optimal Actuator.}  Sort actuators according to decreasing $\ghh$-norm.  The optimal actuator location is the one associated with the largest value of $\|G^i\|_{2'}$.
\end{enumerate}
The same procedure can be applied to analyze the optimal actuator location for controlling separation angle.
To do so, simply substitute $C_\ell\leftarrow \theta$ everywhere above.

\begin{figure}[ht!]
\begin{center}
  \subfloat [Actuator at $x/c=.1$]{
 \includegraphics[width=0.32\textwidth]{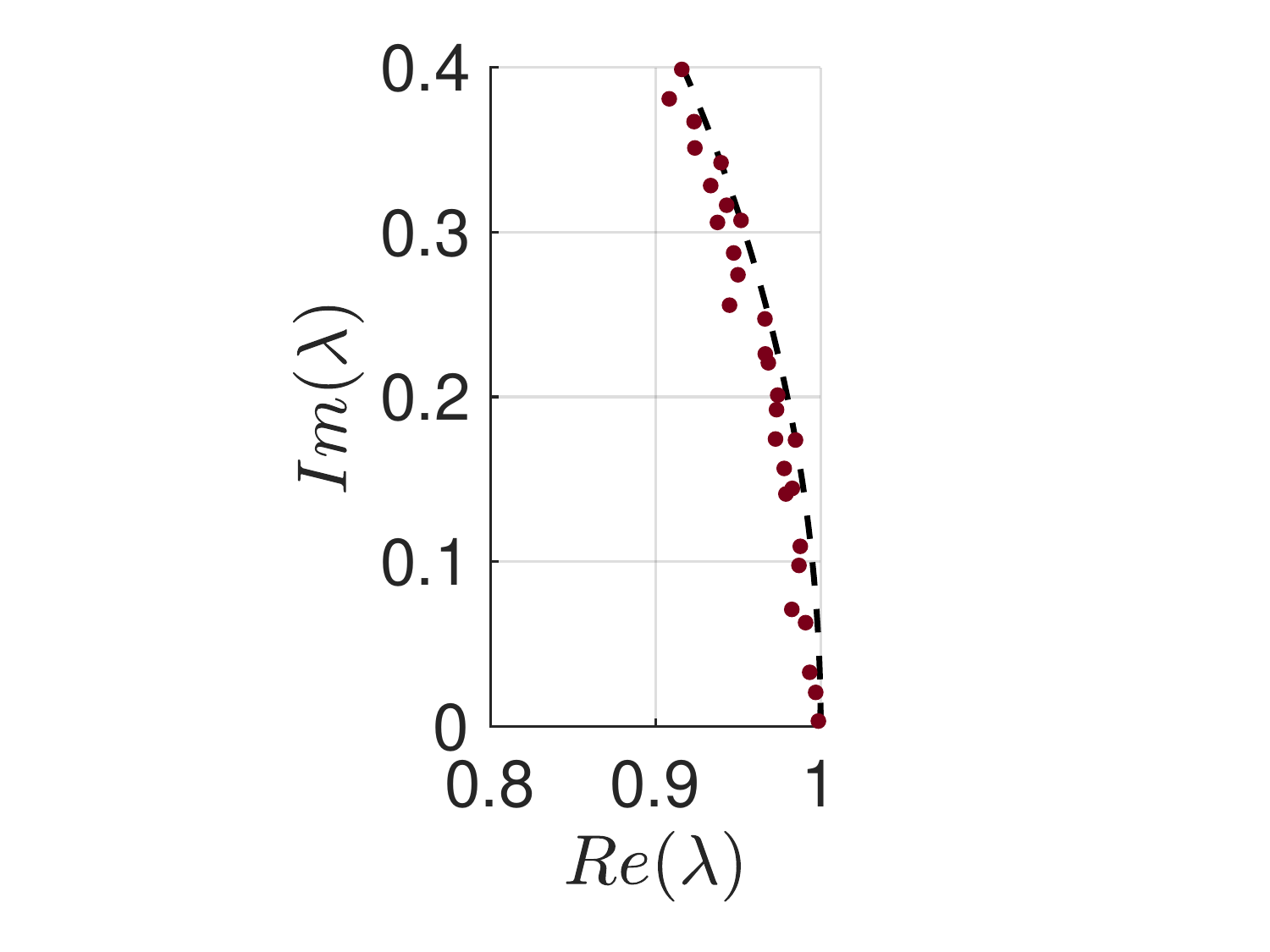}
 }  
  \subfloat [Actuator at $x/c=.2$]{
 \includegraphics[width=0.32\textwidth]{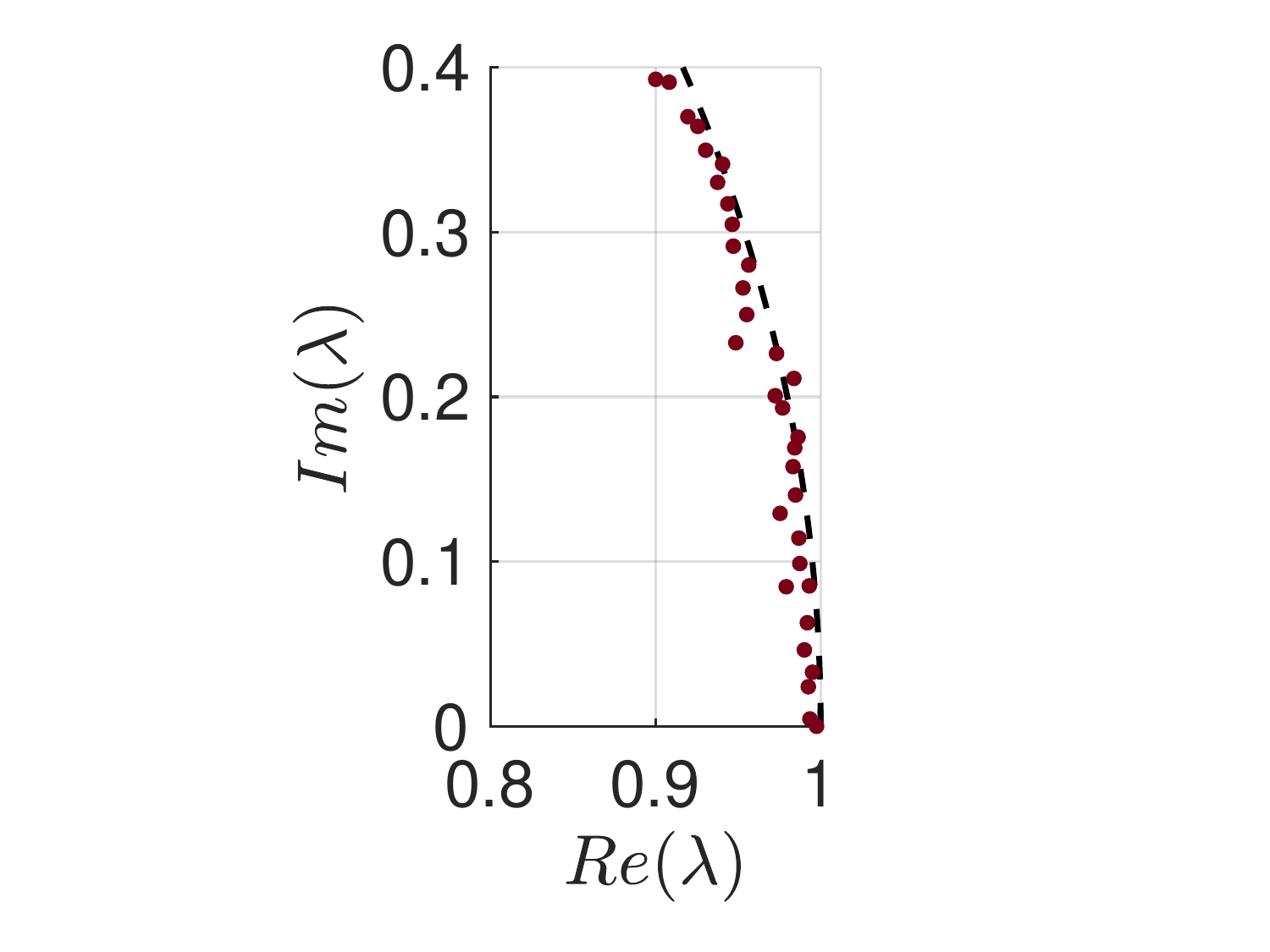}
 }
  \subfloat [Actuator at $x/c=.3$]{
 \includegraphics[width=0.32\textwidth]{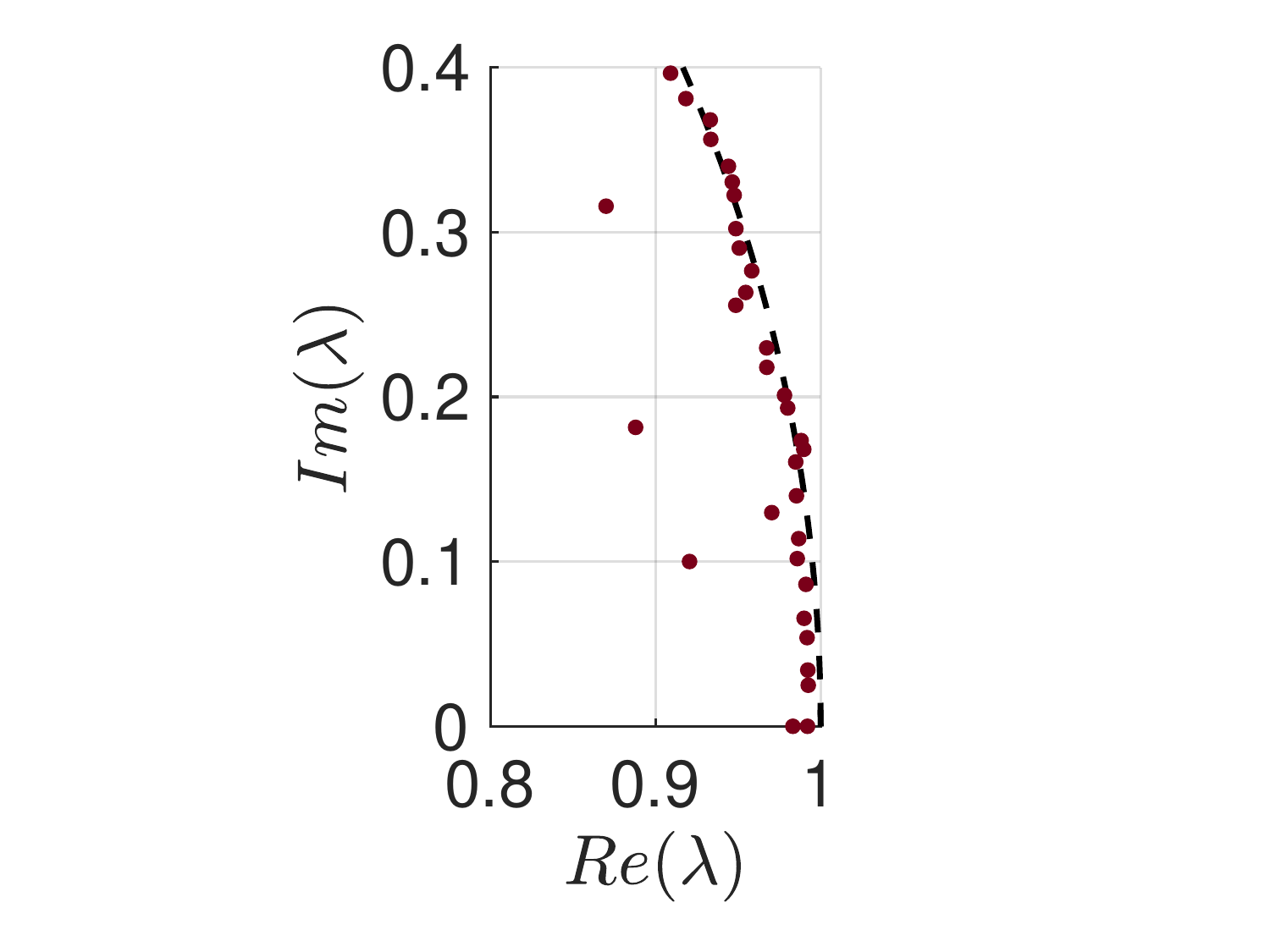}
 }\\
  \subfloat [Actuator at $x/c=.4$]{
 \includegraphics[width=0.32\textwidth]{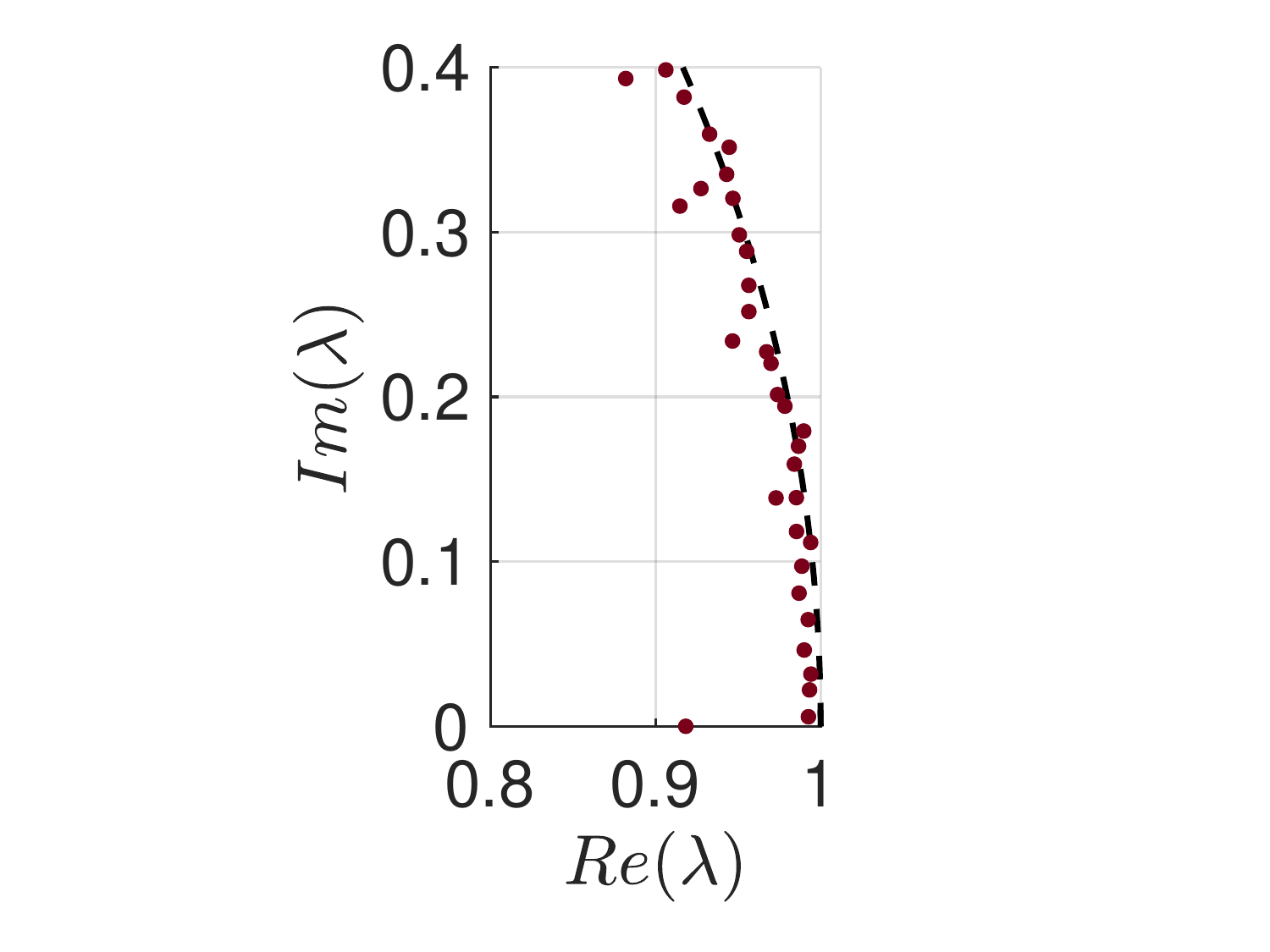}
 }
  \subfloat [Actuator at $x/c=.5$]{
 \includegraphics[width=0.32\textwidth]{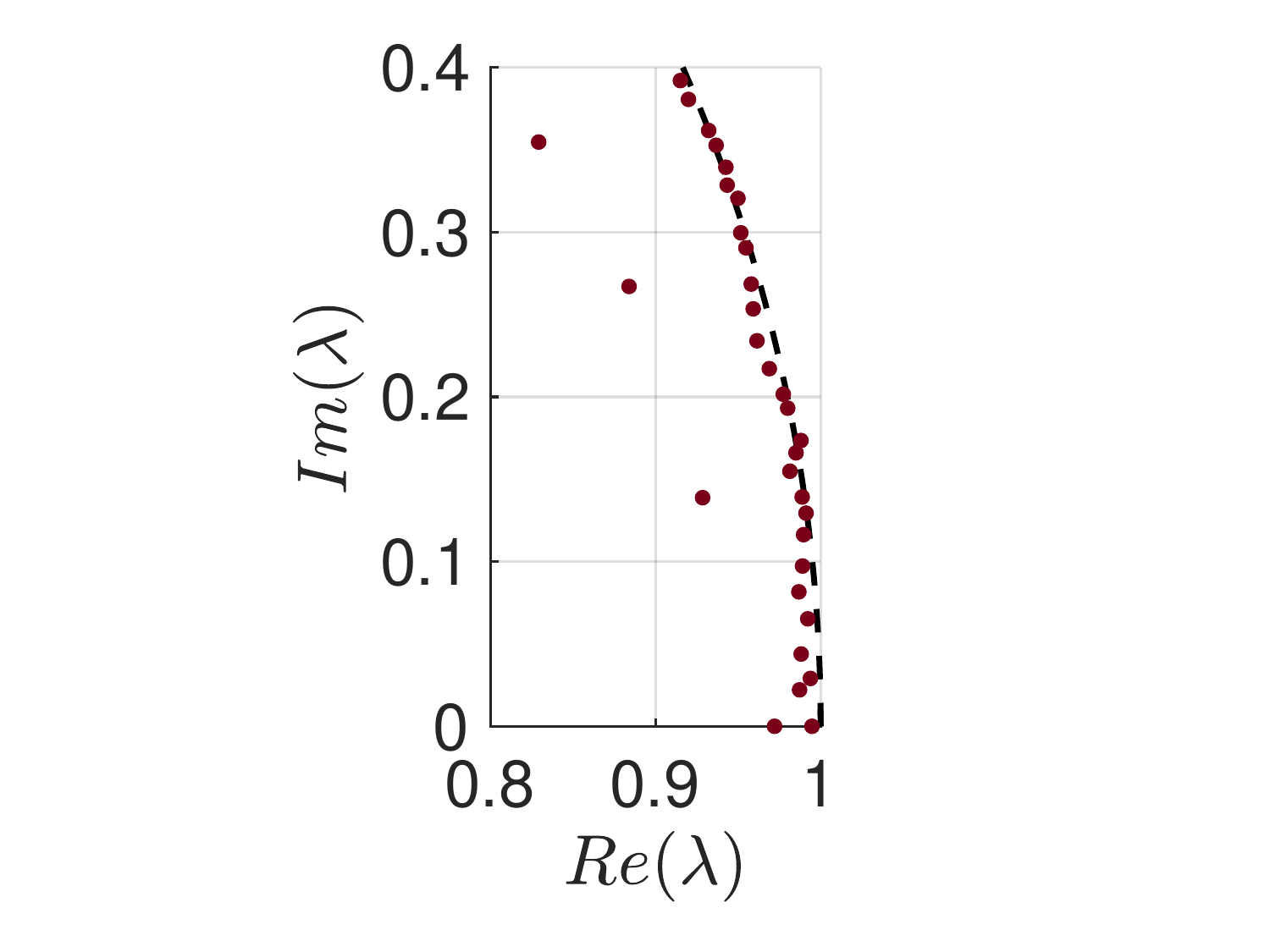}
 }
  \subfloat [Actuator at $x/c=.6$]{
 \includegraphics[width=0.32\textwidth]{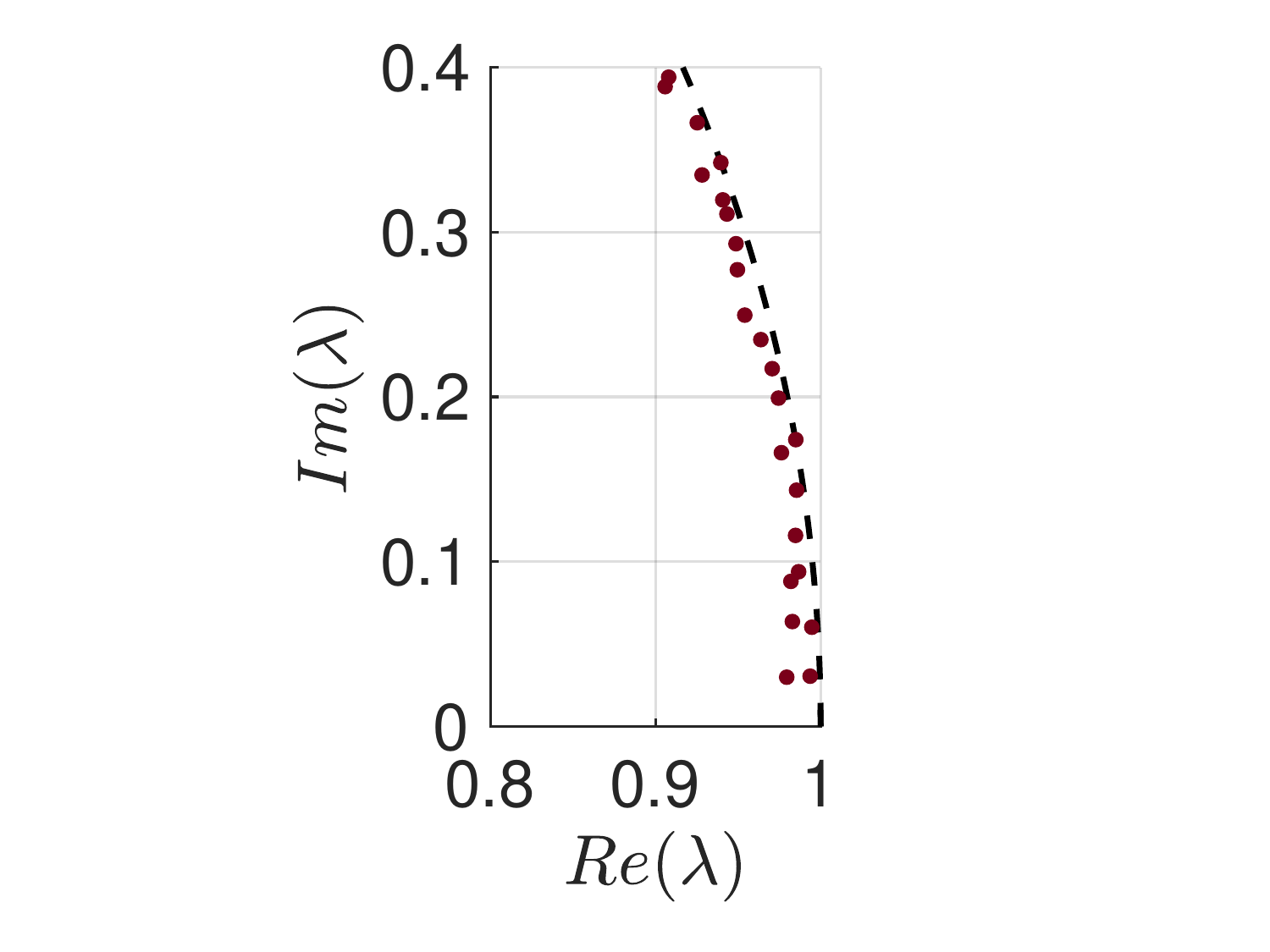}
 }
\caption{System poles of (discrete-time) minimal realization computed from separation angle pulse response data for each actuator location using ERA. Some poles are outside the unit circle for all locations.}
\label{fig:poles_sep}
\end{center}
\end{figure}

We first perform steps 1--3 above for the lift and separation angle responses.
An ERA model of order $r$ is realized for each actuator location and each quantity of interest (see Figures~\ref{fig:pulse} and \ref{fig:pulse_Sep}). 
Here, $r$ is chosen to give the best match in terms of the original data obtained from numerical simulations.
As previously indicated, a number of these realizations exhibit unstable dynamics, with some of the discrete-time poles lying outside the unit circle in Figures~\ref{fig:poles} and \ref{fig:poles_sep}. This motivates the use of generalized controllability Gramians and the associated $\ghh$ for the subsequent analysis.  
We note that the unstable realizations may be related to the slow asymptotic return to the baseline response. %, since the systems have not fully returned to a steady state.
This point is supported by the fact that realizations computed using shorter time-horizons
result in unstable modes that are ``more unstable''.
The minimality and order of ERA-based realizations were sanitized of any potential numerical artificialities by accounting for pole-zero cancellations based on a range of tolerances from $\mathcal{O}(10^{-5})$ to $\mathcal{O}(10^{-7})$. These tolerance values indicate the proximity of poles and zeros required to constitute a numerical pole-zero cancellation.
Tolerances have been selected in conjunction with the system order $r$ to ensure the realization is minimal and able to describe the given response data.
%It was found that decreasing the tolerance any further resulted in realizations that were non-minimal.
%
%
%For the sake of representation, we use figures and tables associated with the tolerance value of $\mathcal{O}(10^{-7})$ in the subsequent sections (Figures~\ref{fig:poles}--\ref{fig:bode_sep} and Tables~\ref{tab:lift_opt_h2}--\ref{tab:sep_opt_h2}). 

We next perform steps 4--5 in the selection process outlined above.
The $\ghh$-norms associated with each actuator location are sorted from most controllable to least controllable and reported in Tables~\ref{tab:lift_opt_h2} and \ref{tab:sep_opt_h2}.
The optimal actuator location for lift control is found to be $x/c=0.2$, whereas for separation angle control it is found to be $x/c=0.3$.
The ranking of actuators and further analysis of these results is presented for lift in Section~\ref{sec:lift} and for separation angle in Section~\ref{sec:sepangle}.
A modal analysis of the flowfield is conducted in Section~\ref{sec:modal} to
help identify physical mechanisms underlying these observations.
% and indicate the degree of controllability among the various candidate locations.
%The ordering of locations indicates their relative rank in terms of optimality for a given output variable.
%

% Similarly, Tables \ref{tab:lift_opt_hinf} and \ref{tab:sep_opt_hinf} are associated with $\hinf$-norms.  This measure corresponds to the largest possible frequency gain associated with a location and can provide insight about frequencies along which maximum output can be expected.
%

%  As is seen in Tables \ref{tab:lift_opt_h2} and \ref{tab:sep_opt_h2}, although the values of the associated norms change with the tolerance used in the minimal realization, the ordering of the actuator positions is relatively constant for a particular output variable.  
%  %
%  This is corroborated by the fact, that the first 2 optimal locations for each tolerance value is always the same for both the lift and separation angle cases. This ordering is also reflected in Tables \ref{tab:lift_opt_hinf} and \ref{tab:sep_opt_hinf}.  
 %
% These values of tolerance are chosen such that the realizations obtained are always minimal.
%

%

%\vspace{cm}
%\clearpage

%

%

%

%\clearpage

\subsection{ Optimal actuator placement for controlling lift}
\label{sec:lift}
\begin{figure}[ht!]
\begin{center}
  \subfloat [Actuator at $x/c=.1$]{
 \includegraphics[width=0.35\textwidth]{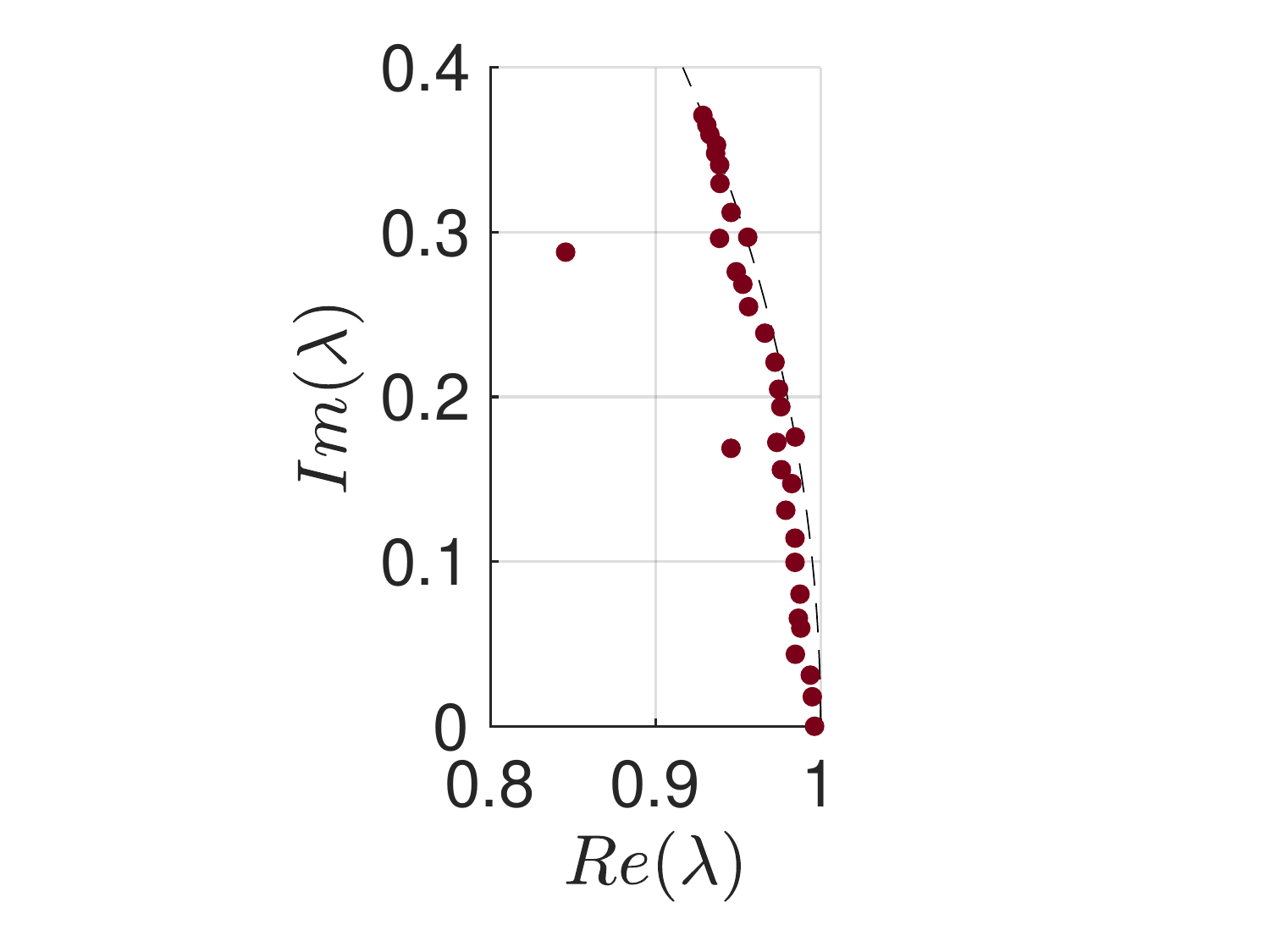}
 }  
  \subfloat [Actuator at $x/c=.2$]{
 \includegraphics[width=0.35\textwidth]{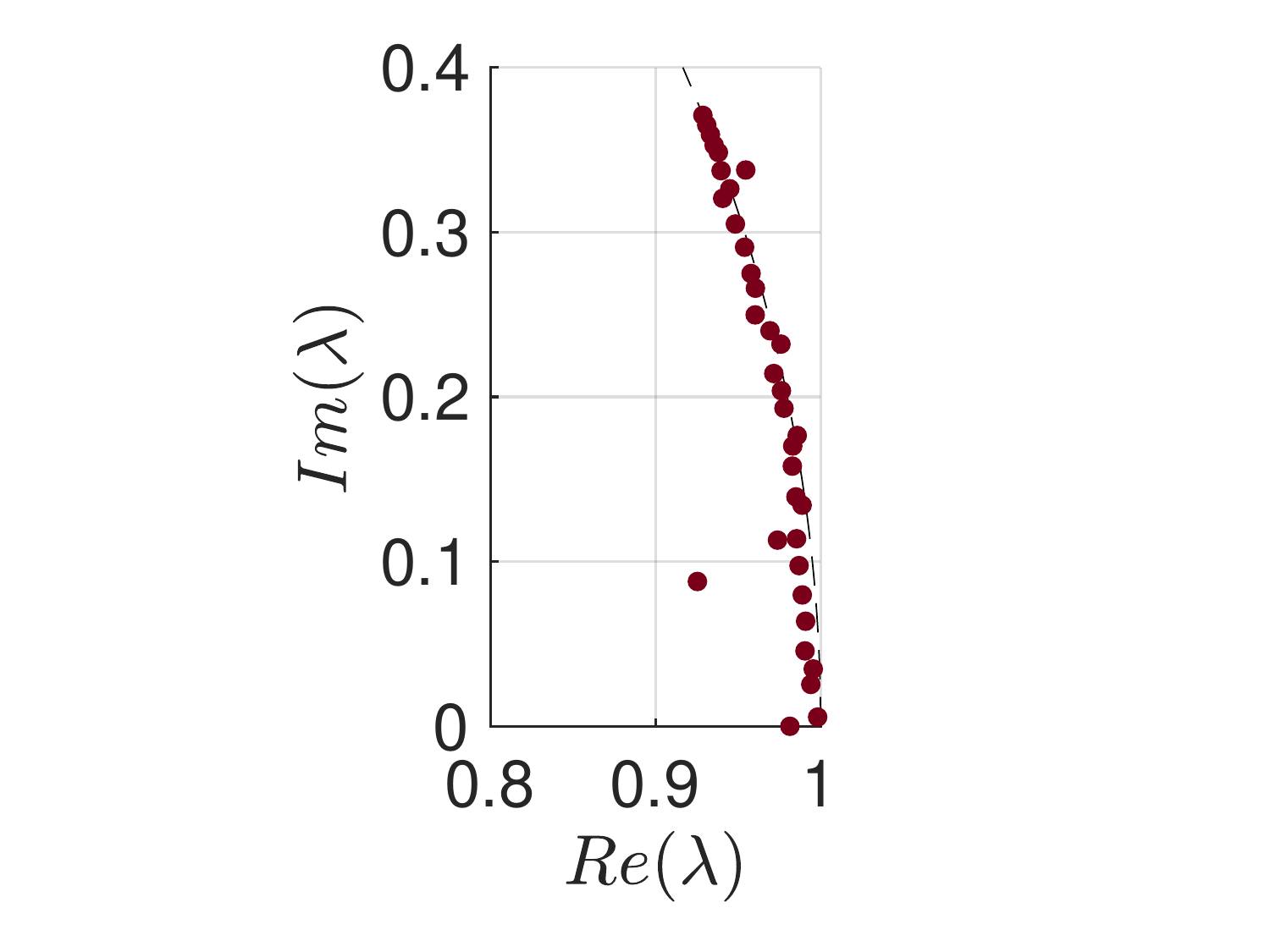}
 }
  \subfloat [Actuator at $x/c=.3$]{
 \includegraphics[width=0.35\textwidth]{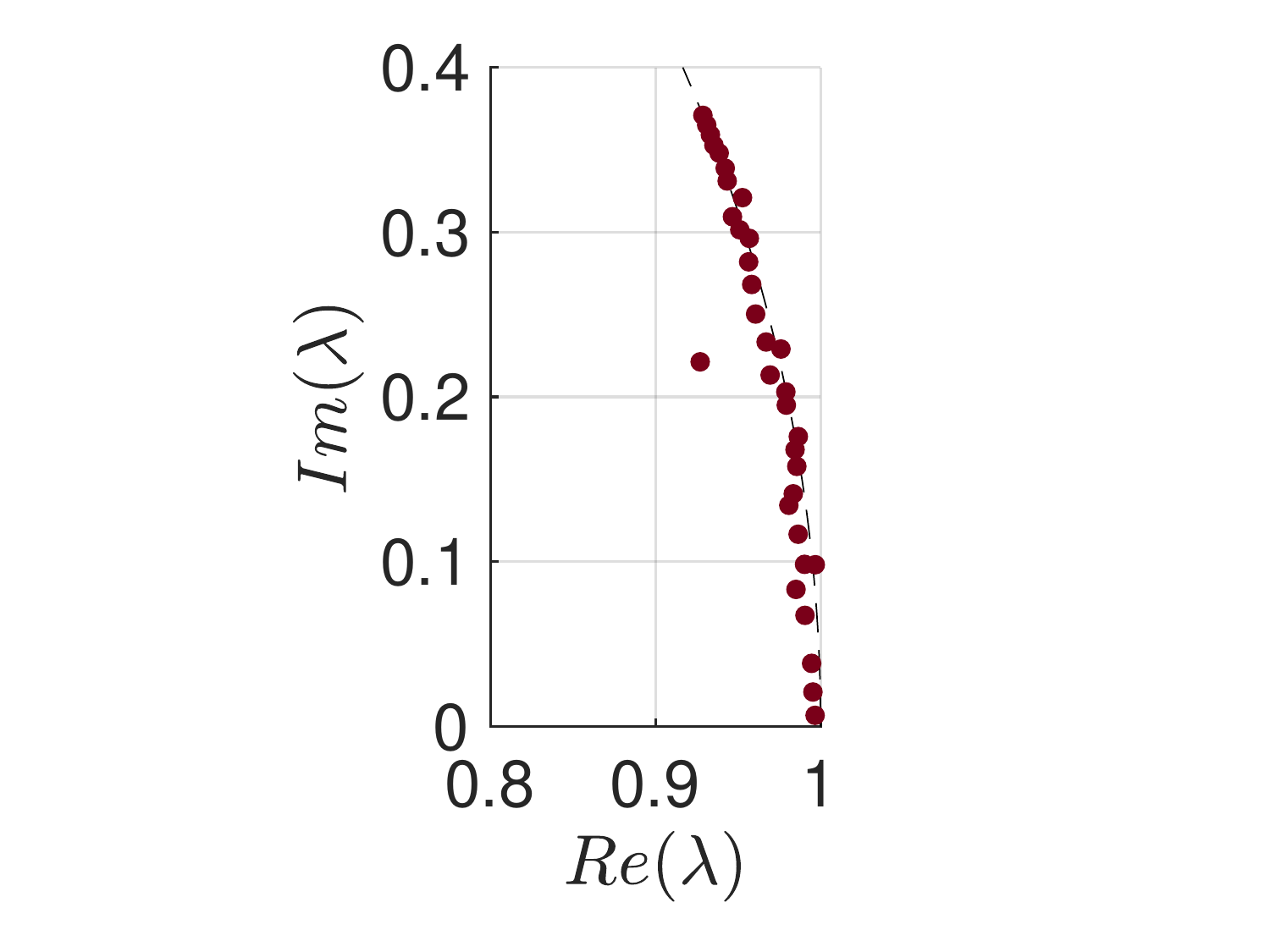}
 }\\
  \subfloat [Actuator at $x/c=.4$]{
 \includegraphics[width=0.35\textwidth]{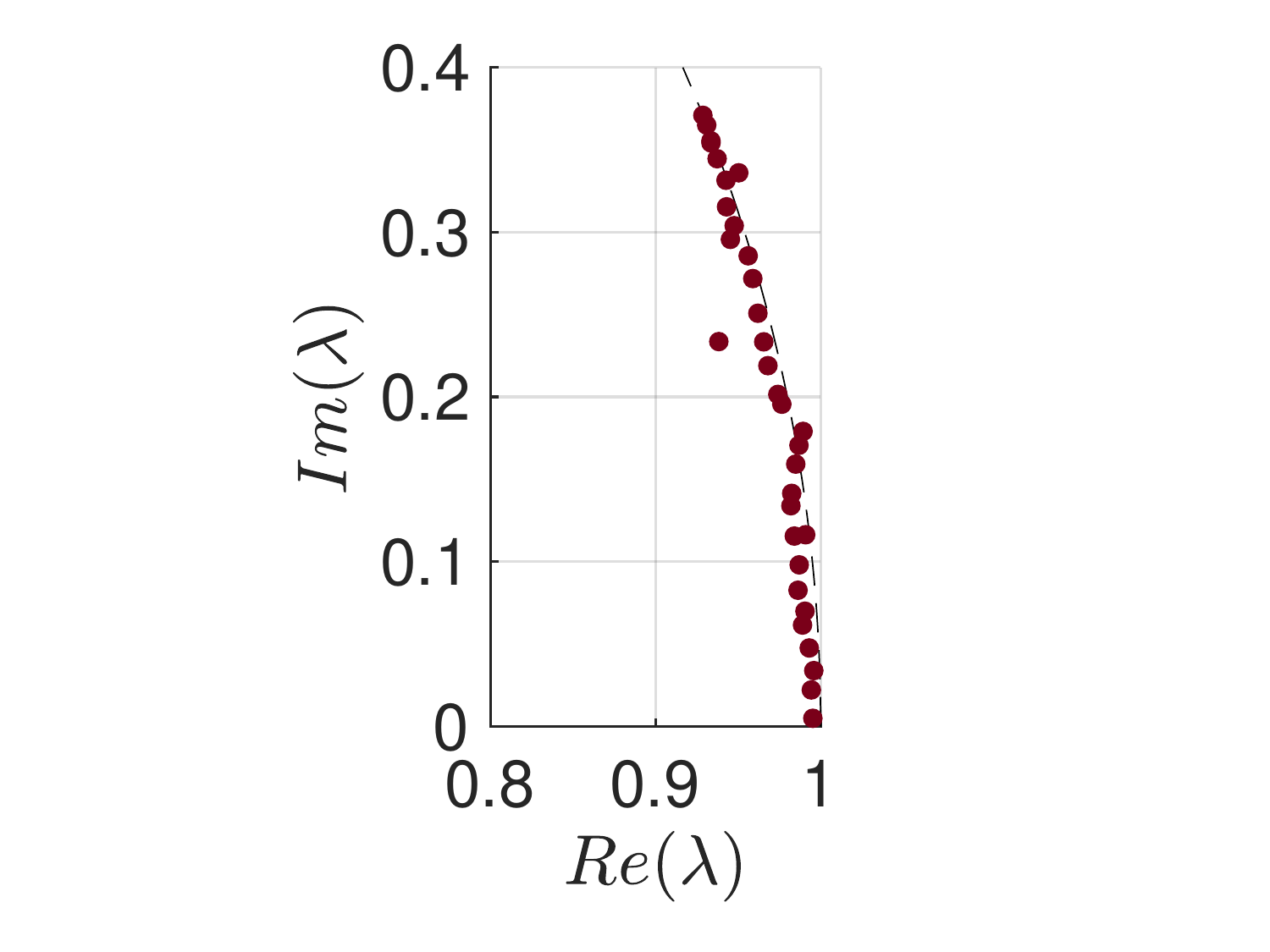}
 }
  \subfloat [Actuator at $x/c=.5$]{
 \includegraphics[width=0.35\textwidth]{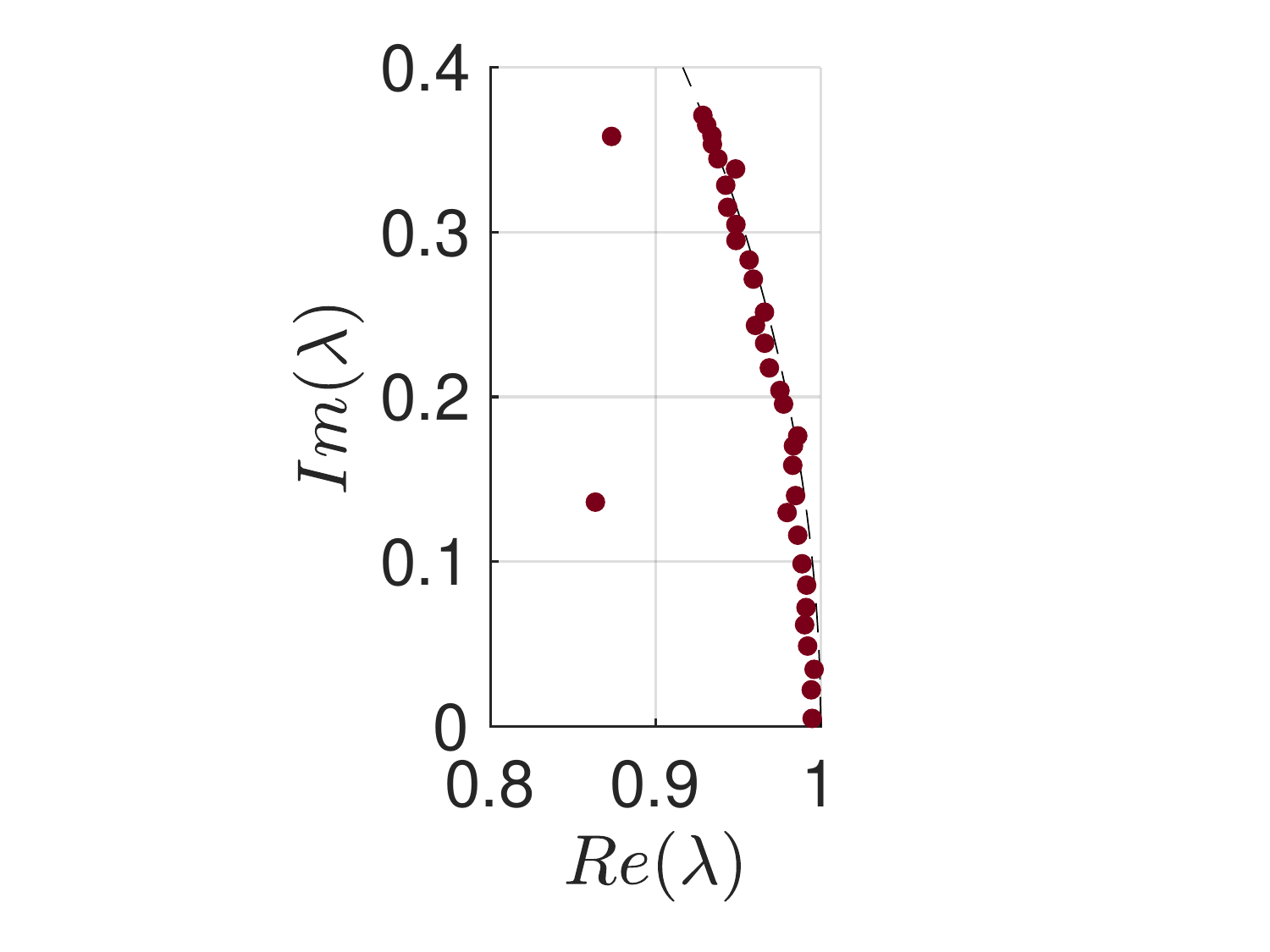}
 }
  \subfloat [Actuator at $x/c=.6$]{
 \includegraphics[width=0.35\textwidth]{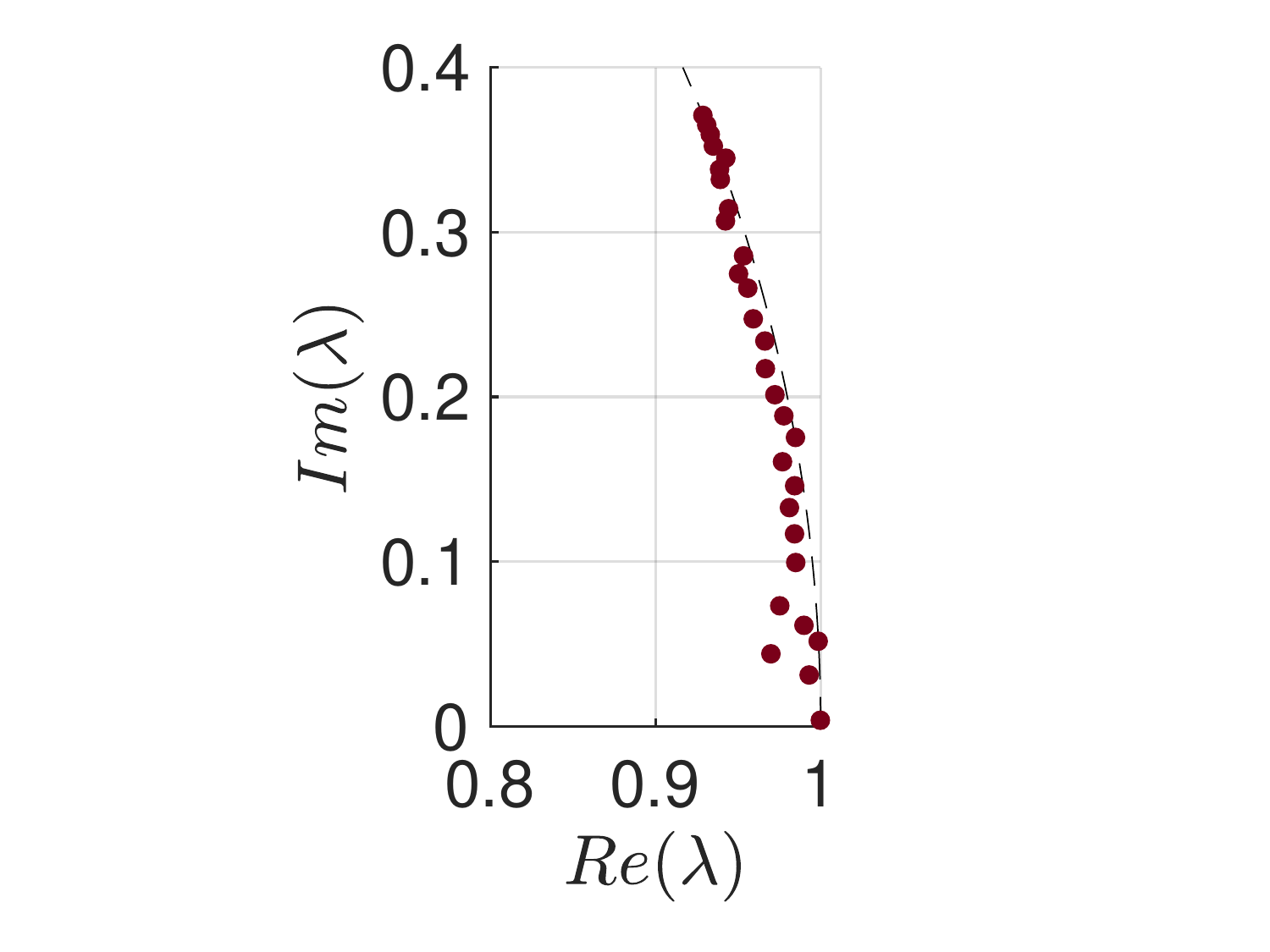}
 }
 \caption{System poles of (discrete-time) minimal realization computed from pulse response data for each actuator location using ERA for lift response data. Some poles are outside the unit circle for all locations.}
\label{fig:poles}
\end{center}
\end{figure}

\begin{figure}[ht!]
\begin{center}

   \subfloat [ Actuator at $x/c=.1$]{
 \includegraphics[width=0.5\textwidth]{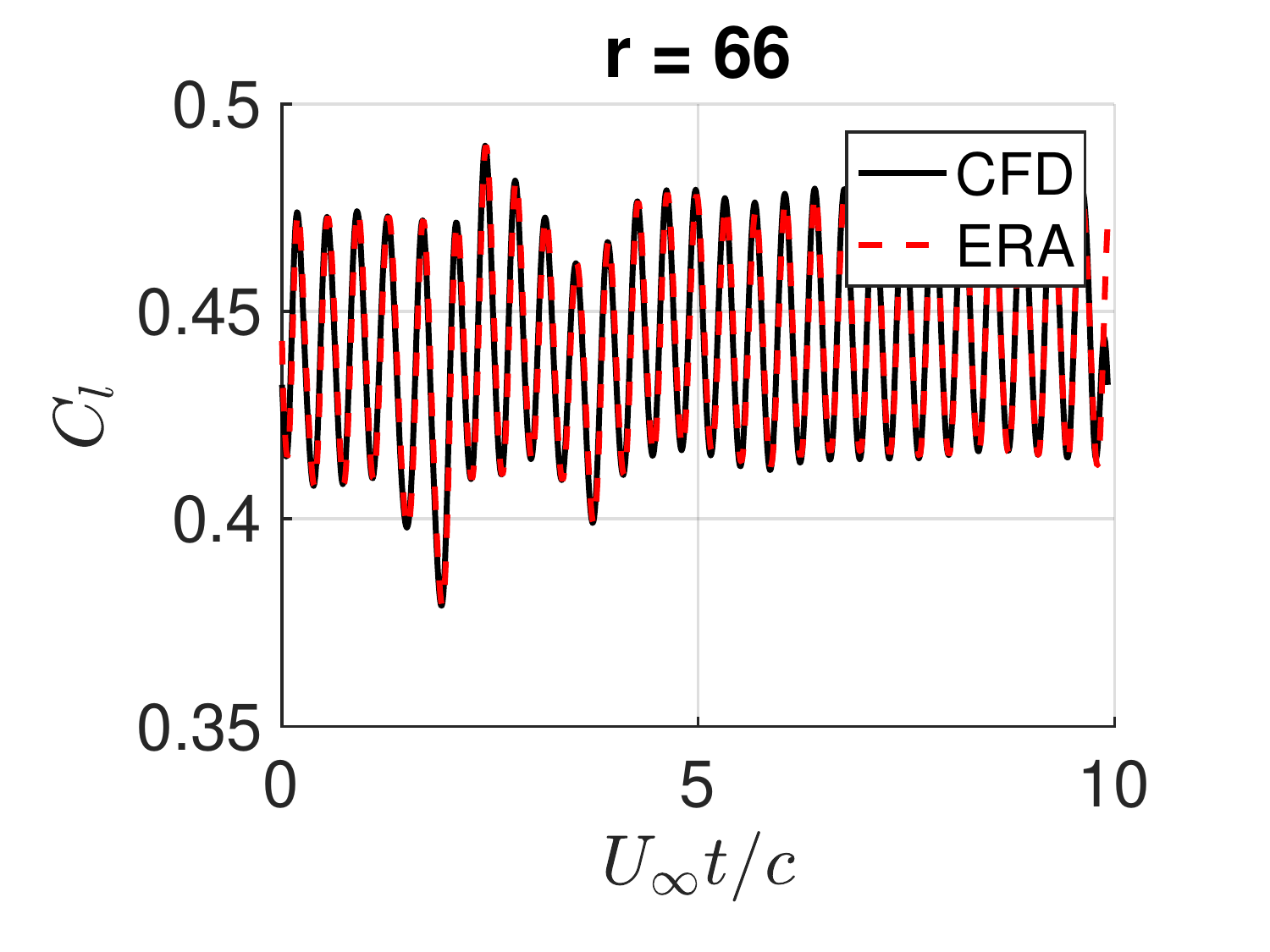}
 }  
  \subfloat [ Actuator at $x/c=.2$]{
 \includegraphics[width=0.5\textwidth]{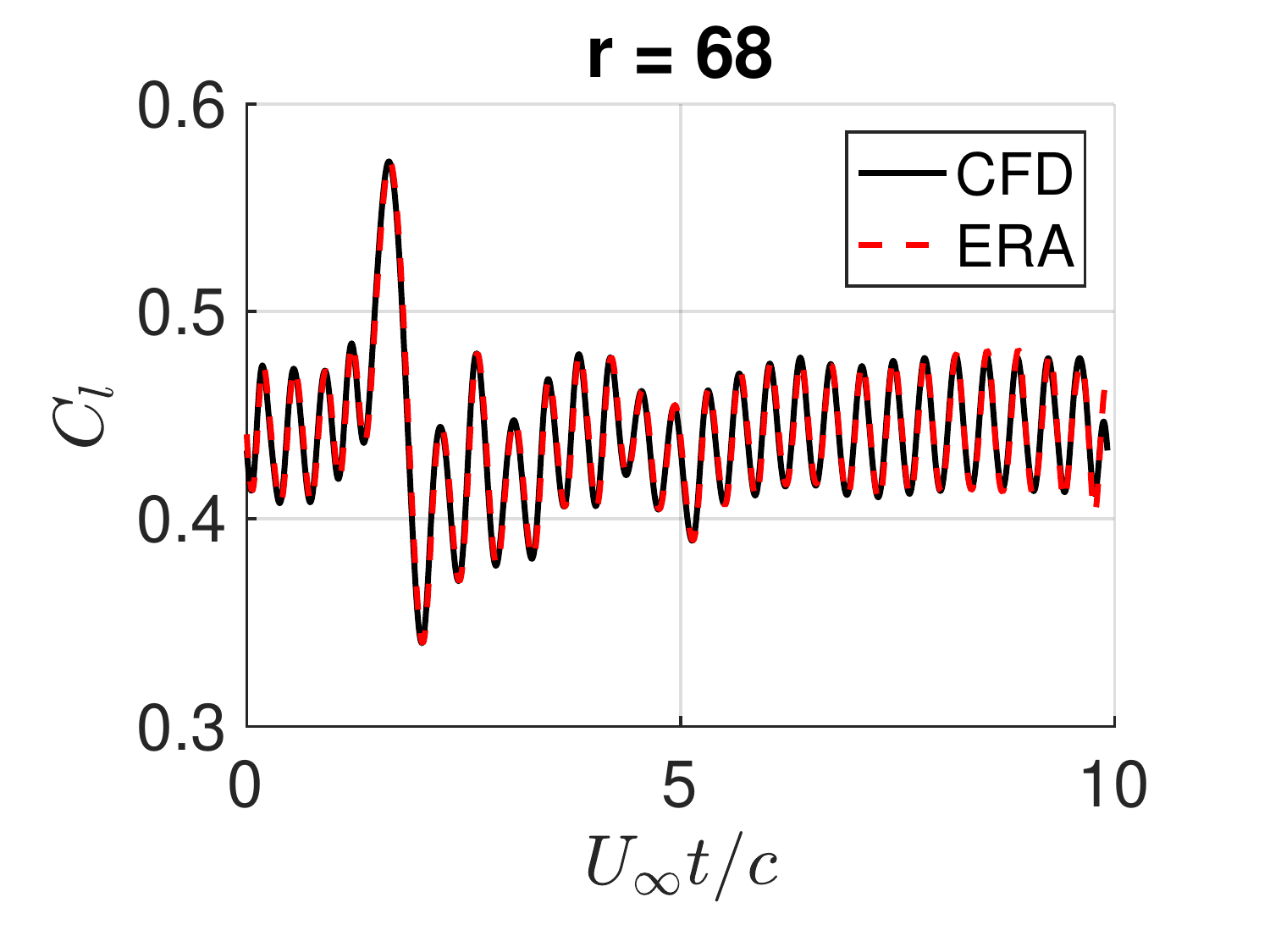}
 }\\
  \subfloat [ Actuator at $x/c=.3$]{
 \includegraphics[width=0.5\textwidth]{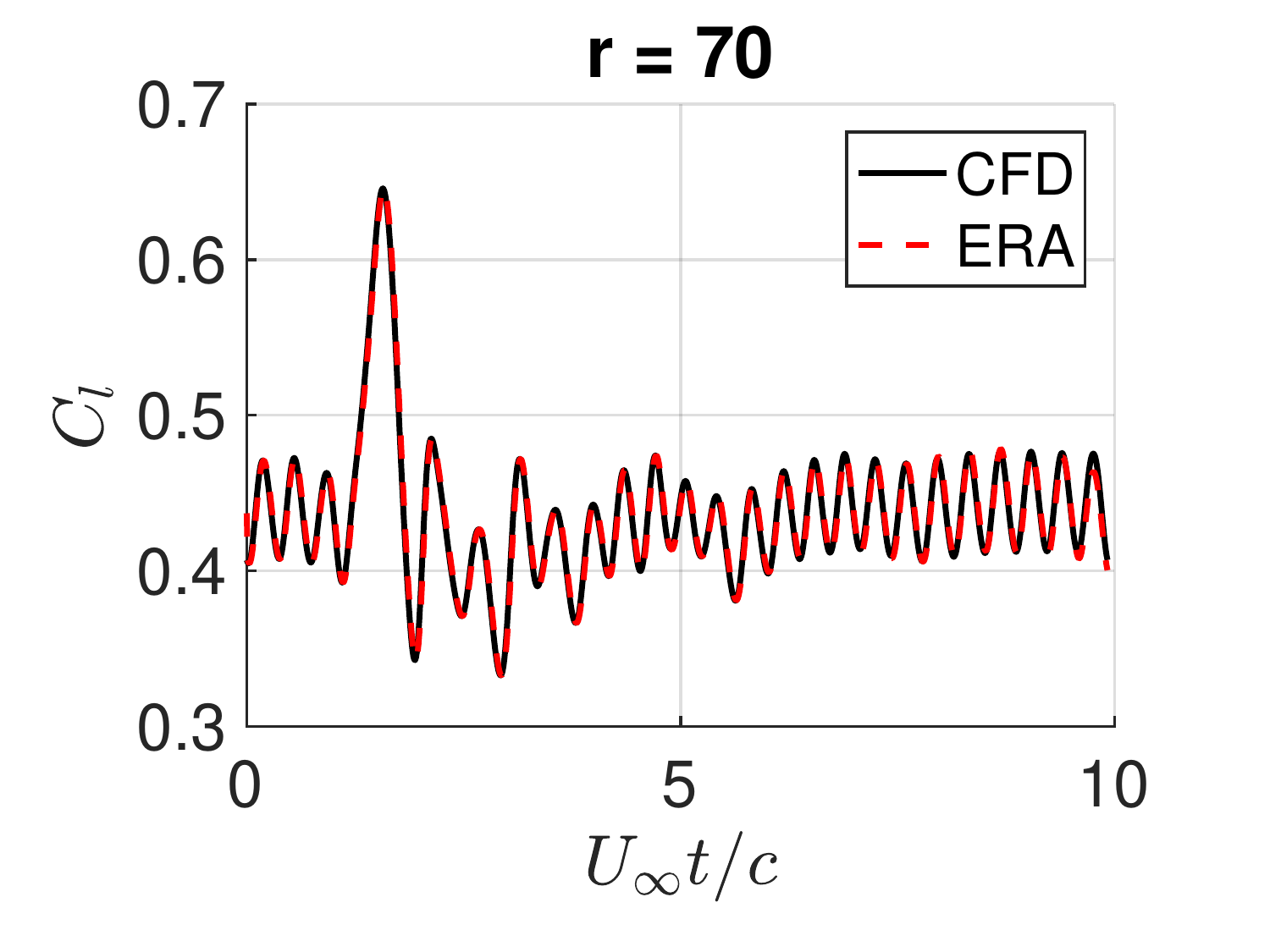}
 }
  \subfloat [ Actuator at $x/c=.4$]{
 \includegraphics[width=.5\textwidth]{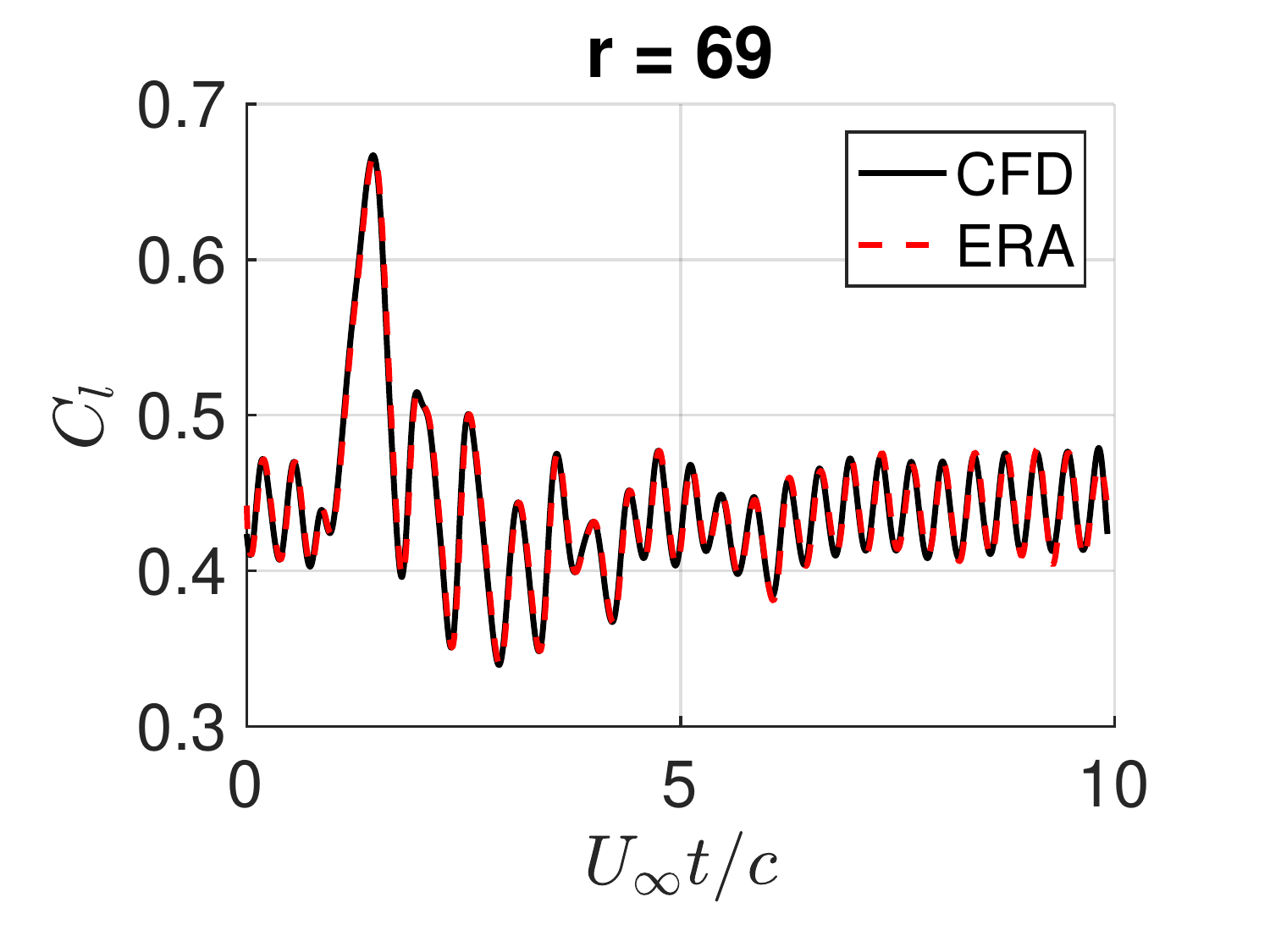}
 }\\
  \subfloat [ Actuator at $x/c=.5$]{
 \includegraphics[width=0.5\textwidth]{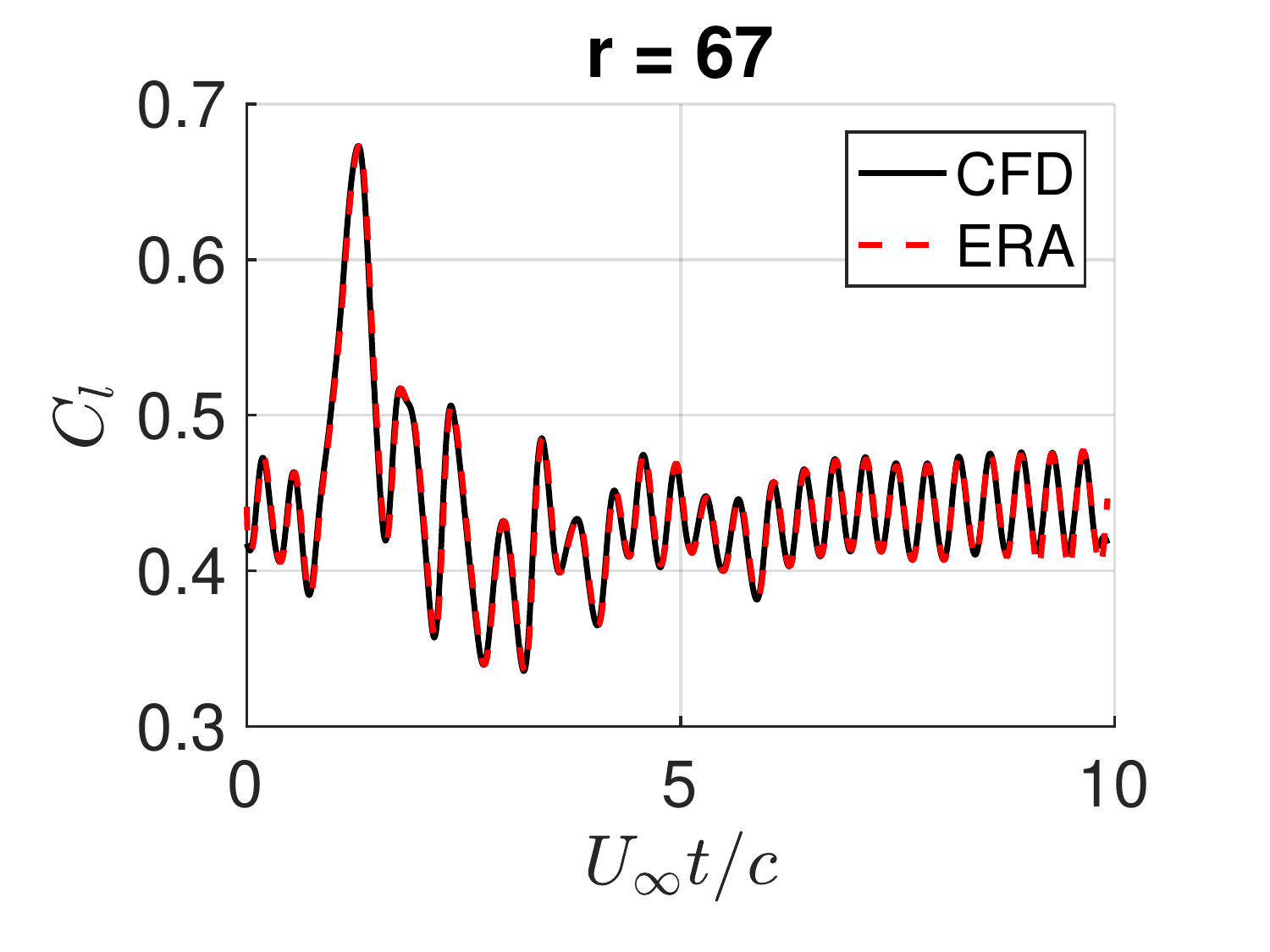}
 }
  \subfloat [ Actuator at $x/c=.6$]{
 \includegraphics[width=0.5\textwidth]{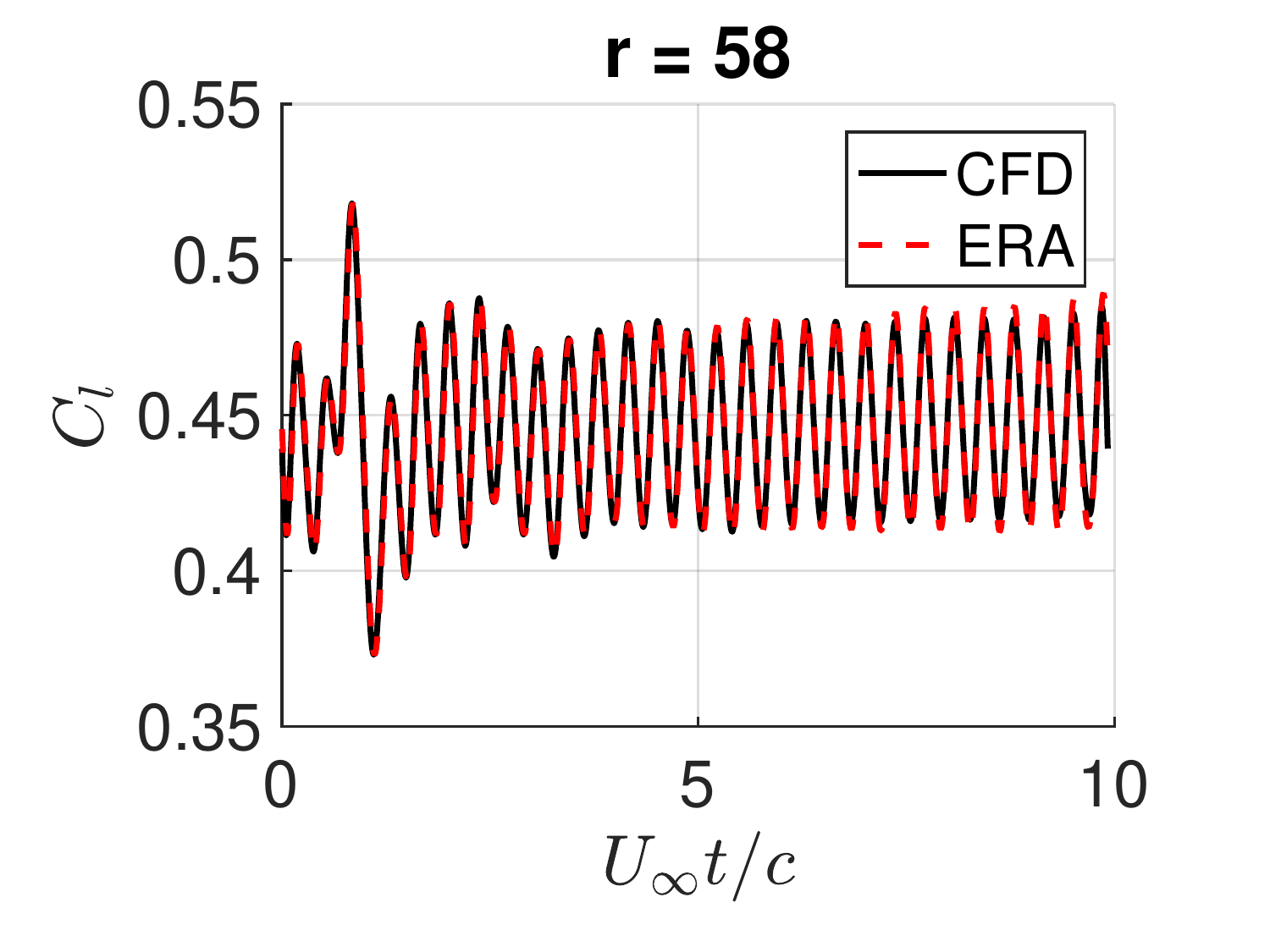}
 }
 \caption{Lift coefficient pulse response data at each actuator location. Each realization is minimal with order $r$.}
\label{fig:pulse}
 \end{center}
\end{figure}

\begin{figure}[ht!]
\begin{center}
  \subfloat [ Actuator at $x/c=.1$]{
 \includegraphics[width=0.4\textwidth]{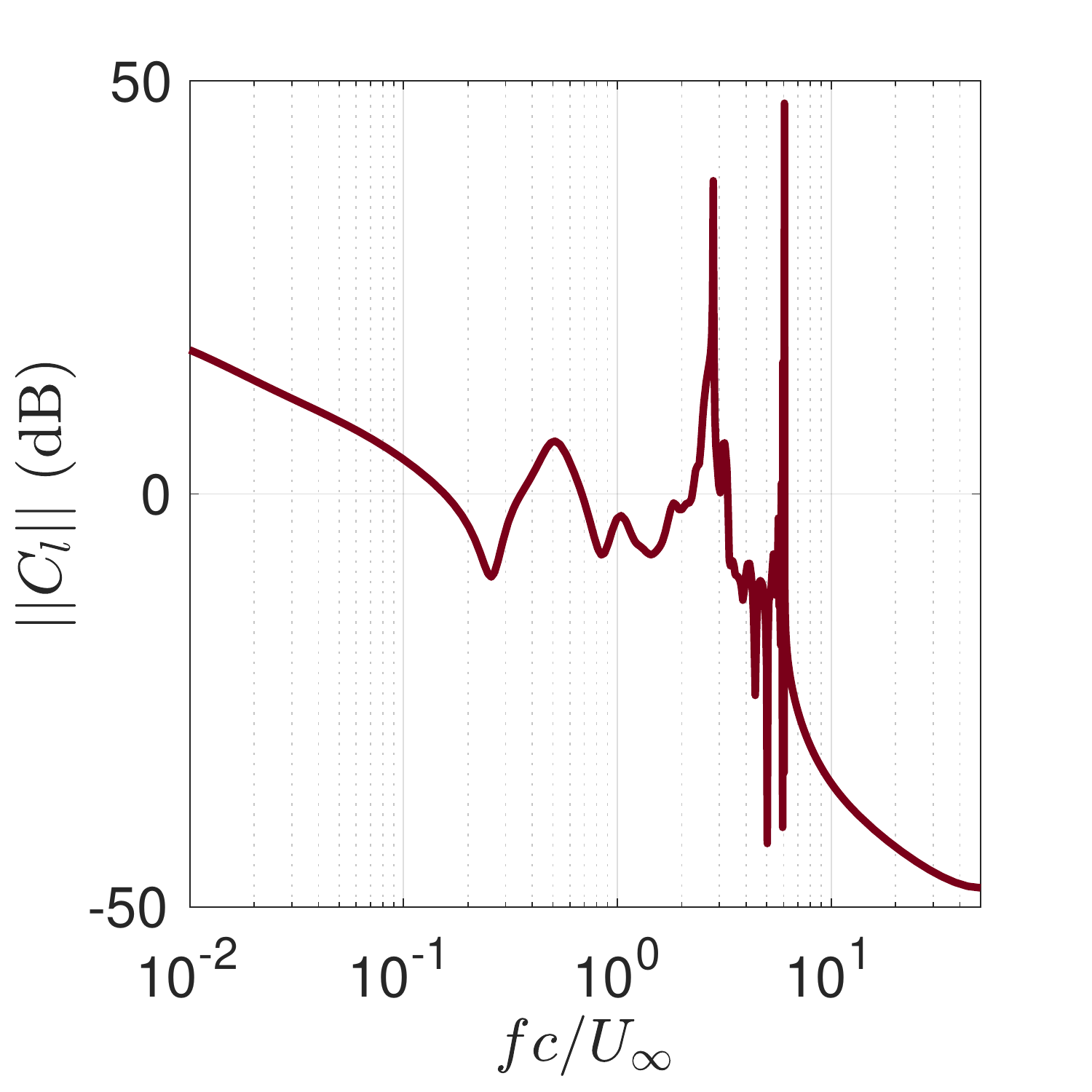}
 }  
  \subfloat [Actuator at $x/c=.2$]{
 \includegraphics[width=0.4\textwidth]{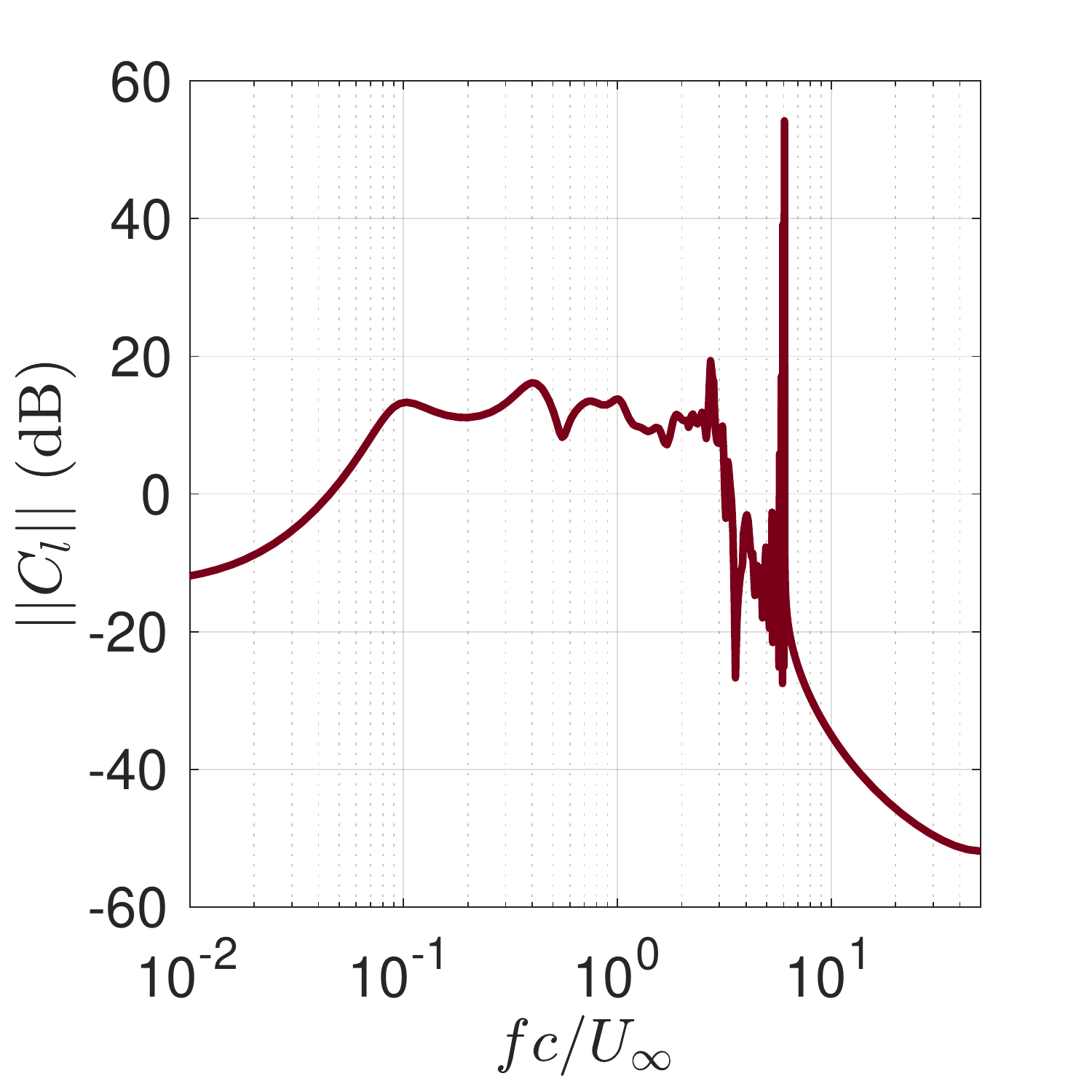}
 }\\
  \subfloat [ Actuator at $x/c=.3$]{
 \includegraphics[width=0.4\textwidth]{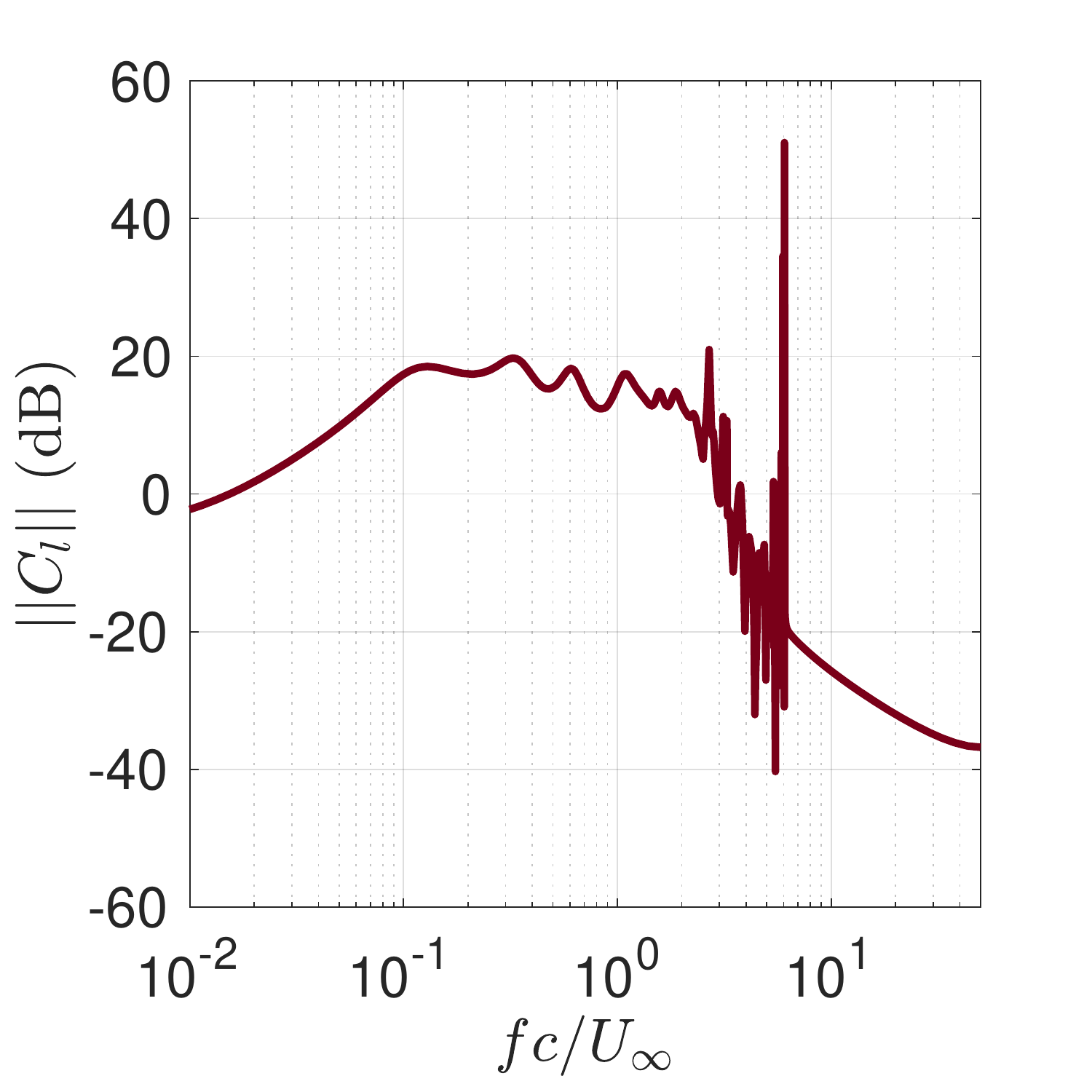}
 }
  \subfloat [ Actuator at $x/c=.4$]{
 \includegraphics[width=0.4\textwidth]{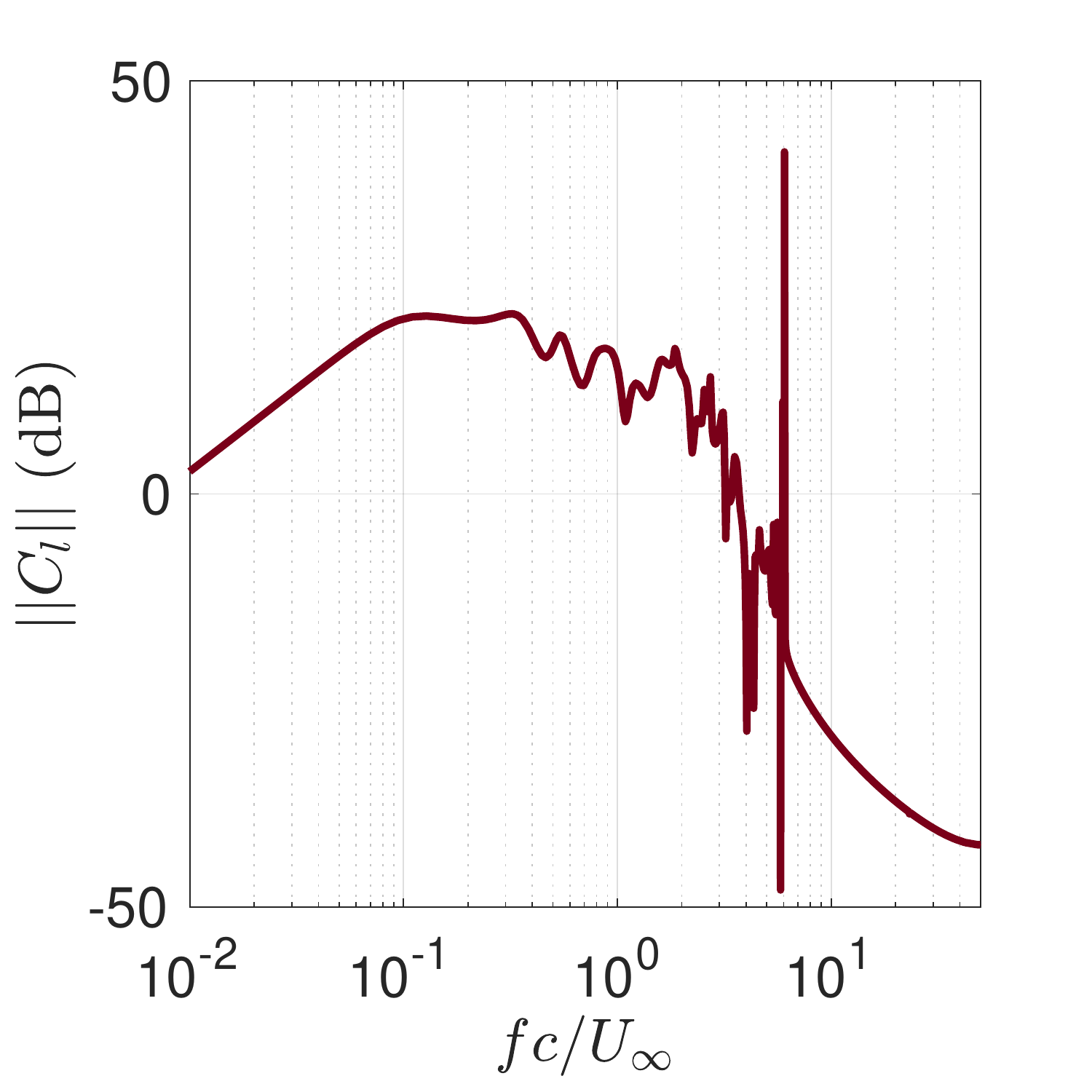}
 }\\
  \subfloat [ Actuator at $x/c=.5$]{
 \includegraphics[width=0.4\textwidth]{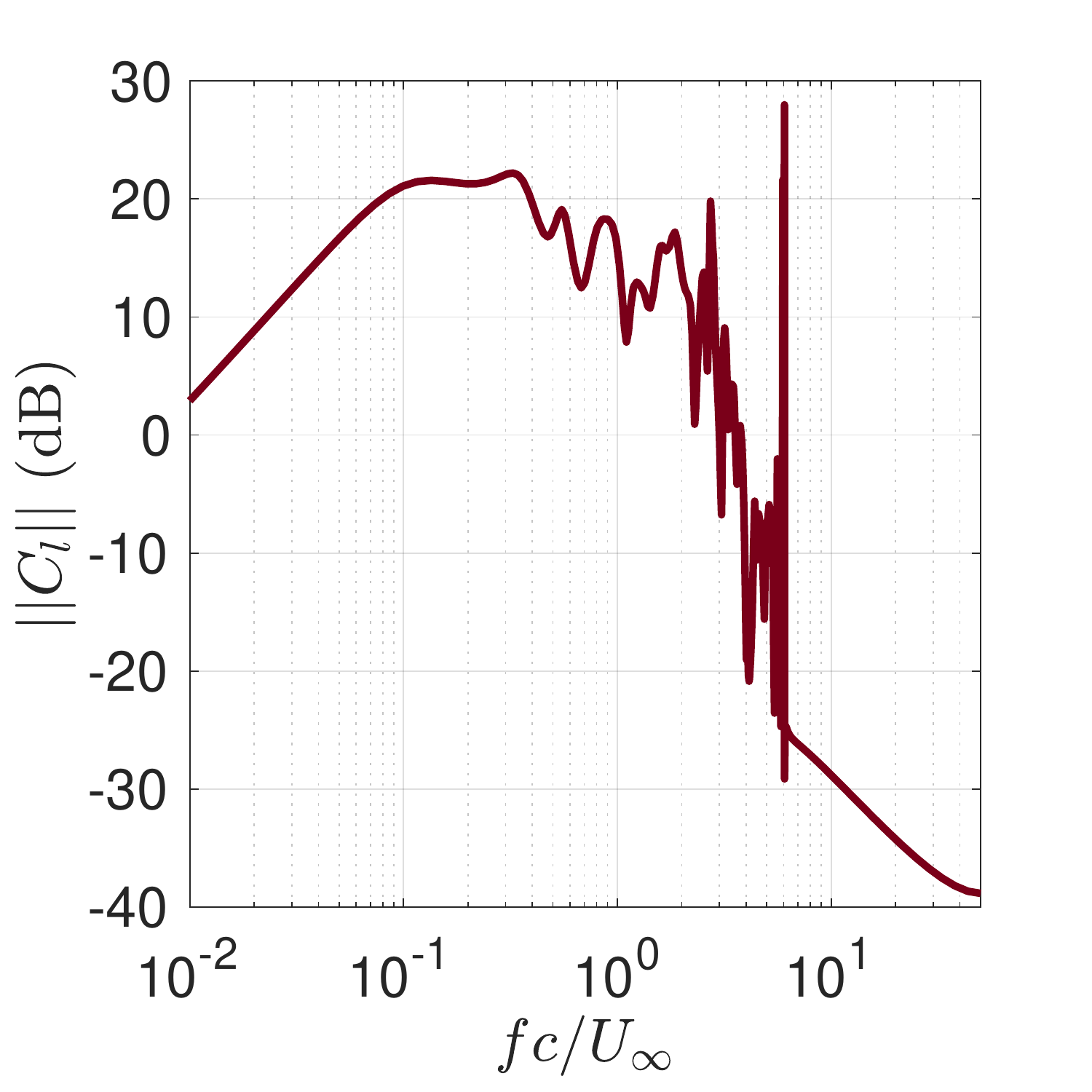}
 }
  \subfloat [ Actuator at $x/c=.6$]{
 \includegraphics[width=0.4\textwidth]{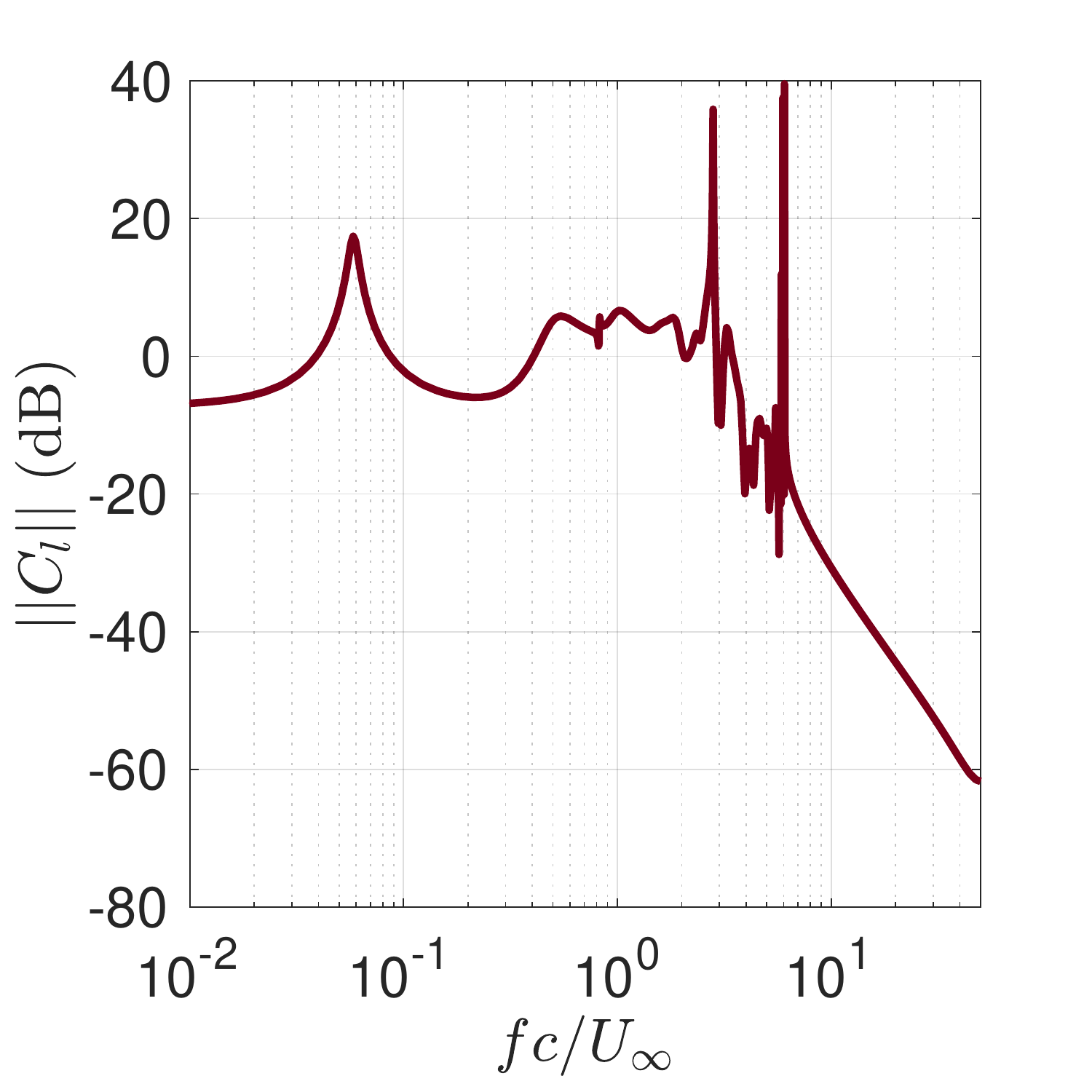}
 }
 \caption{Bode magnitude plot for minimal realization at each actuator location for lift response data.}
\label{fig:bode}
 \end{center}
\end{figure}

Based on the $\ghh$-norm, the optimal actuator location for lift control is  $x/c=0.2$. This location has the highest controllability among all six candidate locations. The optimality study considered tolerance values for pole-zero overlap of $\mathcal{O}(10^{-5})$ to $\mathcal{O}(10^{-7})$. The optimal actuator position is largely constant with these tolerances, although minor variations in the relative ranking of other actuators are observed in the case of lift control.
Some of the eigenvalues of the discrete-time realizations obtained are outside the unit circle for all actuator locations, thereby confirming that the identified systems are unstable (see Figure \ref{fig:poles}).
%
% A similar sort of behavior is obtained while considering the $\hinf$-norm, where actuator location $x/c=0.6$ appears to be the top place-holder.  However, in the case the tolerance value being $10^{-7}$, we see that $x/c=0.2$ has a peak magnitude that is at least an order of magnitude larger than the other actuator locations (see Table~\ref{tab:lift_opt_hinf}).
%
As can be seen in Figure \ref{fig:pulse}, the high order for the obtained minimal realizations, in all likeliness indicates that the system may have some degree of non-linearity in it, which is captured by a larger number of states.
The peak frequency for all actuator locations is $fc/U_\infty=6.12$, as can be seen in Figure \ref{fig:bode} and corresponds to the wake frequency. 
Thus, it appears that among all actuator locations, $x/c=0.2$ is able to induce a resonance by coupling with the flow dynamics at this forcing frequency.
%
%However, some of the frequency response intuition here may have to be reconsidered in a ``non-standard'' manner, since the system dynamics are unstable.
%
%
\begin{table}[ht!]
\begin{center}
\begin{tabular}{|c|c|}
\hline
$x/c$ & $\|G\|_{2'}$ \\ 
\hline
.2 & 51.79 \\
.6 & 31.31 \\
.1 & 17.41 \\
.5 & 15.81 \\
.4 & 15.41 \\
.3 & 13.41\\
\hline
\end{tabular}
\end{center}
\caption{Optimality of actuator locations based on the generalized $\hh$-norm, sorted from most to least optimal for different tolerance values used in minimal realization for lift response data.}
\label{tab:lift_opt_h2}
\end{table}

% \begin{table}[ht!]
% \begin{center}
% \begin{tabular}{|c| c | c | c | c | c|}
% \hline
% \multicolumn{2}{|c|}{Tolerance:$10^{-5}$ } &\multicolumn{2}{|c|}{Tolerance:$10^{-6}$  } & \multicolumn{2}{|c|}{Tolerance:$10^{-7}$ } \\
% \hline
% $x/c$ & $\|G\|_\infty$ & $x/c$ & $\|G\|_\infty$ & $x/c$ & $\|G\|_\infty$\\ 
% \hline
% .6 & 1213.87 & .6 & 1213.87 & .2 & $8.6 \times 10^4$ \\
% .5 & 240.29  & .2 & 196.44  & .6 & 1213.87           \\
% .2 & 196.36  & .1 & 185.68  & .1 & 185.68            \\
% .1 & 185.13  & .3 & 174.06  & .3 & 174.06            \\
% .4 & 157.26  & .4 & 158.58  & .4 & 158.58            \\
% .3 & 143.71  & .5 & 27.27   & .5 & 27.29   \\
% \hline
% \end{tabular}
% \end{center}
% \caption{Optimality of actuator locations based on $\hinf$-norm, sorted from most to least optimal for different tolerance values used in minimal realization.}
% \label{tab:lift_opt_hinf}
% \end{table}

%\clearpage

%%%% SEPARATION ANGLE
\subsection{Optimal actuator placement for controlling separation angle}
\label{sec:sepangle}
\begin{figure}[ht!]
\begin{center}
   \subfloat [ Actuator at $x/c=.1$]{
 \includegraphics[width=0.5\textwidth]{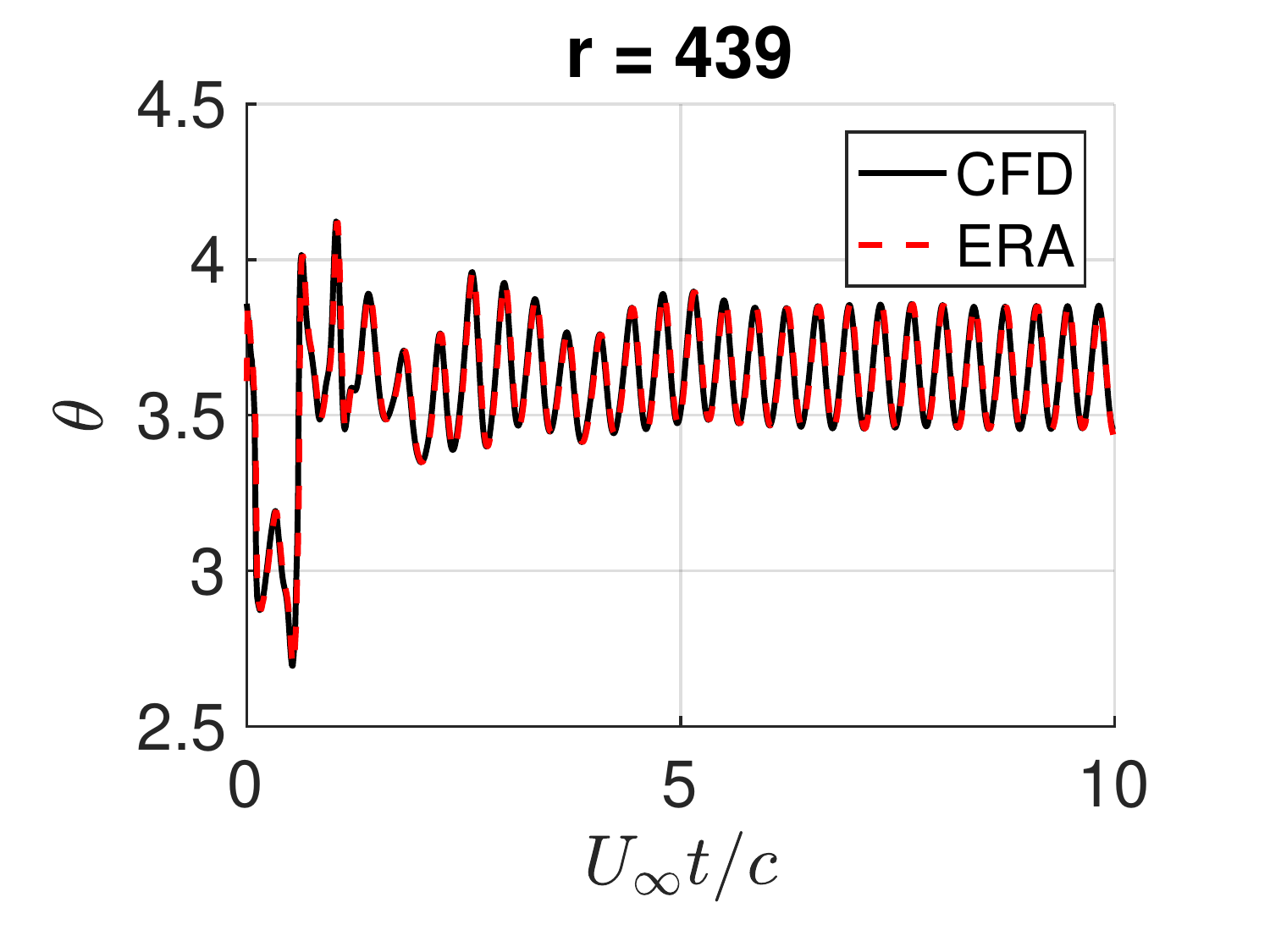}
 }  
  \subfloat [ Actuator at $x/c=.2$]{
 \includegraphics[width=0.5\textwidth]{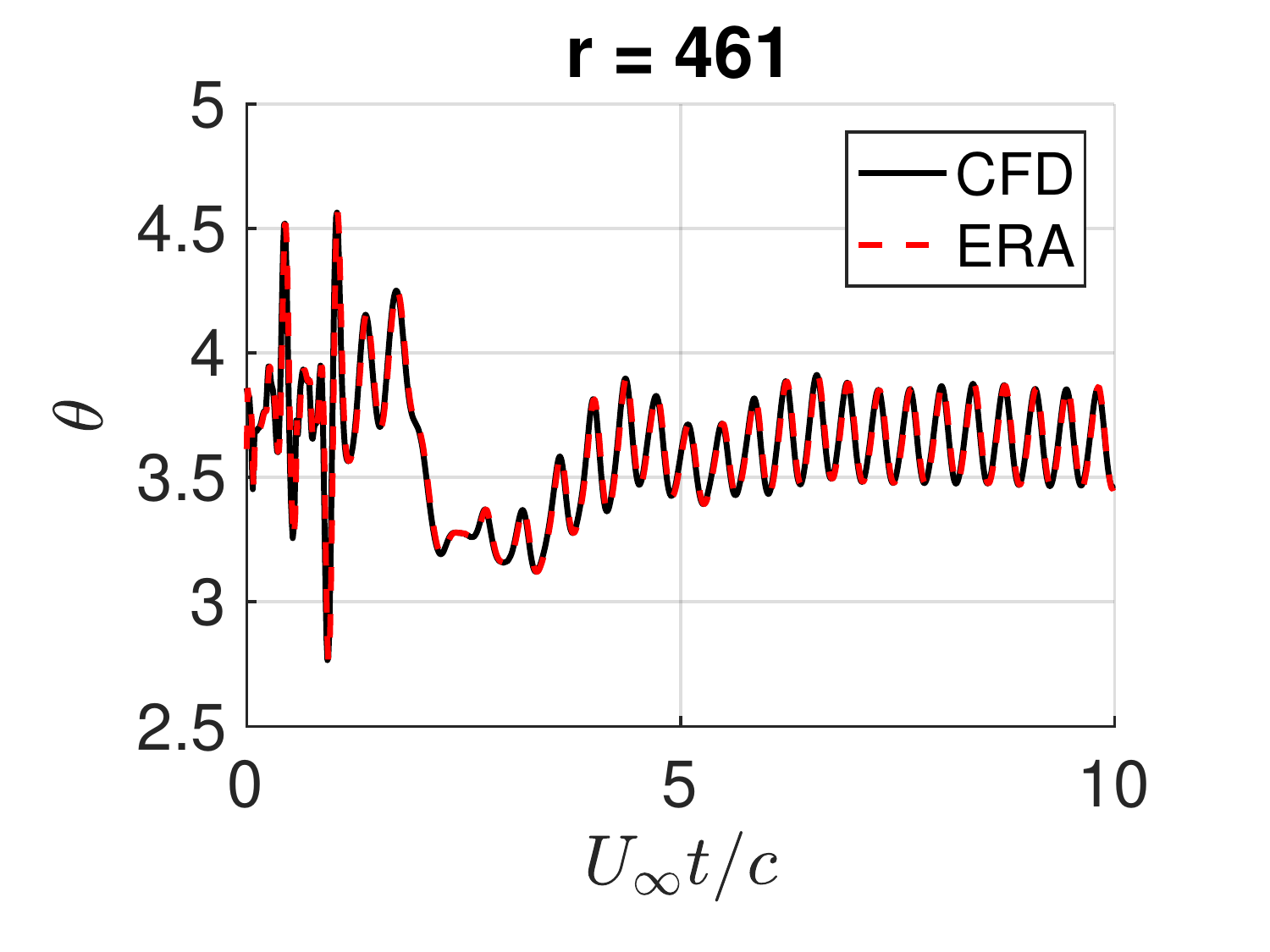}
 }\\
  \subfloat [ Actuator at $x/c=.3$]{
 \includegraphics[width=0.5\textwidth]{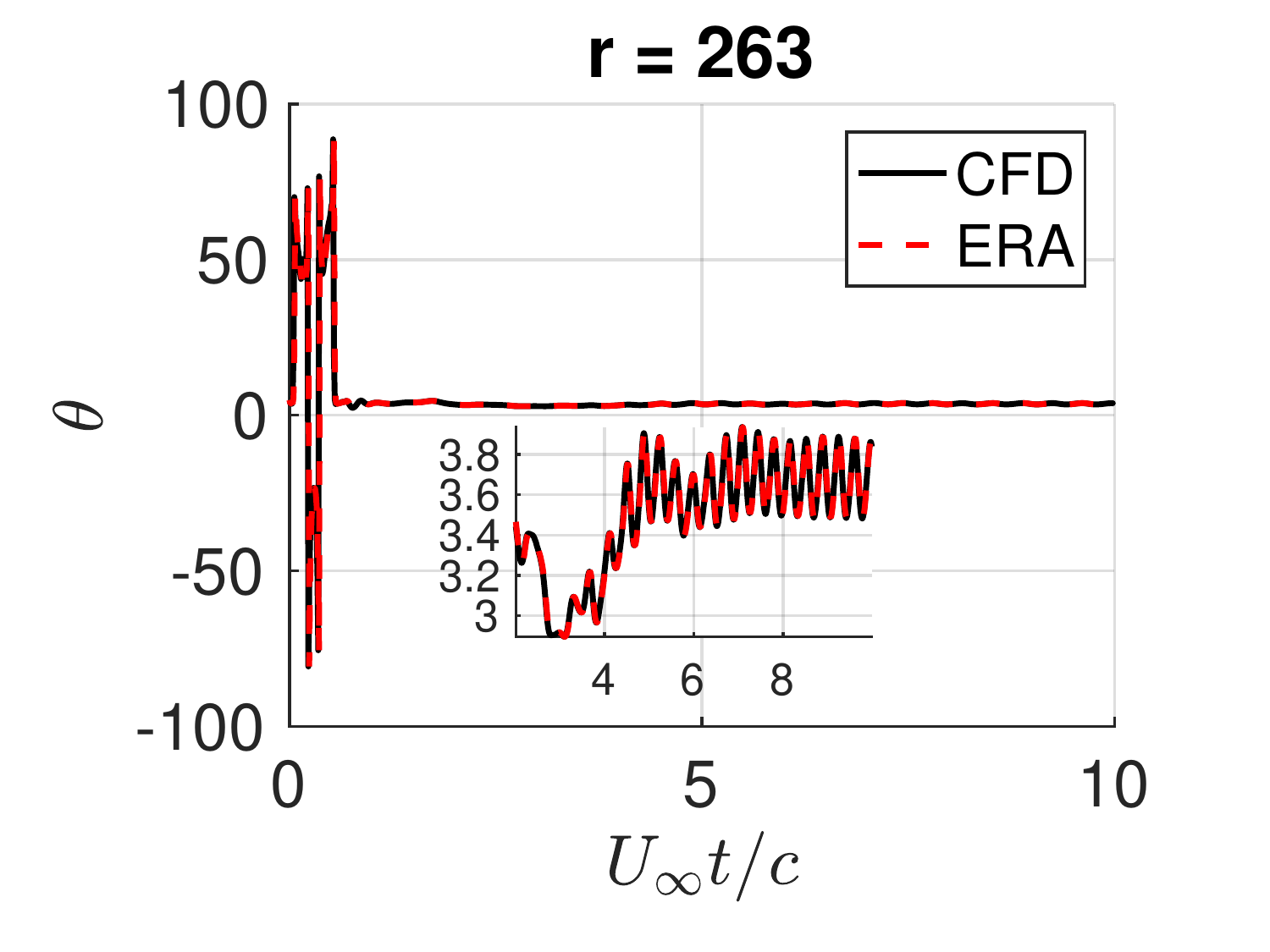}
 }
  \subfloat [ Actuator at $x/c=.4$]{
 \includegraphics[width=.5\textwidth]{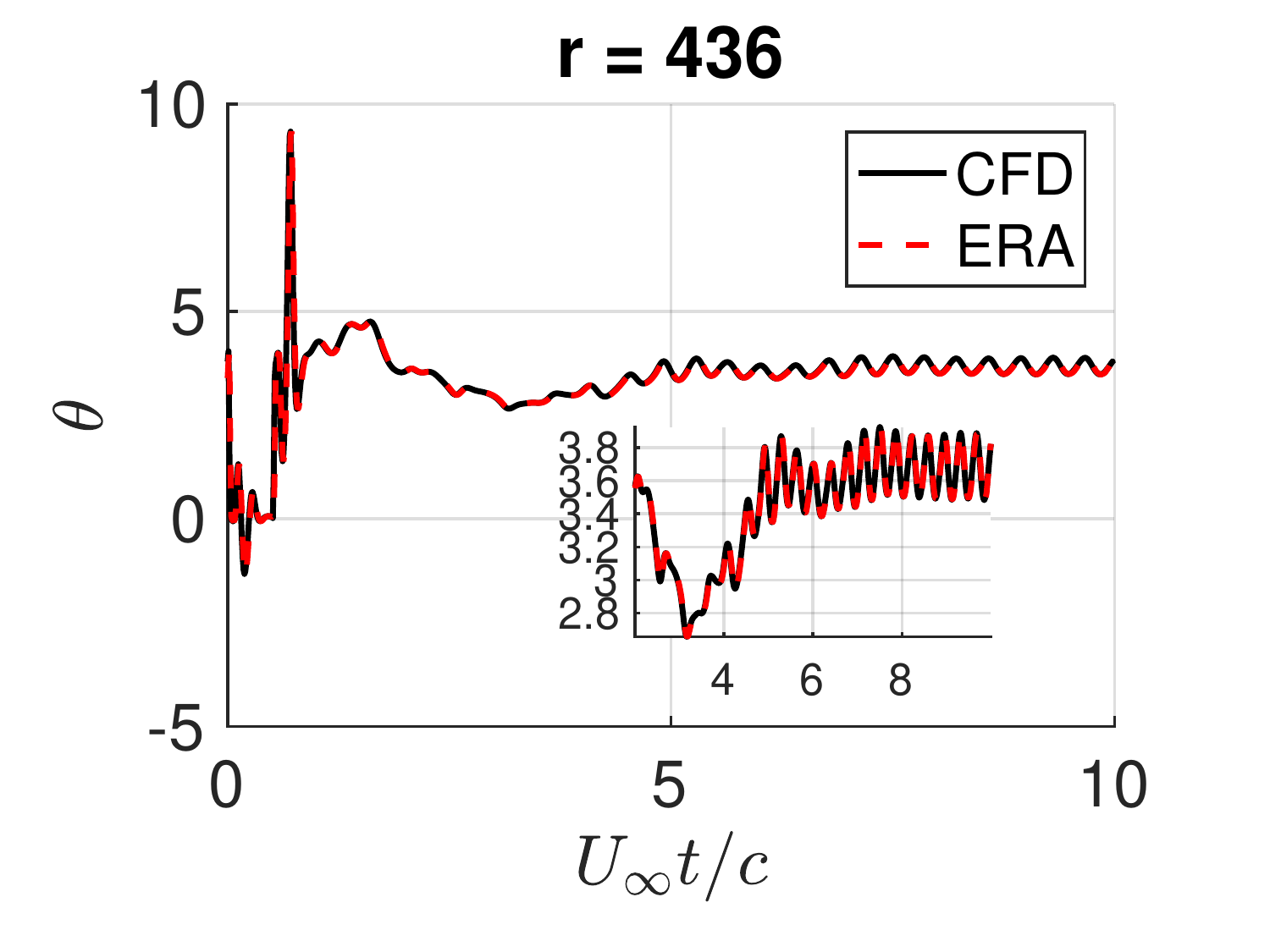}
 }\\
  \subfloat [ Actuator at $x/c=.5$]{
 \includegraphics[width=0.5\textwidth]{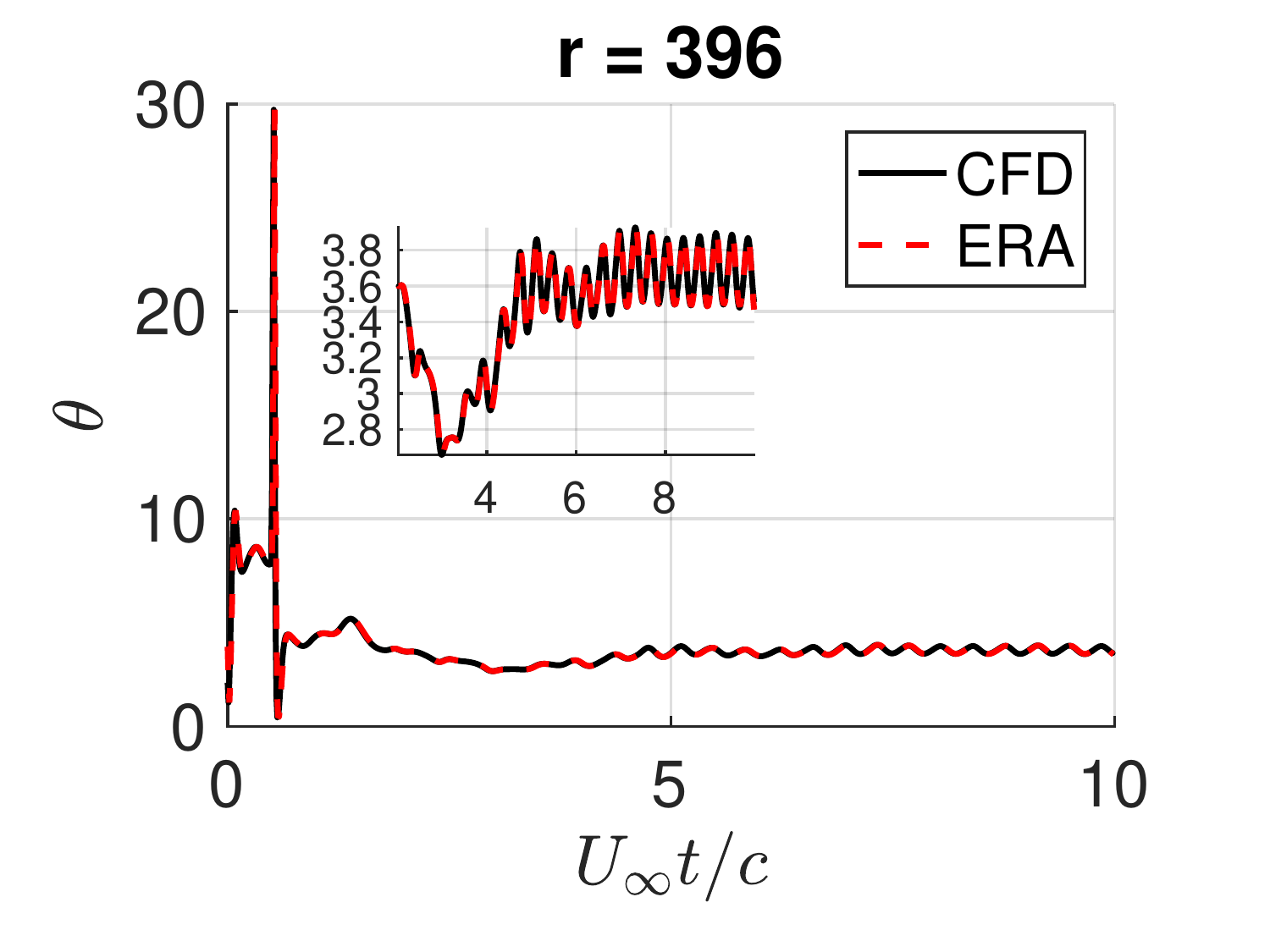}
 }
  \subfloat [ Actuator at $x/c=.6$]{
 \includegraphics[width=0.5\textwidth]{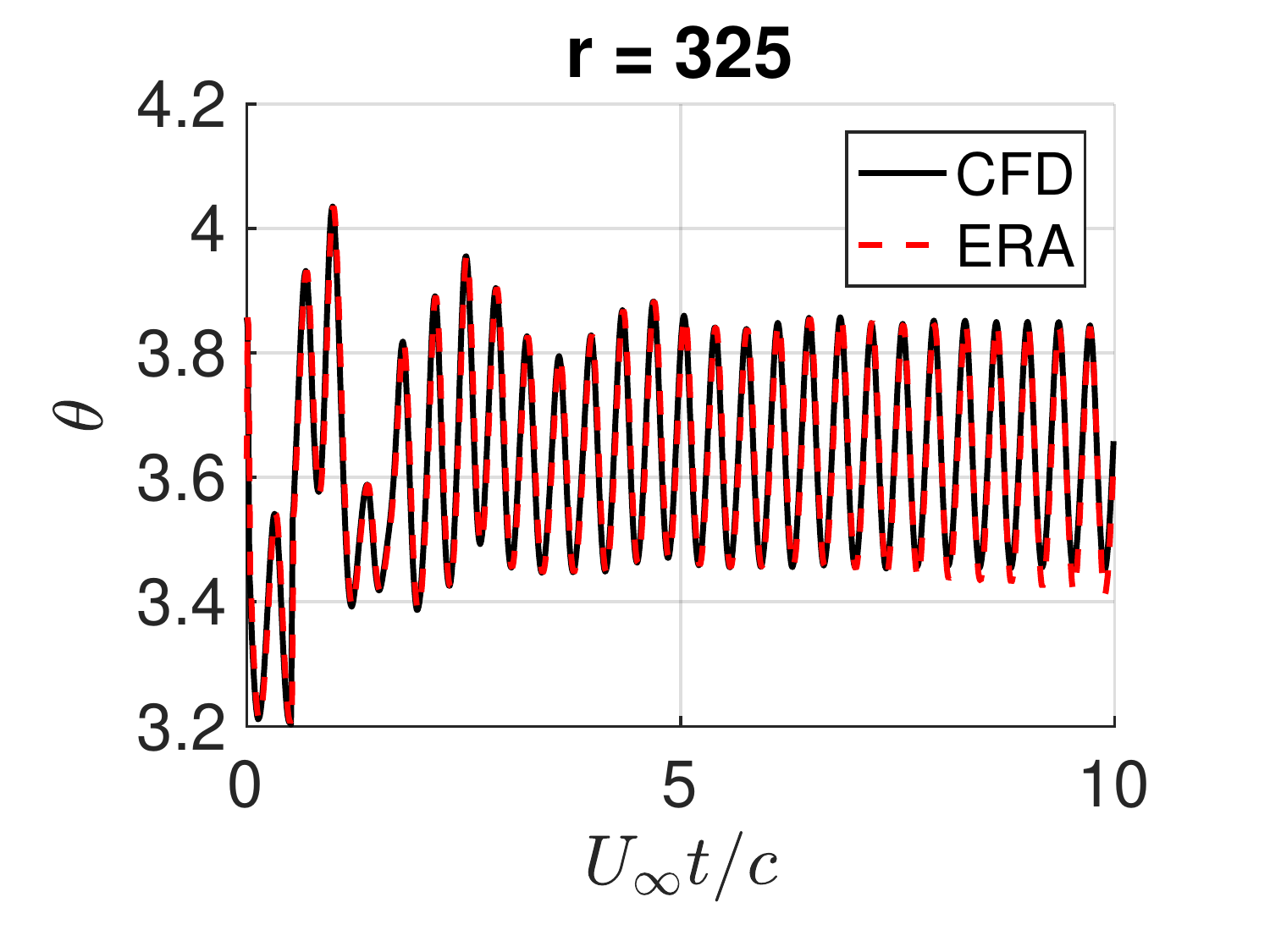}
 }
 \caption{Separation angle pulse response data at each actuator location.  Each realization is minimal with order $r$.}
\label{fig:pulse_Sep}
 \end{center}
\end{figure}

A similar exercise as described above is also undertaken for the separation angle pulse response data. Based on the $\ghh$-norm, the optimal actuator location for separation angle control is $x/c=0.3$ (see Table~\ref{tab:sep_opt_h2}). This location has a degree of controllability which is significantly larger than other locations. The related norm for this actuator location is an order of magnitude greater than the next optimal location. This is also reflected in the very high response peak associated with this location as compared to the other candidate locations (see Figure~\ref{fig:pulse_Sep}). Note that the next optimal location is $x/c = 0.5$,  which coincides with the 
asymptotic
separation point itself~\cite{kamphuis2018pulse}.  

The results provide two interesting observations.  First, the order of the minimal realizations obtained for the separation angle responses are an order of magnitude above the realizations associated with the lift responses. This suggests a greater degree of non-linearity associated with separation angle response than with lift response.  
% Another consequence of this phenomenon can be seen 
Consistently, generalized $\hh$-norms associated with these candidate locations (see Table \ref{tab:sep_opt_h2}) are greater than their lift counterparts. This is especially prominent for the higher ranked locations in the separation angle case.
%No distinct feature is observed from the Bode plots of the system realizations as can be seen in Figure \ref{fig:bode_sep}.
%

% The $\hinf$ norm shows similar behavior as described by the $\ghh$-norm, with actuator location $x/c=0.3$ being the  top contender in all scenarios (see Table~\ref{tab:sep_opt_hinf}).  
%
%
%
\begin{table}[H]
\begin{center}
\begin{tabular}{|c|c|}
\hline
$x/c$ & $\|G\|_{2'}$ \\ 
\hline
.3 & $1.63 \times 10^7$ \\
.5 & $3.99 \times 10^4$ \\
.4 & 1059.57 \\
.1 & 243.24 \\
.6 & 91.95 \\
.2 & 76.74\\
\hline
\end{tabular}
\end{center}
\caption{Optimality of actuator locations based on the generalized $\hh$-norm, sorted from most to least optimal for different tolerance values used in minimal realization for separation angle response data.
%\msh{are the reported numerical values correct?  I don't remember seeing $10^7$ values for norms previously\dots  any idea why the order of magnitude for norms here is so much higher than for the lift case?}
}
\label{tab:sep_opt_h2}
\end{table}

% \begin{table}[ht!]
% \begin{center}
% \begin{tabular}{|c| c | c | c | c | c|}
% \hline
% \multicolumn{2}{|c|}{Tolerance:$10^{-5}$ } &\multicolumn{2}{|c|}{Tolerance:$10^{-6}$  } & \multicolumn{2}{|c|}{Tolerance:$10^{-7}$ } \\
% \hline
% $x/c$ & $\|G\|_\infty$ & $x/c$ & $\|G\|_\infty$ & $x/c$ & $\|G\|_\infty$\\ 
% \hline
% .1 & $1.16 \times 10^4$ & .3 & $1.19 \times 10^4$ & .3 & $1.19 \times 10^4$ \\
% .3 & 1491.20            & .1 & 5393.01            & .1 & 5392.95            \\
% .6 & 451.09             & .5 & 417.56             & .4 & 1480.74            \\
% .5 & 417.20             & .6 & 352.16             & .6 & 443.15             \\
% .4 & 280.01             & .4 & 269.84             & .5 & 418.02             \\
% .2 & 107.49             & .2 & 77.62              & .2 & 77.62  \\
% \hline
% \end{tabular}
% \end{center}
% \caption{Optimality of actuator locations based on $\hinf$-norm, sorted from most to least optimal for different tolerance values used in minimal realization.}
% \label{tab:sep_opt_hinf}
% \end{table}

\begin{figure}[ht!]
\begin{center}
  \subfloat [ Actuator at $x/c=.1$]{
 \includegraphics[width=0.5\textwidth]{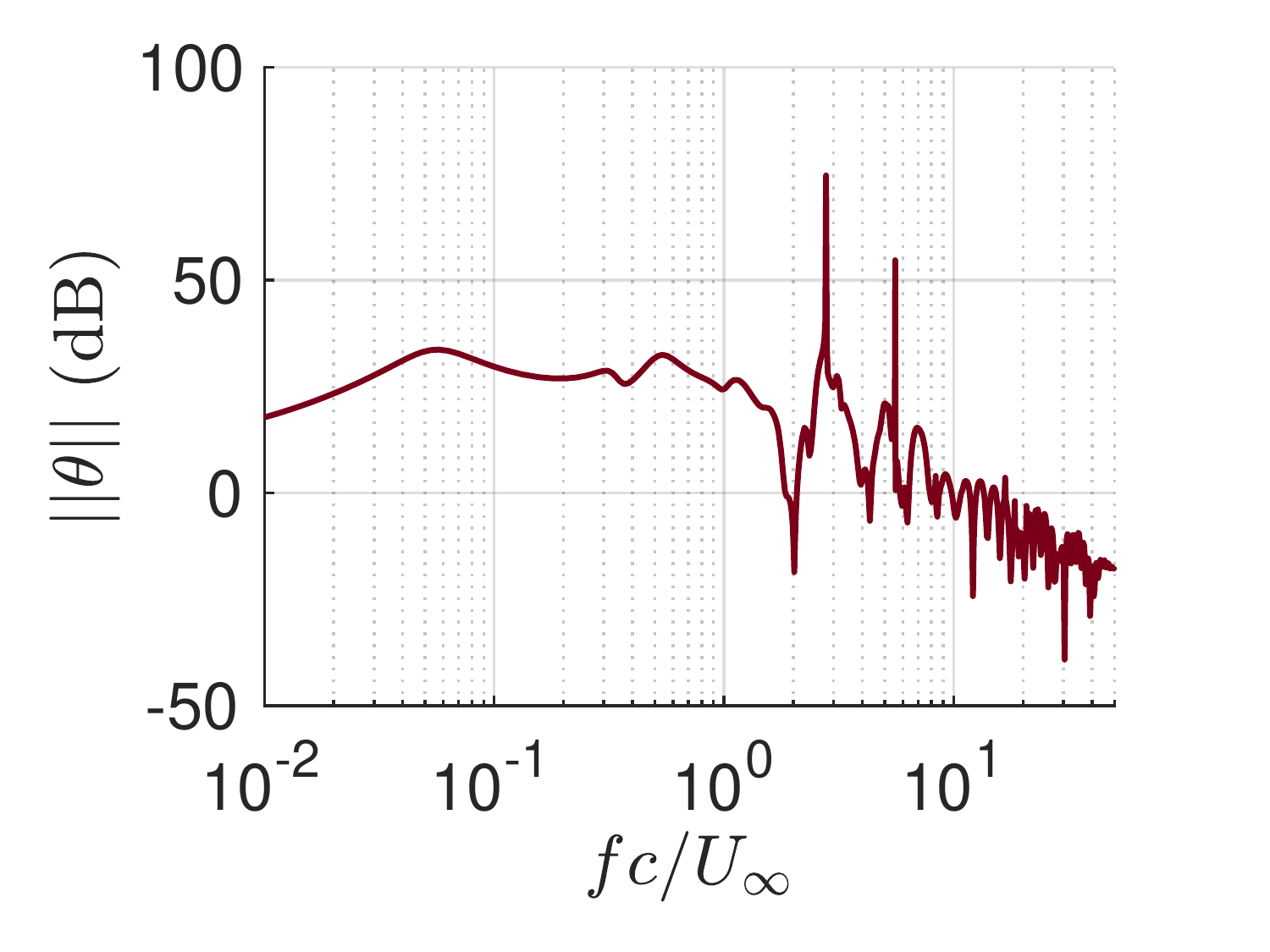}
 }  
  \subfloat [ Actuator at $x/c=.2$]{
 \includegraphics[width=0.5\textwidth]{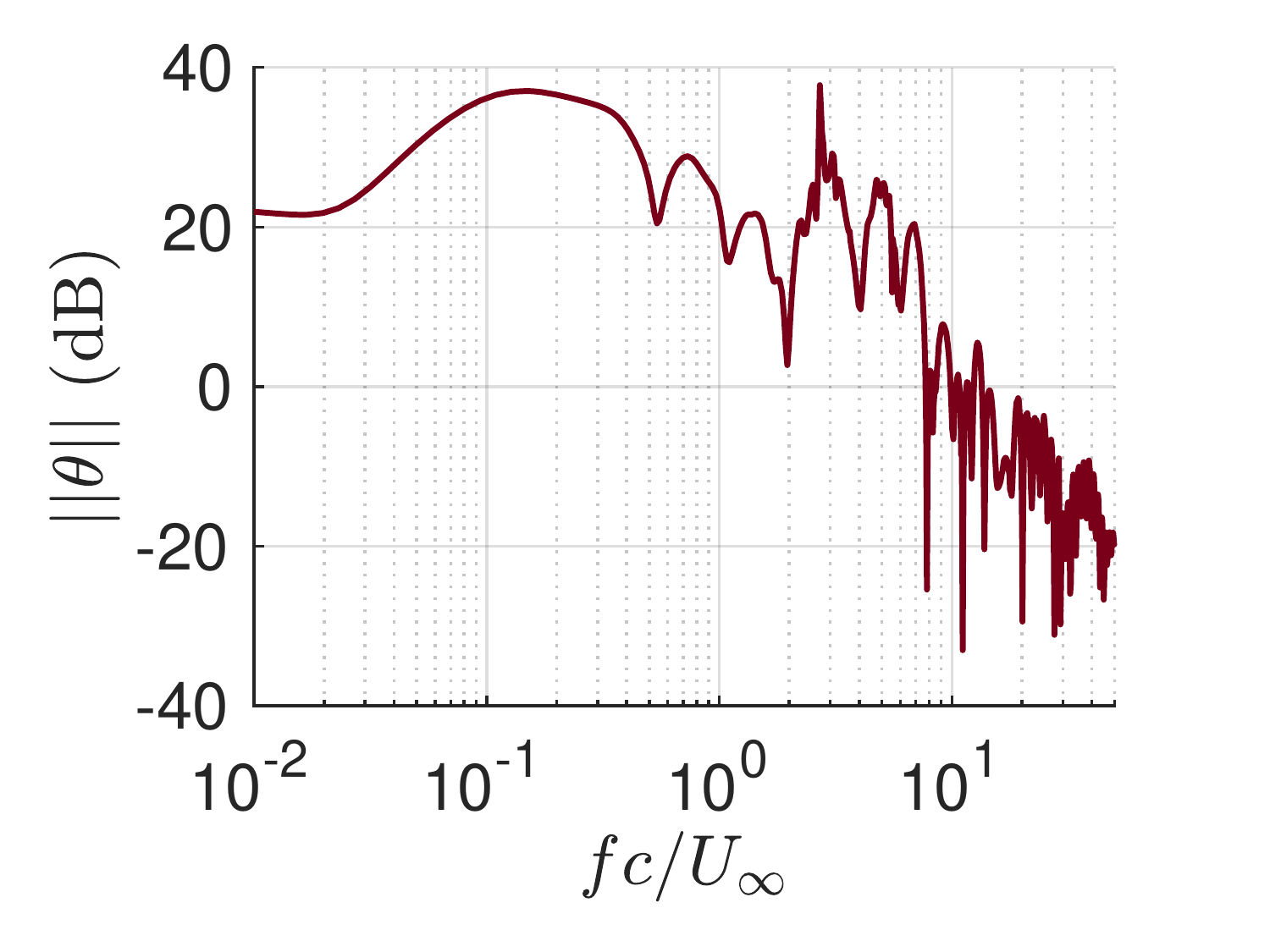}
 }\\
  \subfloat [ Actuator at $x/c=.3$]{
 \includegraphics[width=0.5\textwidth]{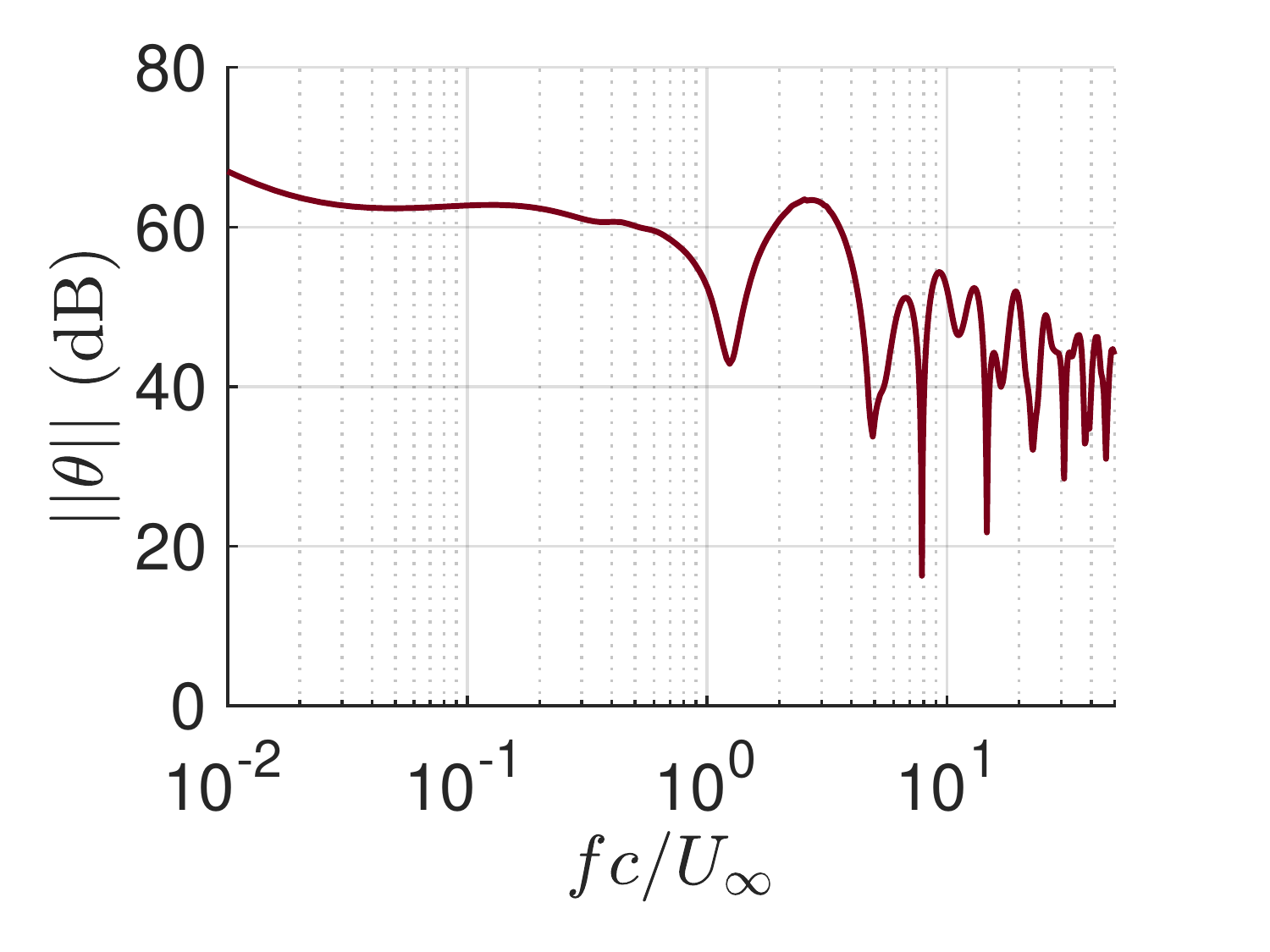}
 }
  \subfloat [ Actuator at $x/c=.4$]{
 \includegraphics[width=0.5\textwidth]{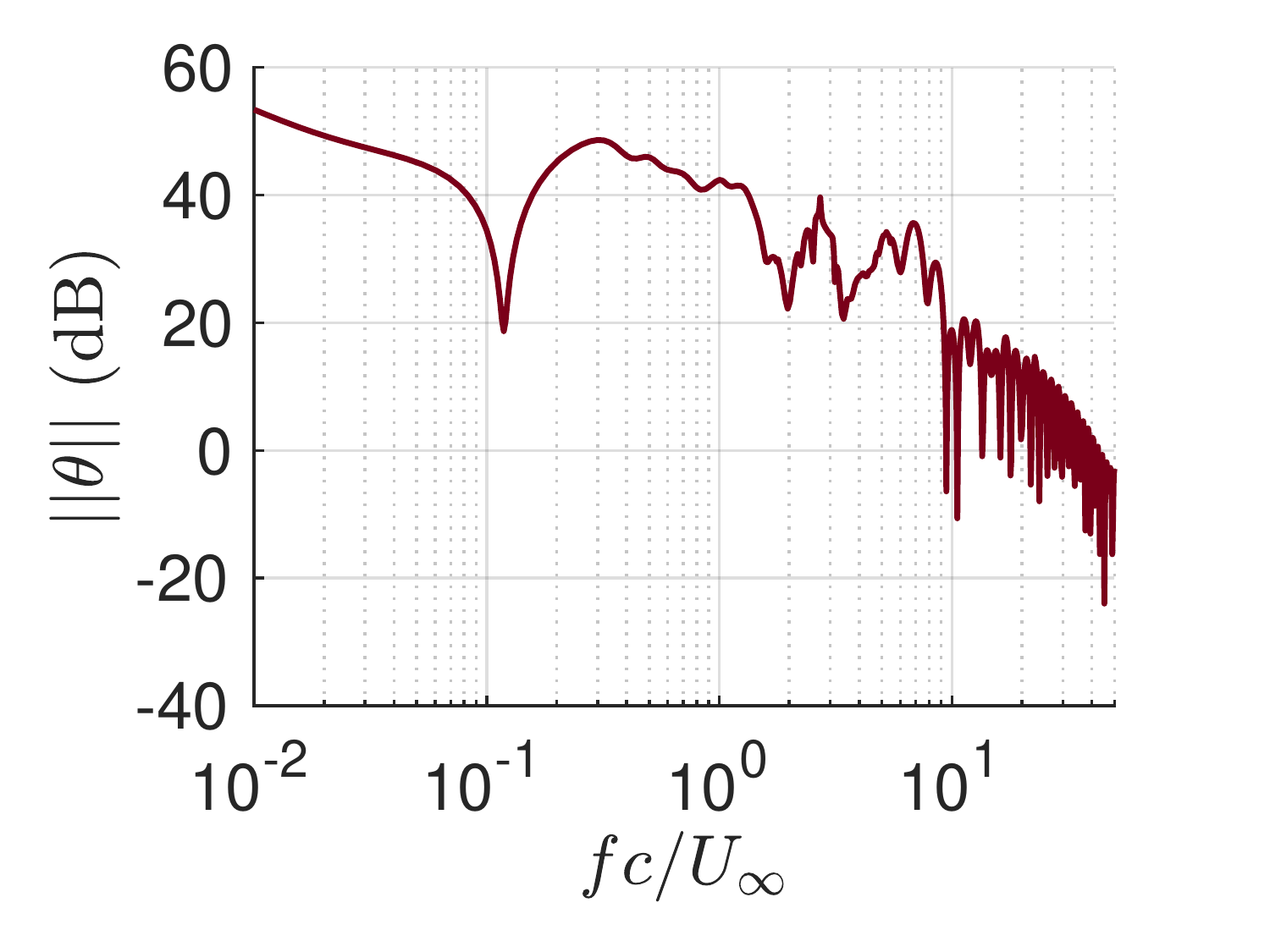}
 }\\
  \subfloat [ Actuator at $x/c=.5$]{
 \includegraphics[width=0.5\textwidth]{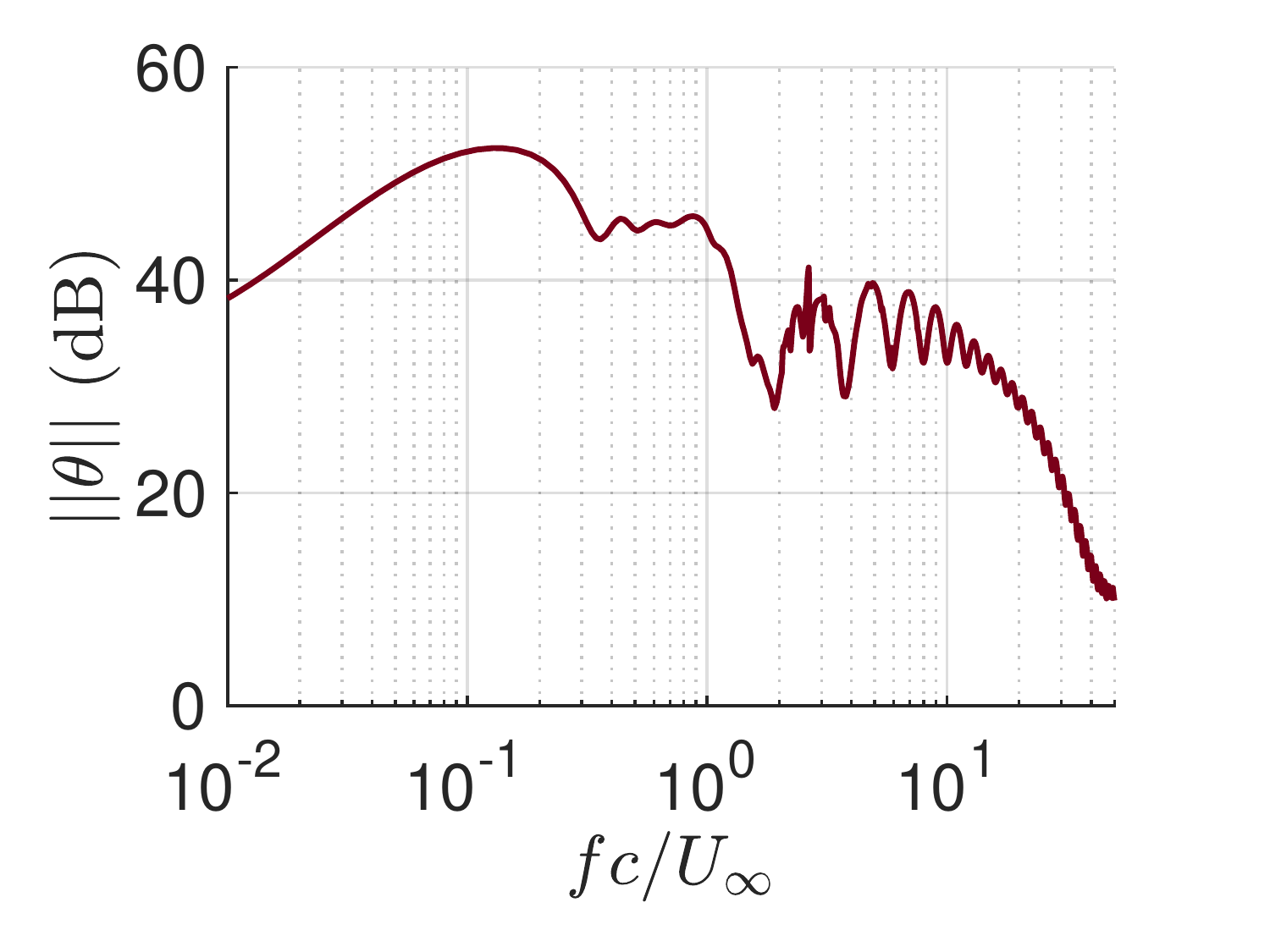}
 }
  \subfloat [ Actuator at $x/c=.6$]{
 \includegraphics[width=0.5\textwidth]{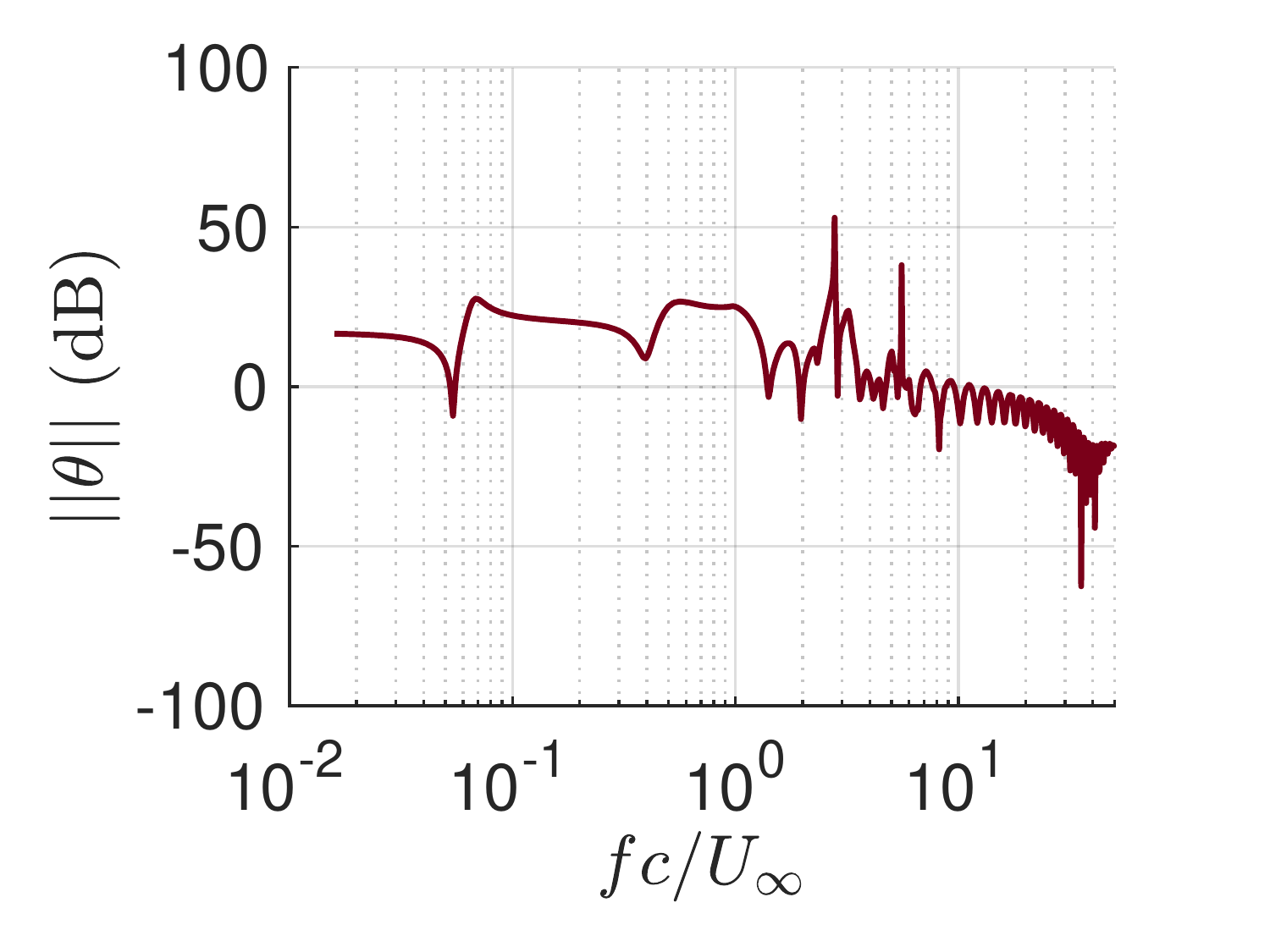}
 }
\caption{Bode magnitude plot for minimal realization at each actuator location for the separation angle response.}
\label{fig:bode_sep}
 \end{center}
\end{figure}

\clearpage

%%%%%%%%%%%% Modal Controllability Results
\subsection{Modal Analysis of the Flowfield Response}
\label{sec:modal}
% In the analysis above, we have found that the optimal actuator location differs for lift control versus separation-angle control.
% %
% %Further, the degree of controllability---as indicated by the $\ghh$-norm---varied significantly between these cases.
% %
% Here, 
To understand \emph{why} the actuator locations for controlling separation
angle and lift are optimal, we perform modal analysis of
the flowfield and attempt to understand the underlying physical
mechanisms.

We leverage the dynamic mode
decomposition~(DMD) to  extract dynamically meaningful
spatio-temporal information from snapshot data of the dynamic response of the
flowfield~\cite{schmid2010dynamic,rowleyJFM2009,tu2014}.
In particular, we use the DMD with control~(DMDc)
algorithm~\cite{proctor2016dynamic} in order to properly account
for the influence of external forcing on the flow from actuation.
DMDc is a data-driven method that is closely related to ERA~\cite{proctor2016dynamic};
however, DMDc assumes access to the full-state output, whereas ERA does not.
Further, DMDc requires additional care when data are gathered from physical experiments,
as measurement noise can introduce bias errors that must be taken into account~\cite{hemati2017biasing,dawsonEXIF2016,askham2018,nonomura2018}.

\begin{figure}[ht!]
\begin{center}
   \subfloat [$x/c=.2$]{
 \includegraphics[width=0.5\textwidth]{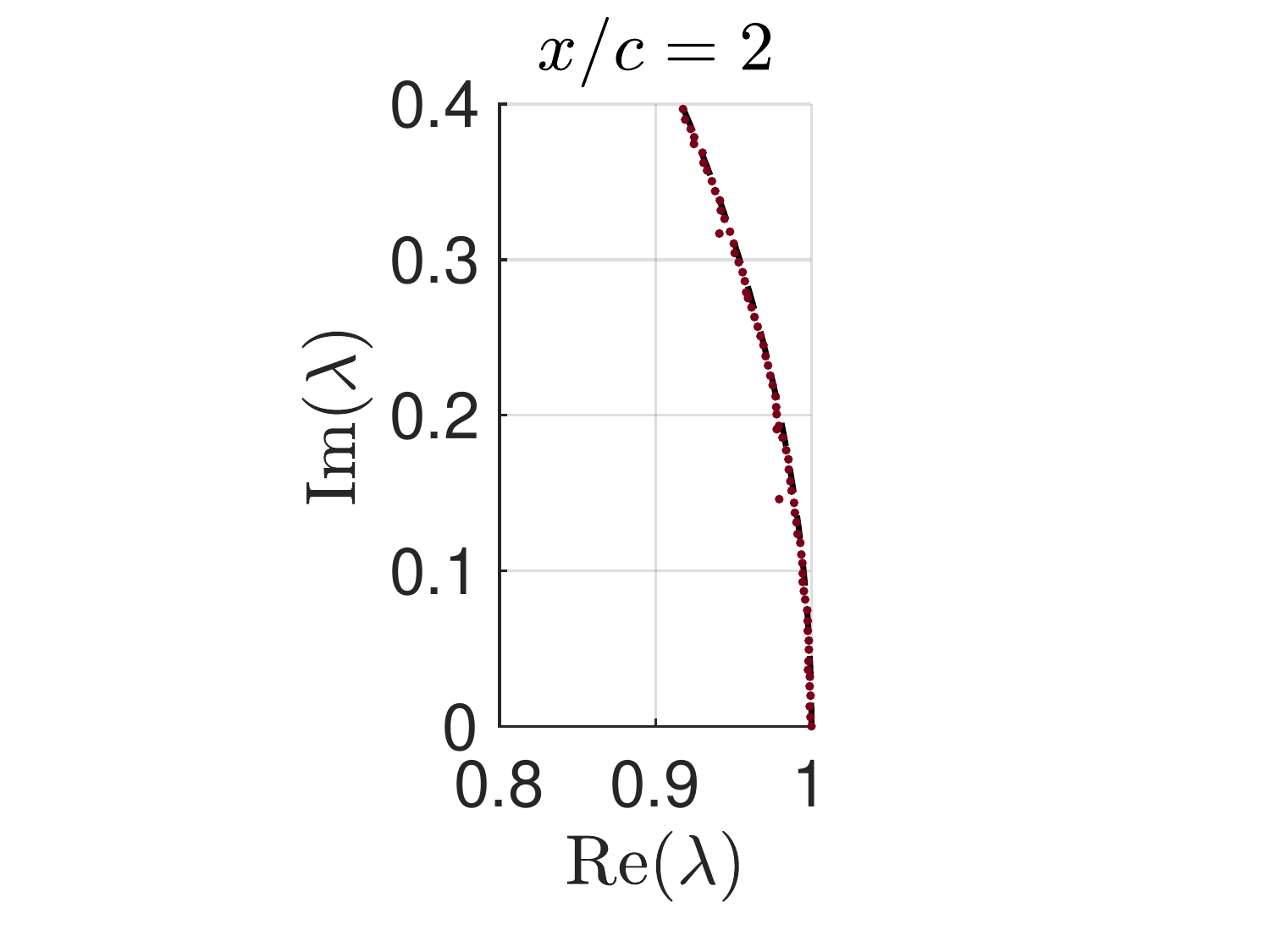}
 }  
  \subfloat [$x/c=.3$]{
 \includegraphics[width=0.5\textwidth]{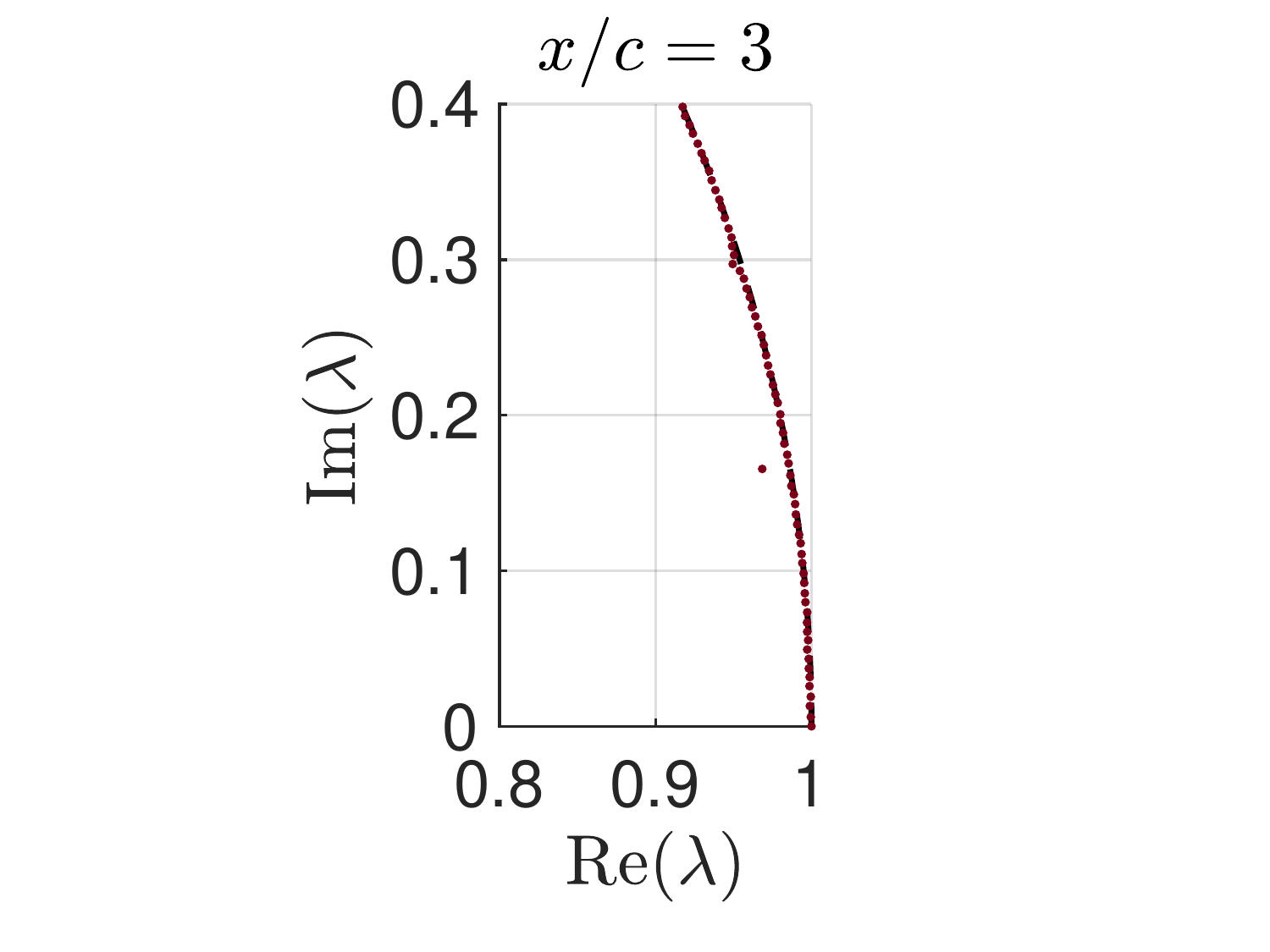}
 }
 \caption{DMD Eigenvalues for actuator locations $x/c=0.2$ and $x/c=0.3$}
\label{fig:dmdc_evals}
 \end{center}
\end{figure}

At its heart, DMDc approximates flow response data with a dynamical system of the form,
\begin{equation}
  x(k+1) = \hat{A}x(k) + \hat{B}u(k).
\end{equation}
Here, we take $x(k)\in\Re^n$ to be a snapshot of the velocity field at time-step $k$
and $u(k)\in\Re$ as the associated input.
For a unit pulse of body-force actuation applied at
a single location on the airfoil at $k=0$,
this corresponds to $u(0)=1$ and  $u(k)=0$ for $k\ge 1$.
Response data are collected and stored in
data matrices of state and input sequences,
\begin{align}
 X &=  \begin{bmatrix}
                     x(1) &  x(2) & \cdots & x(m)  
         \end{bmatrix} \\\label{eq:dmdcX}
 X' &=  \begin{bmatrix}
                      x(2) &  x(3) & \cdots & x(m+1)           \end{bmatrix} \\
 \Upsilon &= \begin{bmatrix}
                        u(1) &  u(2) & \cdots & u(m)           \end{bmatrix}.\label{eq:dmdcUpsilon}
 \end{align}
Then, DMDc approximates the underlying system dynamics $(\hat{A},\hat{B})$ as a least-squares/minimum-norm solution to $X'\approx \hat{A}X + \hat{B}\Upsilon$~\cite{proctor2016dynamic}.
Specifically,
\begin{equation}
\begin{bmatrix}\hat{A}&\hat{B}\end{bmatrix} \approx X'\underbrace{\begin{bmatrix}X\\ \Upsilon\end{bmatrix}}_{\Omega}=\begin{bmatrix}\underbrace{X'\tilde{V}\tilde{\Sigma}^{-1}\tilde{U}_1\trans}_{\bar{A}}&\underbrace{X'\tilde{V}\tilde{\Sigma}^{-1}\tilde{U}_2\trans}_{\bar{B}}\end{bmatrix},
\end{equation}
where the truncated SVD gives a rank-$p$ approximation of $\Omega\approx\tilde{U}\tilde{\Sigma}\tilde{V}\trans$, $\tilde{U}\trans=\begin{bmatrix}\tilde{U}_1\trans & \tilde{U}_2\trans\end{bmatrix}$, $\tilde{U}_1\in\Re^{n\times p}$, $\tilde{U}_2\in\Re^{1\times p}$, $\bar{A}\approx\hat{A}$, and $\bar{B}\approx\hat{B}$.
% DMDc identifies a system model in the form of~\eqref{eqn:dt_lti} from snapshot data $\{x(1),x(2),\dots,x(m+1)\}$ and associated input data $\{u(1),u(2),\dots,u(m)\}$, where $y(k)=x(k)$ is taken to be the full-state output.
% %
% Note that this is in contrast to the eigensystem realization algorithm~(ERA) used for optimal actuator selection, in which full-state information was \emph{not} available for system identification.
%
%  The problem now amounts to solving
% \begin{align*}
%   \argmin_{\hat{A},\hat{B}} \left\|X_2 - \begin{bmatrix}\Tilde{A}&  \Tilde{B}\end{bmatrix}\begin{bmatrix}
% 	X_1 \\  \Upsilon
%  	\end{bmatrix}\right\|_F
%  \end{align*}
%  where $\hat{A}\in\Re^{n\times n}$, $\hat{B}\in\Re^{n\times 1}$.
%  %
 % In fluid applications, the state dimension $n$ tends to be much larger than the number of snapshots $m$.
 % %
Since $n$ is large in fluids applications, DMDc works with a reduced-order representation of the dynamics,
\begin{equation}
  \tilde{x}(k+1) = \tilde{A}\tilde{x}(k) + \tilde{B}u(k),
\end{equation}
where $x=\hat{U}\tilde{x}$, $\tilde{A}=\hat{U}\trans \bar{A}\hat{U}$, $\tilde{B}=\hat{U}\trans \bar{B}$, and $\hat{U}\in\Re^{n\times r}$ is determined from a rank-$r$ approximation of $X'$ computed via the truncated SVD of $X'\approx \hat{U}\hat{\Sigma}\hat{V}\trans$.
It follows that the eigenvectors $v$ and
eigenvalues $\lambda$ of $\tilde{A}$ are related to
the eigenvectors $\phi$ (DMD modes) and
eigenvalues $\lambda$ (DMD eigenvalues) of
$\bar{A}$~\cite{proctor2016dynamic}.
It is also possible to relate the left-eigenvectors $w$ of $\tilde{A}$ to the left eigenvectors $\psi$ (adjoint DMD modes) of $\bar{A}$, as noted in~\cite{tu2014} and~\cite{zhang2017}.
The DMDc algorithm is summarized as,
\begin{enumerate}
  \item Collect data and form the relevant data matrices $X$, $X'$, and $\Upsilon$ defined in Equations \eqref{eq:dmdcX}--\eqref{eq:dmdcUpsilon}, respectively.
 \item Compute the rank-$p$ truncated SVD
   \begin{equation}
     \Omega = \begin{bmatrix}
	X \\  \Upsilon
 	\end{bmatrix} \approx \tilde{U}\tilde{\Sigma}\tilde{V}^*.
      \end{equation}
    \item Compute the rank-$r$ truncated SVD $X'\approx \hat{U}\hat{\Sigma}\hat{V}^*$, where $r<p$.
    \item Compute the reduced-order system realization
      \begin{align}
        \tilde{A} &= \hat{U}^*X'\tilde{V}\tilde{\Sigma}^{-1}\tilde{U}_1^*\hat{U}\\
        \tilde{B} &= \hat{U}^*X'\tilde{V}\tilde{\Sigma}^{-1}\tilde{U}_2
      \end{align}
    \item Compute the eigendecompositions $\tilde{A}v_i=\lambda_i v_i$ and $\tilde{A}\trans w_i=\lambda_i w_i$.  The DMD eigenvalues $\lambda_i$  can be used to determine the associated modal frequencies $\angle \lambda_i/(2\pi\delta t)$ and growth/decay rates $\log |\lambda_i|/\delta t$ , where $\delta t$ is the sampling time. The DMD mode corresponding to each DMD eigenvalue $\lambda_i$ is computed as $\phi_i = X'\tilde{V}\tilde{\Sigma}^{-1}\tilde{U}_1^*\hat{U}v_i$.
    \end{enumerate}   
Further details about DMDc can be found in~\cite{proctor2016dynamic}.
Performing DMDc on velocity field response data due to pulse actuation applied at $x/c=0.2$ and $0.3$ yields realizations $(\tilde{A},\tilde{B})$ of order $r = 300$.
Recall, these locations are found to be optimal for lift and separation angle control, respectively.
DMD eigenvalues for actuator
locations $x/c=0.2$ and $x/c=0.3$ are shown in Figure~\ref{fig:dmdc_evals}.
%
% Using these number of modes retained $95\%$ of the energy of the Singular values for the candidate locations.
%The following two subsections describe methods of modal analysis which can be utilized to develop effective control strategies.
%

We note that $w_i\trans \tilde{B}$ provides information about the controllability of the DMD mode  $\phi_i$.
This
%is fact
%\mshdb{, in fact, is}
, in fact, is
closely related to the well-established modal controllalibity test of Popov, Belevitch, and Hautus~(PBH)~\cite{brogan_modern}.
Here, we invoke a measure of modal controllability for linear systems proposed in~\cite{hamdan_measures},
\begin{equation}
  \gamma_i=\frac{|w_i\trans \tilde{B}|}{\|w_i\|\|\tilde{B}\|}
  \label{eqn:hamdan_measures}
\end{equation}
where $\tilde{B}$ is a column vector in this study because there is only one input channel per realization.
Then, the measure $\gamma_i$ corresponds to the cosine of the (acute) angle between the two one-dimensional subspaces defined by $w_i$ and $B$.
If the two subspaces are orthogonal,
then $\gamma_i=0$, indicating that
the DMD mode $\phi_i$ is uncontrollable
from the input.
On the other hand, if the two subspaces are perfectly aligned, then $\gamma_i=1$, indicating that DMD mode $\phi_i$ is maximally controllable.
We note that for multi-input systems, one must consider modal controllability from each available input channel.  In such instances, a measure of gross modal controllability can be defined to account for the relative norms of columns in $\tilde{B}$.
Further details can be found in~\cite{hamdan_measures}.

\begin{figure}[h!]
\begin{center}
% 
     %  \subfloat [$fc/U_{\infty}$ = 0, $\theta_{m1}$ = 0.2891]{
     % \includegraphics[width=0.5\textwidth]{Modal/ctrb_2/Act_2_Cont_1}
     % }  
      \subfloat [$x/c=0.2$, $fc/U_{\infty}$ = 2.3560, $\gamma$ = 0.2786]{
     \includegraphics[width=0.9\textwidth]{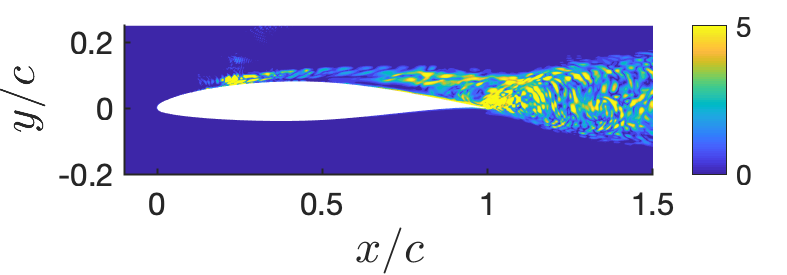}
   }\\
   \subfloat [$x/c=0.3$, $fc/U_{\infty}$ = 2.6935, $\gamma$ = 0.3598]{
     \includegraphics[width=0.9\textwidth]{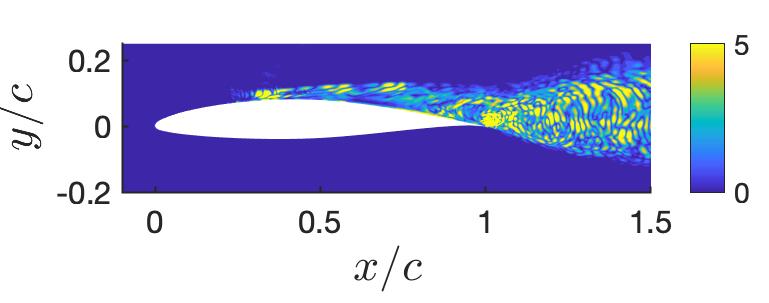}
     }
     
\caption{The magnitude of the most controllable DMD mode associated with actuation at $x/c=0.2$ and $x/c=0.3$, visualized using vorticity.}
\label{fig:act2}
\end{center}
\end{figure}

Using this procedure, we sort
DMD modes according to their
relative controllability measures.
The magnitude of the most controllable DMD modes for actuator locations $x/c=0.2$ and $x/c=0.3$
are plotted in Figure \ref{fig:act2}. % and \ref{fig:act3}.
For $x/c=0.2$---the optimal location for lift control---the most controllable DMD mode is strongly active within the separation bubble and into the wake near the trailing edge of the airfoil.
This suggests that the separation bubble and near-wake are most receptive to actuation that benefits lift control.
In contrast, the most controllable DMD mode for $x/c=0.3$---optimal for separation angle control---is most active in the shear layer and in the wake.
Indeed, this mode shows evidence of vortical structures within the shear layer
that are effectively manipulated via control at $x/c=0.3$.
This observation suggests that vortex roll-up within the shear layer provides
a mechanism that benefits separation angle control.

Since DMD modes are single-frequency flow structures,
the frequencies associated with the most controllable DMD modes
may serve as good candidate frequencies for open-loop control using sinusoidal forcing.
Further, we note that the frequencies associated with the most controllable DMD modes
are consistent with the peaks in the separation angle frequency response determined via ERA (see Figure~\ref{fig:bode_sep}).
This finding supports our earlier claim that controlling the separation angle
may prove to be more effective than controlling lift directly.
%
%determined from ERA.
% The most controllable DMD modes associated with actuator
% location $x/c = .3$ are found to be significantly more controllable than the DMD modes associated with actuator location $x/c = .2$.
%
% \msh{is this correct? not really an indication of one observable v another\dots}
% These results are consistent with our earlier findings from the optimal actuator selection analysis: separation angle is more sensitive to control inputs than lift.
%, suggesting that is important variable when compared with lift in terms of formulating active control strategies for flow separation and reinforces what was mentioned in \cite{bhattacharjee_optimal}.
%
% As has been mentioned in Section \ref{sec: cont_dyn}, this analysis can be particularly helpful while developing closed-loop control strategies where specific frequencies in the flow need to be targeted. The control input matrix, B, can be designed appropriately to dispense higher controllability to a particular dynamic mode.
%

%% Power analogy

%%%%%%%%%%%%%%%%%%%%%%%%%%%

% \begin{figure}[h!]
% \begin{center}
% % 
%      %  \subfloat [$fc/U_{\infty}$ = 0, $\theta_{m1}$ = 0.4520]{
%      % \includegraphics[width=0.5\textwidth]{Modal/ctrb_3/Act_3_Cont_1}
%      % }  
% %      \subfloat [$fc/U_{\infty}$ = 2.6935, $\gamma$ = 0.3598]{
%      \includegraphics[width=0.9\textwidth]{Modal/ctrb_3/Act_3_Cont_2}
% %     }
     
% \caption{The magnitude of the most controllable DMD mode associated with actuation at $x/c=0.3$, visualized using vorticity. ($fc/U_{\infty}$ = 2.6935, $\gamma$ = 0.3598)}
% \label{fig:act3}
% \end{center}
% \end{figure}

Next, we leverage the generalized controllability Gramian~$P$ to determine the most
controllable directions in state-space.
To do so, we first transform the discrete-time DMDc system realization $(\tilde{A},\tilde{B})$ to the associated continuous-time realization, then compute 
the generalized controllability Gramian directly
from Eq.~\eqref{eq:gengram1} and \eqref{eq:gengram2}.
The principal directions of~$P$ can be used to reveal the
flow structures that are most sensitive to control action.
In particular, the most controllable flow structures are associated with the one-dimensional subspace spanned by
\begin{equation}
    \xi_P = X'\tilde{V}\tilde{\Sigma}^{-1}\tilde{U}_1\trans\hat{U}v_P,
    \label{eqn: flow_structure}
\end{equation}
where $v_P$ is the eigendirection associated with the largest eigenvalue of $P$.

\begin{figure}[h!]
\begin{center}
      \subfloat [$x/c=0.2$]{ %, $\lambda_P$ = 855.3]{
     \includegraphics[width=0.9\textwidth]{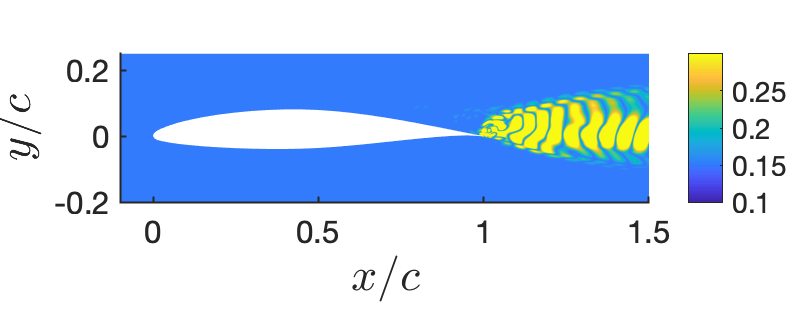}
      }\\
           \subfloat [$x/c=0.3$]{ %, $\lambda_P$ = 796.9]{
     \includegraphics[width=0.9\textwidth]{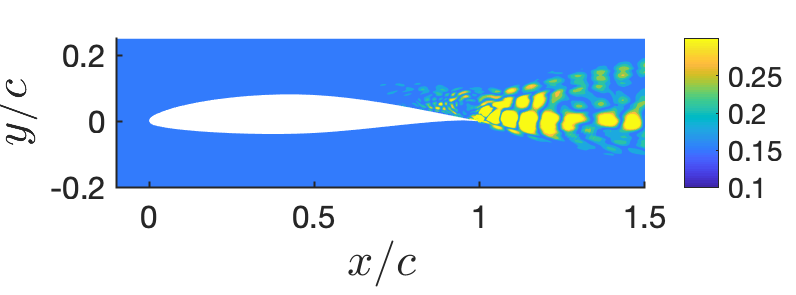}
     }  
     %  \subfloat [$\sigma$ = 819.89]{
     % \includegraphics[width=0.5\textwidth]{DMDc/ctrb_2/Act_2_Cont_2}
     % }
     
\caption{The magnitude of the first principal direction of the generalized controllability Gramian associated with actuation at $x/c=0.2$ and $x/c=0.3$, visualized using vorticity.}
\label{fig: dyn_act2}
\end{center}
\end{figure}

The most controllable flow structures for actuator location $x/c=0.2$ and $x/c=0.3$ are shown in Figure~\ref{fig: dyn_act2}.
Unlike the most controllable DMD modes, the most controllable flow structures
identified by this Gramian-based analysis are not associated with
just a single-frequency;
rather, these structures can exhibit rich dynamics that are associated with evolution along the most controllable direction in state-space.
As such, the controllable subspace reveals a different
description of control mechanisms than the modal controllability analysis.
The optimal actuator location for lift control appears to activate vortex shedding in the wake, starting immediately at the trailing edge of the airfoil.
This is consistent with the modal controllability analysis for actuation at $x/c=0.2$.
The fact that the wake is most sensitive to actuation at $x/c=0.2$ is also consistent with physical intuition,
since the transfer of bound vorticity into free vortcity in the wake is the
physical mechanism for lift production.
For separation angle control, the Gramian-based analysis
reveals complex dynamics between the rear of the shear layer, the separation bubble, and the near-wake.
The fact gives a slightly different picture than what was observed in the modal controllability analysis.
These differences suggest that the dynamics governing the separation angle
response are highly nonlinear compared to the dynamics governing
the lift response.
Further, both the modal controllability analysis and
this Gramian-based analaysis suggests that actuation at $x/c = 0.3$
can make regions in the shear layer more controllable.
As pointed out by several other studies, the shear layer plays
a key role in the optimal control of fluid-flows. Several studies
have suggested actuating the flow at the shear layer frequency for
better control of coherent structures. The results presented here
provide evidence to believe that the actuator placed at $x/c = 0.3$ is
able to excite the fluid flow more effectively in these regions and hence
provide greater controllability.

\section{Conclusion}
\label{sec:conclusion}
%\msh{clean up}

%In this work, we presented a data-driven approach for optimal actuator selection.
%
We presented a data-driven approach for determining the actuator location
requiring the minimum control energy to drive an output quantity-of-interest.
Given input-output response data for a candidate set of actuator locations,
the eigensystem realization algorithm was used to extract state-space system descriptions
suitable for solving a minimum input energy optimal control problem and computing the generalized $\hh$-norm for each location.
The method only requires access to input-output response data, making it relevant for numerical and experimental studies alike.
%The the generalized $\hh$-norm for each system realization was computed to determine the actuator location requiring the minimum control energy to drive the output quantity-of-interest.
%
%
%This is desirable as system models for large-scale complex systems are often unavailable.
%
The method was used to investigate the optimal actuator location for
airfoil separation control using data from high-fidelity numerical simulations of
a NACA 65(1)-412 airfoil, with $\alpha=\ang{4}$ and $Re_c=20,000$.
% the airfoil separation control problem is cast in the framework of optimal actuator selection.   
Lift and separation angle response data to a pulse of localized body force actuation were used to determine the optimal location among a candidate set of six locations on the upper surface of the airfoil. 
%
% The optimality measure is related to the minimum energy required to drive the system output to an arbitrary value. Minimal realizations computed from pulse response data were found to exhibit unstable behavior. Therefore, we proposed a formulation of the optimal actuator selection problem to accommodate stable and unstable systems alike.  
% %
% For the optimality procedure to be insensitive to algorithmic parameters, several precautions were presented. This resulted in a consistent yield in terms of the rankings of the various actuator locations. 
% For the NACA 65(1)-412 airfoil,
It was found that the location $x/c = 0.2$ was optimal for controlling lift, whereas the location $x/c =0.3$ was found to be optimal for controlling separation angle. 
The analysis also revealed separation angle to be more sensitive than lift to
actuation from the associated optimal location,
making separation angle the more attractive quantity
to regulate in separation control applications.

In order to identify physical mechanisms underlying these results,
we presented a data-driven framework for conducting controllability analysis of
the flowfield using dynamic mode decomposition with control~(DMDc).
A controllability analysis of the dominant single-frequency DMD modes
confirmed greater controllability for the actuator placed at $x/c =0.3$,
which was the optimal location for separation angle control.
Actuation from this location was found to excited flow structures within the shear layer,
corroborating previous findings on the effectiveness of shear layer
excitation for separation control.
A complementary analysis of the controllable subspaces in the flowfield dynamics
confirmed that coherent structures in shear layer were most sensitive to
actuation applied at the optimal location for separation control ($x/c=0.3$).
In contrast, coherent structures in the wake were most sensitive to actuation applied at the optimal location for lift control ($x/c=0.2$).

The methods introduced in this paper are generally applicable for optimal actuator selection and controllability analysis.

A distinctive feature of the proposed optimal actuator selection method is that it is entirely data-driven.
The approach does not require access to primal or adjoint simulations,
which are often required to conduct similar analyses.
This makes for a convenient analysis procedure that can be used to objectively assess the optimal actuator location from available or easy-to-acquire response data.
Further, the data-driven nature of the method also makes it generally applicable, and should benefit investigations of other flow control configurations as well.

\section*{Acknowledgements}
This material is based upon work supported by the Air Force Office of Scientific Research under awards
FA9550-16-1-0392, FA9550-17-1-0252, and FA9550-19-1-0034 monitored by Drs.~Douglas~R. Smith and Gregg Abate.
The authors thank Dr.~Kevin~K. Chen for initial discussions related to optimal actuator selection.
%
% An earlier version of this work was presented as AIAA Paper 2018-3692 at the AIAA Flow Control Conference in 2018~\cite{bhattacharjee_optimal}.

%\vspace{-.35in}
%===== REFERENCES =====
\bibliographystyle{aiaa}
\bibliography{mainTCFD}

\end{document}